\documentclass[12pt]{iopart}
\usepackage{amssymb}
\usepackage{bm}
\usepackage{epsfig}
\usepackage{color}

\newcommand{\eps}{\varepsilon}
\newcommand{\sw}{\sin^2\theta_W}

\begin{document}

\title{Parton Momentum and Helicity Distributions in the Nucleon}

\author{P.~Jimenez-Delgado$^1$,
	W.~Melnitchouk$^1$,
	J.~F.~Owens$^2$}
\address{$^1$	Jefferson Lab, Newport News,
		Virginia 23606, USA	\\
	 $^2$	Department of Physics,
		Florida State University, Tallahassee,
		Florida 32306, USA}
\ead{wmelnitc@jlab.org}

\begin{abstract}
We review the current status of spin-averaged and spin-dependent parton
distribution functions (PDFs) of the nucleon.  After presenting the
formalism used to fit PDFs in modern global data analyses, we discuss
constraints placed on the PDFs by specific data types.  We give
representative examples of unpolarized and polarized PDFs and their
errors, and list open questions in global QCD fitting.  Finally,
we anticipate how future facilities, with fixed-target and collider
experiments, may impact our knowledge of PDFs and reduce their
uncertainties.
\end{abstract}


\maketitle

\tableofcontents

\newpage
\section{Introduction}

Quantum Chromodynamics, the theory describing the interactions of
quarks and gluons, has been shown to provide excellent descriptions of
a wide variety of phenomena ranging from hadron spectroscopy via lattice
field theory to hard scattering phenomena via perturbation theory.
Yet QCD is a theory whose degrees of freedom are the quarks and gluons
-- quanta that can not be observed directly due to their confinement in
hadrons.  On the other hand, experiments deal with hadrons and leptons.
In order to calculate observables measured in high energy scattering
processes a technique is needed that allows conversion from a
description in terms of hadrons to one in terms of quarks and gluons
(collectively referred to as partons).  For hadrons in the initial state
this conversion is provided by parton distribution functions (PDFs),
while for final state hadrons a similar role is played by parton
fragmentation functions.  The PDFs, denoted by $f_i(x)$, for each
type of quark and antiquark and for the gluon ($i = q, \bar q, g$),
allow one to essentially describe a beam of hadrons as an effective
beam of quarks and gluons.  In the infinite momentum frame, the PDFs
can be interpreted as probability densities describing how the parent
hadron's momentum is shared amongst the different types of partons,
as a function of the hadron's momentum fraction $x$.  With knowledge of
the PDFs and the Feynman rules for QCD, one can calculate hard scattering
cross sections and differential distributions that can then be compared
to data.

PDFs are traditionally determined by simultaneously fitting a wide
variety of data for large momentum transfer processes.  The parameters
of the fits describe the PDFs at some initial momentum transfer scale,
while evolution equations are then used to calculate the PDFs at all
other scales needed for the calculations.  The data sets used for such
fits often include deep inelastic scattering (DIS) of charged leptons
on proton and deuterium targets, or neutrinos on heavy nuclear targets,
lepton pair production on proton and deuterium targets, and the
production of photons, vector bosons, or jets at large values of
transverse momentum.

While the PDFs provide us with a detailed description of the partonic
substructure of hadrons, they contain only partial information.
Partons (both quarks and gluons) have nonzero spin, so that the
fundamental distributions in nature are the PDFs for a specific
helicity (spin projection along the direction of motion), 
$f_i^\uparrow$ and $f_i^\downarrow$, corresponding to parton spins
aligned and antialigned with that of the hadron, respectively.
Experiments with unpolarized beams and targets are therefore sensitive
only to the sums of the helicity PDFs,
        $f_i = f_i^\uparrow + f_i^\downarrow$,
while information on the differences,
        $\Delta f_i = f_i^\uparrow - f_i^\downarrow$,
can be obtained from measurements involving polarized beams
and/or targets.

It is the purpose of this topical review to give a current ``snapshot''
of the status of our knowledge of PDFs, both unpolarized and polarized.
To this end, brief reviews of both the theory and the types of data
used in the fits will be presented, and theoretical issues that need
to be addressed in the perturbative calculations will be outlined.
Throughout this review we shall attempt to answer the question
``what do we know and how do we know it?''
Recently there have been several excellent reviews of PDFs
\cite{Forte:2013wc, Blumlein:2012bf, Accardi:2013}, updating the
early treatises of PDFs in Refs.~\cite{Buras:1979yt, Owens:1992hd},   
which have focused on various aspects of unpolarized PDFs, such as  
the impact on the phenomenology of the Large Hadron Collider (LHC).   
The spin structure of the nucleon, including spin-dependent PDFs,
was also reviewed in Refs.~\cite{Lampe:1998eu, deFlorian:2011ia,
Aidala:2012mv}.
The present pedagogical review should be viewed as complementary to
these efforts, emphasizing the different aspects of hadron structure
that can be revealed through the study of unpolarized and polarized
PDFs, at both small and large parton momentum fractions $x$.

There is an old saying in experimental particle physics that   
``today's discovery is tomorrow's calibration.''  A similar effect
occurs for theoretical calculations.  What was once touted as a test
of QCD, now often plays the role of providing a description of
backgrounds for searches for new types of particles.  Of course,
one wants these calculations to be as accurate as possible and this,  
in turn, requires precise knowledge of the PDFs that enter into
the calculation.  The error estimates on PDFs will be discussed
at some length for this reason.
On the other hand, there are many aspects of hadron structure that
are currently not understood well, or at all, either theoretically
or experimentally, and clearly identifying and delineating the limits
of our knowledge in the context of global PDF fits serves a valuable
purpose.

The plan of the review is as follows.  Sec.~2 contains an overview of
the necessary QCD theory required to understand the results of global
fits.  Sec.~3 contains the review of unpolarized PDFs, while Sec.~4
summarizes the status of polarized PDFs.  Finally, Sec.~5 provides
some speculations as to how our knowledge of PDFs may be improved
through anticipated results of ongoing and future experiments.

\section{QCD Analysis}
\label{sec:QCD}

As noted in the introduction, PDFs are necessary ingredients for
obtaining predictions for hard scattering hadron--hadron and
lepton--hadron processes.  The cross section for a typical
hadron--hadron process involving collisions of hadrons $A$ and $B$,
with momenta $p_A$ and $p_B$, respectively, producing a state $C$
in addition to other hadrons (collectively denoted by $X$),
can be written in the form
\begin{eqnarray}
\hspace*{-0.5cm}
\sigma_{A B \to C X}(p_A, p_B)
&=& \sum_{a,b} \int dx_a \, dx_b \,
    f_{a/A}(x_a, \mu_f)\, f_{b/B}(x_b, \mu_f)		\nonumber\\
& & \hspace*{-0.0cm} \times
    \sum_n \alpha_s^{n}(\mu_r)\,
    \hat\sigma^{(n)}_{a b \to C X}
	\left( x_a p_A, x_b p_B, Q/\mu_f, Q/\mu_r \right),
\label{eq:xsec}
\end{eqnarray}
where $\hat\sigma^{(n)}$ denotes an $n$-th order parton--parton
``cross section'' that produces the desired final state, and the
functions $f_{a/A}$ and $f_{b/B}$ are the PDFs of flavor $a$
in hadron $A$ and flavor $b$ in hadron $B$, respectively.
The state $C$ might denote a high-$p_T$ jet, a lepton pair, a photon
or weak vector boson, for example, and $Q$ is a scale that characterizes
the hard scattering.  For high-$p_T$ jet or photon production the scale
$Q \sim p_T$.  The form (\ref{eq:xsec}) can also be applied to the
case when the hadrons $A$ and $B$ are polarized; in the remainder of
this section, however, we shall for illustration focus primarily on
the unpolarized case.

The parton-level cross sections will generally possess ultraviolet 
singularities that must be renormalized.
Doing so leads to the introduction of the running coupling
$\alpha_s(\mu_r)$ which depends on a renormalization scale $\mu_r$.
There can also be infrared singularities associated with loop graphs;
these will cancel corresponding singularities from the emission of
real soft gluons, provided that the observable is suitably defined,
{\it i.e.,} that it is ``infrared safe.''
Finally, there will be collinear singularities associated with, in this
case, the initial partons emitting additional partons at zero angle.
These collinear configurations correspond to internal propagators
going on-shell and, as such, correspond to long-distance physics.
Such singular terms can be {\it factorized} and absorbed into the PDFs.
This process introduces a factorization scale, $\mu_f$, that separates
the long-distance and short-distance hard scattering physics.

As shown in Eq.~(\ref{eq:xsec}), the partonic hard scattering cross
section can be expanded in powers of the running coupling.
The dependence on $\mu_r$ that enters via the running coupling
cancels that which appears in the partonic cross sections.
Similarly, the dependence on $\mu_f$ in the PDFs cancels against
the $\mu_f$ dependence in the partonic cross sections.  These
cancellations represent the fact that the physical cross section,
if calculated to all orders in perturbation theory, should not
depend on the two scales introduced to control the ultraviolet and
collinear singularities.  However, at any fixed order in perturbation
theory these cancellations will only be approximate.  Indeed, at order
$\alpha_s^m$ one has the relations
\begin{eqnarray}
\hspace*{-0.5cm}
\mu_f^2\, \frac{\partial \sigma_{A B \to C X}}{\partial \mu_f^2}
&=& 0 + {\cal O}(\alpha_s^{m+1}), \ \ \ \ \ \
\mu_r^2\, \frac{\partial \sigma_{A B \to C X}}{\partial \mu_r^2}\
 =\ 0 + {\cal O}(\alpha_s^{m+1}).
\label{eq:scale}
\end{eqnarray}
These results suggest that the dependence on the renormalization and
factorization scale choices will decrease as one carries out the 
perturbative calculations to higher order.  This is indeed the case
for the processes typically used in global fits for PDFs.
(If a new kinematic configuration opens up at a higher order, then the
corrections at that order can be large.  This happens when going from
LO to NLO for heavy quark production, for example, in which case the
scale dependence can actually increase at that order.  The reduction
then occurs at the next order.)
These processes are all known to at least NLO and some to NNLO, as will
be discussed below.  In principle, the two scales $\mu_r$ and $\mu_f$
can be chosen independently.  However, in processes where there is one
large scale characterizing the hard scattering it is often the case 
that the choice $\mu_f = \mu_r \equiv \mu$ is used.  This simplifying
choice will be utilized in the results discussed in the sections to
follow.

\subsection{Running coupling}

The dependence on the renormalization scale of the running coupling is   
governed by the QCD $\beta$ function via the renormalization group 
equation,
\begin{equation}
\mu_r^2\, \frac{\partial \alpha_s(\mu_r^2)}{\partial \mu_r^2}
= \beta(\alpha_s(\mu_r^2)).
\end{equation}
The QCD $\beta$ function is calculable in perturbation theory and
can be written as 
\begin{equation}
\beta(\alpha_s)
= - b_0 \alpha_s^2 - b_1 \alpha_s^3 - b_2 \alpha_s^4
+ {\cal O}(\alpha_s^5),
\label{eq:alphas}
\end{equation}
where the coefficients for $n_f$ flavors are
\begin{eqnarray}
b_0 &=& \frac{33 - 2 n_f}{12 \pi},		\\
b_1 &=& \frac{153-19 n_f}{24 \pi^2},		\\
b_2 &=& \frac{77139-15099 n_f +325 n_f^2}{3456 \pi^3}.
\end{eqnarray}
The strong coupling $\alpha_s$ calculated using the expression
for $\beta$ in Eq.~(\ref{eq:alphas}) with terms up to $b_n$ is
referred to as the $(n+1)-$loop running coupling.
LO calculations retain only $b_0$, while NLO calculations retain
$b_1$, and NNLO calculations retain also $b_2$.  Note that the
$\beta$ function is negative, resulting in a decrease of $\alpha_s$
with increasing scale, a phenomenon known as asymptotic freedom.
This behavior is essential for the applicability of perturbation
theory for hard scattering processes.

\subsection{$Q^2$ evolution}
\label{ssec:Q2evol}

In Eq.~(\ref{eq:xsec}) both the PDFs and the partonic cross sections
depend on the factorization scale $\mu_f$.  As seen in
Eq.~(\ref{eq:scale}), this scale dependence cancels up to the order
of perturbation theory used.
This fact allows one to calculate the $\mu_f^2$ dependence of the PDFs.
Indeed, the singularities associated with collinear parton emission
are universal and factorize from the hard scattering subprocesses.
The factorization scale dependence of the parton distributions is
then also universal, reflecting the independence of physical
quantities on the scale $\mu_f$.  This leads to the $Q^2$ evolution,
or renormalization group equations (RGE),
\begin{equation}
\mu_f^2\, \frac{\partial f_i(x,\mu_f^2)}{\partial \mu_f^2}\
=\ \sum_j\int_x^1\frac{dy}{y} P_{ij}
   \left( \frac{x}{y},\mu_f^2 \right) f_j(y,\mu_f^2)\
=\ \sum_{j} P_{ij} \otimes f_j,
\label{eq:RGE}   
\end{equation}
where the Bjorken $x$ variable is (related to) the longitudinal
momentum fraction of the partons. The evolution kernels $P_{ij}$
or ``splitting functions'' stem from the collinear divergences
absorbed in the distributions and represent the (collinear) resolution
of a parton $i$ in a parton $j$.  They are calculated perturbatively
as a series expansion in $\alpha_s(\mu_f^2)$,
\begin{equation}
P_{ij} \left( \frac{x}{y},\mu_f^2 \right)\
=\ \sum_{m=0}^{\infty}
   \left( \frac{\alpha_s(\mu_f^2)}{2\pi} \right)^{m+1} P_{ij}^{(m)}
   \left( \frac{x}{y} \right).
\label{eq:Pexp}
\end{equation}
At present these have been computed up to 3 loops ($m\!=\!2$),
which is the necessary accuracy for a NNLO analysis.

The evolution of spin-dependent PDFs $\Delta f_i$ follows in
a similar manner, with a corresponding set of spin-dependent
splitting functions $\Delta P_{ij}$.
In this case the splitting functions have been computed to
2 loops, enabling analyses to be performed at NLO accuracy.

\subsection{Hard scattering processes}

\subsubsection{Lepton--hadron deep-inelastic scattering.}
\label{sssec:DIS}  

Traditionally, information on the PDFs of the nucleon has come from
the process of deep-inelastic scattering (DIS) of leptons from protons
or nuclei, beginning with the pioneering experiments at SLAC in the
late 1960s.
In the one-boson exchange approximation, the differential DIS
cross section can be written as a product of lepton and hadron
tensors,
\begin{equation}
{ d^2\sigma \over d\Omega dE' }
= { \alpha^2 \over Q^4 } {E' \over E} 
  \sum_j \eta_j\, L_{\mu\nu}^j\, W^{\mu\nu}_j,
\label{eq:sigtens}
\end{equation}  
where $\alpha$ is the fine structure constant,
$\Omega = \Omega(\theta,\phi)$ is the laboratory solid angle
of the scattered lepton, and $E (E')$ is the incoming (outgoing)
lepton energy.
For neutral-currents, the summation is over $j=\gamma, Z$ and the
interference $\gamma Z$, while for charged-currents only $W^\pm$
exchange contributes.
The leptonic tensor $L_{\mu\nu}$ depends on the charge $e = \pm 1$
and helicity $\lambda = \pm 1$ of the lepton,
\begin{eqnarray}
L_{\mu\nu}^\gamma
&=& 2 \left( k_\mu k'_\nu + k'_\mu k_\nu - g_{\mu\nu} k \cdot k'
	     - i \lambda \eps_{\mu\nu\alpha\beta} k^\alpha k'^\beta
      \right),						\nonumber\\
L_{\mu\nu}^Z
&=& (g_V^e + e \lambda g_A^e)\, L_{\mu\nu}^{\gamma Z}\
 =\ (g_V^e + e \lambda g_A^e)^2\, L_{\mu\nu}^\gamma,	\\
L_{\mu\nu}^W
&=& (1+e\lambda)^2\, L_{\mu\nu}^\gamma,			\nonumber
\end{eqnarray}
where $k$ and $k'$ are the initial and final electron momenta,
and $g_V^e = -1/2 + 2 \sw$ and $g_A^e = -1/2$ are the electron
vector and axial-vector charges, respectively.
The factors $\eta_j$ in Eq.~(\ref{eq:sigtens}) denote the ratios of the 
corresponding propagators and couplings to the photon propagator and
coupling squared \cite{PDG12}
\begin{eqnarray}
\eta_\gamma
&=& 1, \hspace*{1cm}
\eta_Z\
 =\ \eta_{\gamma Z}^2\
 =\ \left( {G_F M_Z^2 \over 2\sqrt{2} \pi \alpha} \right)^2
    { 1 \over (1 + M_Z^2/Q^2)^2 },			\nonumber\\
\eta_W
&=& {1\over 2}
    \left( {G_F M_Z^2 \over 2\sqrt{2} \pi \alpha} \right)^2
    { 1 \over (1 + M_W^2/Q^2)^2 },
\end{eqnarray}
where $G_F$ is the Fermi weak interaction coupling constant,
and $M_W$ is the $W$ boson mass.

The hadronic tensor $W_{\mu\nu}$ contains all of the information
about the structure of the hadron target.
Using constraints from Lorentz and gauge invariance, together with  
parity conservation, it can be decomposed into spin-independent
and spin-dependent contributions,
\begin{eqnarray}
W_{\mu\nu}
&=& -\widetilde{g}_{\mu\nu}\, F_1(x,Q^2)\
 +\ {\widetilde{p}_\mu \widetilde{p}_\nu \over p\cdot q}\, F_2(x,Q^2)\
 +\ i \eps_{\mu\nu\alpha\beta}\, p^\alpha q^\beta\, F_3(x,Q^2)\
							\nonumber\\
& &
 +\ i \epsilon_{\mu\nu\alpha\beta} {q^\alpha \over p \cdot q}
  \left[ s^\beta\, g_1(x,Q^2)
       + \left( s^\beta - {s \cdot q \over p \cdot q}\, p^\beta
         \right) g_2(x,Q^2)
  \right],
\label{eq:Wmunu}
\end{eqnarray}
where $p_\mu$ and $q_\mu$ are the nucleon and exchanged boson
four-momenta,
$\widetilde{g}_{\mu\nu} = g_{\mu\nu} - q_\mu q_\nu/q^2$, and
$\widetilde{p}_\mu = p_\mu - (p\cdot q/q^2) q_\mu$.
The nucleon polarization four-vector $s^\beta$ satisfies
$s^2=-1$ and $p \cdot s=0$.

For spin-averaged scattering, the nucleon structure is parametrized
in terms of the vector $F_1$ and $F_2$ structure functions, and
the vector-axial vector interference $F_3$ structure function,
which requires weak currents.
These are generally functions of two variables (such as $x$ and $Q^2$),
but become functions of $x$ only in the Bjorken limit, in which both
$Q^2$ and $\nu \to \infty$, but $x$ is fixed.
In this limit the $F_1$ and $F_2$ structure functions become
proportional, according to the Callan-Gross relation,
$F_2(x) = 2 x F_1(x)$, and in the parton model are given in terms
of quark $q$ and antiquark $\bar q$ distribution functions,
\begin{equation}
\label{eq:f22xf1}
F_1(x) = {1 \over 2} \sum_q e_q^2 \left[ f_q(x) + f_{\bar q}(x) \right],
\end{equation}
where $f_q(x)$ is interpreted as the probability to find a quark of
flavor $q$ and charge $e_q$ in the nucleon with light-cone momentum
fraction $x$.
At finite energies, the logarithmic $Q^2$ dependence from the evolution
equations described in Sec.~\ref{ssec:Q2evol}, as well as residual $Q^2$
dependence associated with power corrections (see Sec.~\ref{ssec:power}
below), give corrections to the simple parton model expectations.

The spin-dependent structure functions $g_1$ and $g_2$ can be extracted
from measurements where longitudinally polarized leptons are scattered
from a target that is polarized either longitudinally or transversely
relative to the electron beam.
For longitudinal beam and target polarization, the difference between
the cross sections for spins aligned and antialigned is dominated at
high energy by the $g_1$ structure function.
The $g_2$ structure function can be determined with additional
measurement of cross sections for a nucleon polarized in a direction
transverse to the beam polarization.
In practice one often measures the polarization asymmetry $A_1$,
which is given as a ratio of spin-dependent and spin-averaged
structure functions,
\begin{eqnarray}
A_1(x,Q^2) =
{ 1 \over F_1(x,Q^2) }
\left[ g_1(x,Q^2) - {4M^2x^2 \over Q^2} g_2(x,Q^2) \right].
\end{eqnarray} 
At small values of $x^2/Q^2$, one then has $A_1 \approx g_1/F_1$.
If the $Q^2$ dependence of the polarized and unpolarized structure
functions is similar, the polarization asymmetry $A_1$ will be
weakly dependent on $Q^2$.

In analogy with the unpolarized $F_1$ structure function, in the parton
modelthe structure function $g_1$ can be expressed at LO in terms of
differences between quark distributions with spins aligned
($q_f^\uparrow$) and antialigned ($f_q^\downarrow$) relative to that
of the nucleon, $\Delta f_q(x) = f_q^\uparrow(x) - f_q^\downarrow(x)$,
\begin{equation}
g_1(x) = {1 \over 2} \sum_q e_q^2
         \left[ \Delta f_q(x) + \Delta f_{\bar q}(x) \right].
\label{eq:g1quark}
\end{equation}
The $g_2$ structure function, on the other hand, does not have
a simple parton model interpretation.  However, its measurement
provides important information on the subleading, higher twist
contributions which parametrize long-range nonperturbative
parton-parton correlations in the nucleon (see Sec.~\ref{ssec:power}).

\subsubsection{Hadron--hadron scattering.}
\label{sssec:h-h}

As discussed in Sec.~\ref{sec:QCD}, the PDFs are universal, the
collinear singularities associated with initial parton collinear
emission having been absorbed into the PDFs.  Therefore, the parton
distributions appearing, for example, in the expressions for DIS
structure functions, are the same that describe the structure of
the incoming hadrons in hadronic production.  Repeating the general
form given in Eq.~(\ref{eq:xsec}) (with $\mu_f=\mu_r \equiv \mu$),
one has, for spin-unpolarized scattering,
\begin{eqnarray}
\hspace*{-0.5cm}
\sigma_{A B \to C X}(p_A, p_B)
&=& \sum_{a,b} \int dx_a \, dx_b \,
    f_{a/A}(x_a, \mu)\, f_{b/B}(x_b, \mu)               \nonumber\\
& & \hspace*{0.0cm} \times
    \sum_n \alpha_s^{n}(\mu)\,
    \hat\sigma^{(n)}_{a b \to C X}
        \left( x_a p_A, x_b p_B, Q/\mu \right),
\label{eq:xsec2}
\end{eqnarray}
and an analogous expression for the spin-dependent cross section,
involving the difference of cross sections with hadrons $A$ and $B$
polarized in the same and opposite directions,
\begin{eqnarray}
\hspace*{-0.5cm}
\Delta\sigma_{\vec A \vec B \to C X}(p_A, p_B)
&=& \sum_{a,b} \int dx_a \, dx_b \,
    \Delta f_{a/A}(x_a, \mu)\, \Delta f_{b/B}(x_b, \mu)	\nonumber\\
& & \hspace*{0.0cm} \times
    \sum_n \alpha_s^{n}(\mu)\,
    \Delta\hat\sigma^{(n)}_{\vec a\, \vec b \to C X}
        \left( x_a p_A, x_b p_B, Q/\mu \right).
\label{eq:xsec_pol}
\end{eqnarray}  
Here $\Delta f_{a/A}$ and $\Delta f_{b/B}$ are the spin-dependent
PDFs for flavor $a$ in a hadron $A$ and flavor $b$ in a hadron $B$,
respectively, and $\Delta\hat\sigma^{(n)}$ is the corresponding
spin-dependent partonic cross section.
As indicated in Eqs.~(\ref{eq:xsec2}) and (\ref{eq:xsec_pol}), the
partonic cross sections are calculable in fixed--order perturbation
theory as series expansions in $\alpha_s(\mu^2)$, which starts
with different powers depending on the process.

It will be useful for subsequent sections to describe the parton
kinematics for several examples of hadron--hadron processes.
Although all processes used in global fits are known at least to NLO,
the LO parton kinematics nevertheless serves as a useful guide to
what region of parton momentum fraction and which combinations of
parton flavors will be constrained by a given set of data.

Consider high-$p_T$ dijet production in lowest order where two jets
are produced with approximately balancing transverse momenta $p_T$
and rapidities denoted by $y_1 {\rm \ and\ } y_2$.
Two-body phase constrains the momentum fractions to be
\begin{equation}
x_a = \frac{x_T}{2} \left(e^{y_1} \, + \, e^{y_2}\right)
\ {\rm \ and\ \ }
x_b = \frac{x_T}{2} \left(e^{-y_1} \, + \, e^{-y_2}\right),
\label{eq:dijet}
\end{equation}
where the dimensionless ratio $x_T = 2 p_T/\sqrt{s}$. 
For centrally produced dijets one sees that the parton momentum
fractions are both approximately $x_T$.  However, if one or both
of the jets is produced at far forward or backward rapidity values,
then the $x$ range can be considerably expanded towards large or
small values thereby allowing a ``tuning'' of the probed $x$ range.

For the case of inclusive production of a jet with transverse momentum 
$p_T$ and rapidity $y$ there is, at LO, an integration over one of the 
parton momentum fractions, say $x_a$, given by
\begin{equation}
\frac{x_T e^y}{2 - x_T e^{-y}} \le x_a \le 1 {\rm \ \ with \ \ }
x_b = \frac{x_a x_T e^{-y}}{2 x_a - x_T e^y}.
\label{eq:singlejet}
\end{equation}
Again, one sees that for $y$ near 0 the ranges of the momentum
fractions are centered near $x_T$.

Another important class of experiments involves the production of
a system of mass $M_B$, rapidity $y$, and transverse momentum $p_T$.
Examples of such systems include $W^{\pm}$, $Z^0$, and charged
lepton pairs $l^+l^-$.
The four--momentum of the produced system is conveniently given as
\begin{equation}
(E;\, p_x, p_y, p_z) \equiv
\big( m_T\cosh{y};\, p_T\sin{\phi},\, p_T\cos{\phi},\, m_T\sinh{y}
\big),
\end{equation}
where $m_T^2 \equiv M_B^2 + p_T^2$, and $\phi$ is the azimuthal angle.
The rapidity $y$ and, equivalently, the Feynman--$x$ variable $x_F$
are defined by
\begin{equation}
y = \frac{1}{2} \ln{\left(\frac{E+p_z}{E-p_z}\right)},\quad\ \ \ 
x_F = \frac{p_z}{p_z^{\rm max}} \simeq \frac{2p_z}{\sqrt{s}},
\end{equation}
which lead to the relations \cite{Yndurain, BargerPhillips, Pink}
\begin{equation}
x_F = 2 \sqrt{(\tau + p_T^2/s)}\, \sinh{y}\
\Longleftrightarrow\ \
y = {\rm arcsinh} \frac{x_F}{\sqrt{4 \tau + x_T^2}},
\end{equation}
where $\tau \equiv M_B^2/s \simeq x_a x_b$.

Depending on the detected final state it is possible to define
different cross sections for hadron--hadron collisions.  In the
Drell-Yan mechanism \cite{Drell:1970yt} a quark from one hadron
and an antiquark from the other one annihilate into an intermediate
vector boson ($\gamma^*$, $Z^0$ or $W^\pm$) which subsequently decays
into a lepton pair.  For lepton pairs of invariant mass $M_B \ll M_Z$,
the process is dominated by virtual photon exchange.
The experiments usually consist of proton\,(beam)--nucleon
(proton or some other nucleus) collisions and the cross sections
are extracted from the detection of muon pairs (dimuon production)
produced in the decay of the virtual photons.

The double differential cross section for lepton pair production
at LO is given by
\begin{eqnarray}
\frac{d^2\sigma}{dM_B^2\, dy}
&=& \frac{4 \pi \alpha^2}{9 s M_B^2}
    \left[ \sum_q e_q^2 f_q(x_a,\mu^2)\, f_{\bar q}(x_b,\mu^2)
           + (q \leftrightarrow \bar q)
    \right],
\label{eq:DY}
\end{eqnarray}
where the two parton momentum fractions are given by
\begin{equation}
x_a = \frac{M_B}{\sqrt{s}} e^y {\rm \ \ and\ \ }
x_b = \frac{M_B}{\sqrt{s}} e^{-y}.
\end{equation}
From Eq.~(\ref{eq:DY}) one can see that for $pp {\rm \ or \ }pd$
collisions this process is sensitive separately to sea and valence
distributions, in contrast to the neutral current DIS cross sections,
where they enter as $q+\bar{q}$.
Inclusion of the Drell--Yan data in global QCD analyses of PDFs
is in fact instrumental in fixing ratio
$\bar{d}/\bar{u} \approx \sigma^{pd}/\sigma^{pp} - 1$
for $x_a \gg x_b$ at LO (see Eq.~(\ref{eq:sig_pd_pp}) below).
Another alternative is the use of charged current neutrino DIS
structure functions, although this process entails the use of
model dependent corrections for the heavy nuclear targets typically
employed in such measurements.  Nuclear corrections can be avoided
if one instead uses the time reversed processes such as
$e^{\pm} p \to \bar \nu_e (\nu_e)\, X$.

A process closely related to Drell--Yan dilepton production is the
hadronic production of electroweak bosons \cite{BargerPhillips,
Pink, Reya:1979zk}.  For $pp$ scattering, the inclusive production
cross section for $W^\pm$ bosons, for example, is given at LO by
\begin{eqnarray}
\frac{d\sigma}{dy}
&=& \frac{2\pi G_F}{3\sqrt{2}}
    \sum_{q,\bar q'} \left| V_{q \bar q'} \right|^2\,
    x_a x_b\, f_q(x_a, M_W^2)\, f_{\bar q'}(x_b, M_W^2),
\label{eq:Wprod}
\end{eqnarray}
where $V_{q \bar q'}$ are the CKM matrix elements, and the sum
runs over all light quark and antiquark flavors in both hadrons.
The production of $W^+$ bosons is then sensitive primarily to
the products
	$u(x_a) \bar d(x_b) + \bar d(x_a) u(x_b)$,
while $W^-$ production is sensitive to
	$d(x_a) \bar u(x_b) + \bar u(x_a) d(x_b)$.
At large $W$ boson rapidity, or equivalently $x_a \gg x_b$,
$W^\pm$ production is sensitive to the $u$ and $d$ quark PDFs
in the proton, respectively.  Provided the antiquark distributions
at $x_b$ are known to sufficiently accuracy, the $W^-/W^+$ cross
section ratio would then be a clean probe of the $d/u$ PDF ratio,
which is presently not well determined at large $x_a$.
Conversely, if the quark PDFs are known in particular regions of
$x_{a,b}$, then the $W^+/W^-$ ratios can also be used to probe
the antiquark distributions.

Similar considerations apply when one or both of the protons
are polarized, as for the case of the spin program at RHIC.
For collisions of longitudinally polarized protons from unpolarized
protons, $\vec p\, p \to W^\pm\, X$, the generalization of
Eq.~(\ref{eq:Wprod}) simply involves replacing one of the quark
or antiquark distributions by the corresponding polarized quark
(antiquark) PDF, $\Delta f_{q (\bar q)}$.
Single-spin asymmetries can then constrain ratios of polarized
to unpolarized PDFs, $\Delta f_{q (\bar q)} / f_{q (\bar q)}$,
for various flavors, over specific regions of $x_{a,b}$
(see Sec.~\ref{ssec:pol_sea}).
Double-spin asymmetries, formed by combinations of inclusive
hadron or jet production cross sections in the scattering of
polarized protons from polarized protons,
	$\vec p\, \vec p \to {\rm (hadron\ or\ jet)}\, X$,
have also been studied in order to constrain the polarized
gluon distribution $\Delta g$ (see Sec.~\ref{ssec:pol_gluon}).

\subsection{Heavy quarks}
\label{ssec:HQ}

The treatment of heavy quarks in high-energy processes generally
requires more care than for light quarks.
The production mechanism of heavy quarks in DIS is predominantly
the photon--gluon fusion (PGF) process, which makes it particularly
sensitive to the gluon distribution in the nucleon.  Although in
principle there could exist an ``intrinsic'' (initial state) heavy
quark content of the nucleon, existing measurements are well described
through ``extrinsic'' (generated in the hard scattering) heavy quark
production only ({\it cf.} \cite{Abramowicz:1900rp} and references
therein), so that any intrinsic heavy quark content of the nucleon
is small and restricted to large values of Bjorken $x$ (see
Sec.~\ref{ssec:HQresults} below).

The natural framework for describing the PGF mechanism is known
as the ``fixed flavor number scheme'' (FFNS).  In this approach
the light-quark flavors ($u,d,s$) and the gluon are considered as
massless partons within the nucleon.  This scheme allows one to
compute the contributions of heavy quarks ($c,b,t$) perturbatively
as final-state quantum fluctuations, taking into account the full
dependence of the production cross section on the mass $m$ of the
heavy quark.
As is common in the $\overline{\rm MS}$ renormalization scheme,
the evaluation of the strong running coupling $\alpha_s(\mu^2)$
is nevertheless based on the usual variable flavor number scheme
for the $\beta$--function governing its scale dependence.
This procedure automatically resums the contributions to the strong
coupling from heavy quarks, and consequently improves the stability
of the perturbative expansion \cite{Gluck:2006ju}.
Complete calculations of (heavy--quark) DIS structure functions
are presently known at LO \cite{Witten:1975bh, Babcock:1977fi,
Leveille:1978px, Gluck:1980cp} and at NLO \cite{Laenen:1992zk,
Riemersma:1994hv}.  Beyond this, approximations based on threshold
resummation have been made \cite{Laenen:1998kp, Kawamura:2012cr}
(NNLO* in Sec.~\ref{ssec:HQresults}), but thus far the NNLO
coefficient functions are only partially known
\cite{Bierenbaum:2009mv, Bierenbaum:2009zt, Ablinger:2012sm},
which constitutes a major drawback of any precision QCD analysis
at NNLO accuracy.
Another aspect of these calculations is the mass definition used;
while the original calculations of the amplitudes employ the pole
mass definition, it has been reported \cite{Alekhin:2010sv} that
the use of the running mass definition improves the stability of
the perturbative series.

In many situations calculations within the (fully massive) FFNS scheme
become unduly complicated, and in many cases even impossible due to
the unknown massive ($m \neq 0$) matrix elements at NNLO or even NLO.
For this reason it is common to generate parton distributions in the
so-called ``variable flavor number scheme'' (VFNS), where the heavy
quarks are also considered to be massless constituents of the nucleon.
Here, the required NLO and NNLO cross sections have been computed
for a variety of important production processes, in particular at
hadron colliders.
In the VFNS scheme the effective heavy quark distributions are generated
perturbatively from the nonperturbative distributions of light quarks
and gluons using the boundary conditions of Ref.~\cite{Buza:1996wv},
typically at the unphysical ``thresholds'' $Q^2 = m^2$.
From there one proceeds to their renormalization group evolution
with an increased number of flavors $n = 4, 5$ and (eventually) 6.

For situations where the (threshold) invariant mass of the produced
system far exceeds the mass of the interacting heavy flavor in the FFNS,
the VFNS predictions have been shown \cite{Gluck:2008gs} to deviate
from the FFNS typically by about 10\%, although the exact amount
depends on the particular process and energy scale.  This is usually
within the margins of the renormalization and factorization scale
uncertainties and other theoretical ambiguities related to PDFs.
Eventually, one nevertheless has to {\em assume} that these massless
``heavy'' quark distributions are relevant asymptotically, in that
they can correctly describe the relevant cross sections at scales
much larger than the heavy quark masses involved.

Unfortunately this is not the case for DIS structure functions,
for which the strictly massless approach, known in this context as
the zero--mass VFNS, or ZM-VFNS, is well known to be experimentally
inadequate, particularly near the heavy quark production thresholds.
To remedy this the heavy-quark mass effects are re--inserted in
what are known as general-mass VFNSs (GM-VFNS), for which several
prescriptions are available in the literature \cite{Buza:1996wv,
Aivazis:1993pi, Thorne:1997ga, Steffens:1999hx, Forte:2010ta}
(see also Ref.~\cite{Collins:1998rz}).
These schemes interpolate (in a model-dependent way) between the
FFNS results near production threshold and the asymptotic results
of the ZM-VFNS.  However, since the interpolating schemes are based
on the same massive matrix elements as the FFNS, they do not have
complete information on the ${\cal O}(\alpha_s^3)$ heavy flavor
Wilson coefficients (even though they are sometimes referred to as
``NNLO GM-VFNSs'').

The importance for global PDF analysis of finite heavy quark mass
effects in the calculation of DIS structure functions had been
previously emphasized in Refs.~\cite{Gluck:1993dpa, Gluck:1994uf,
Gluck:1998xa}, but was only universally recognized after the
analysis of \cite{Kretzer:2003it}.  Currently the ABM
\cite{Alekhin:2012ig} and JR \cite{Gluck:2007ck, JimenezDelgado:2009tv}
collaborations use the FFNS for the calculation of DIS structure
functions in their global analyses, while the MSTW \cite{Martin:2009iq},
CT \cite{Lai:2010vv, Gao:2013xoa}, HERAPDF \cite{Aaron:2009aa}
and NNPDF \cite{Ball:2010de} groups use variants of GM-VFNS.
The calculations of hadron collider cross sections are carried out
in the VFNS in all cases.

\subsection{Power corrections}
\label{ssec:power}

The elegant machinery that has been developed within the framework of
perturbative QCD to analyze leading twist PDFs is, strictly speaking,
valid only at high values of $Q^2$ and $W$ where all hadron mass scales
are suppressed, $M^2/Q^2, M^2/W^2 \ll 1$.  In real experiments performed
at a finite beam energy $E$, however, the maximum values of $Q^2$ and
$W$ are limited, which inevitably restricts the available coverage in
Bjorken $x$.
This is especially relevant at large $x$, where in DIS the invariant mass
squared of the produced hadronic system is given by
\begin{equation}
W^2 = M^2 + Q^2 \left( \frac{1-x}{x} \right),\ \ \ \ \ \
Q^2 < Q^2_{\rm max} = 2 M E x,
\end{equation}
with $M$ the mass of the target nucleon.
For fixed $Q^2$, as $x \to 1$ the final state hadron mass $W$ decreases
as one descends into the region dominated by nucleon resonances at
$W \lesssim 2$~GeV.
The resonance region may be treated using the concept of quark-hadron
duality \cite{Melnitchouk:2005zr}, although this goes beyond the scope
of the usual pQCD analysis.

In the region of low $Q^2$, power corrections to the Bjorken limit
results that scale as powers of $\Lambda_{\rm QCD}^2/Q^2$ become
increasingly important.
In the operator product expansion, these are associated with higher
twist corrections, which arise from multi-parton correlations and
characterize the long-range nonperturbative interactions between quarks
and gluons.  Of tremendous interest in their own right as providing
glimpses into the dynamics of quark confinement, the power corrections
are viewed as troublesome backgrounds to efforts aimed solely at
extracting leading twist PDFs.

To avoid the complications from the higher twist corrections,
the usual strategy in global PDF analyses is to apply cuts specifying
minimum values of $Q^2$ and $W^2$.  In many analyses of unpolarized
scattering data, the cuts are of the order $Q^2 \gtrsim 4$~GeV$^2$
and $W^2 \gtrsim 14$~GeV$^2$ \cite{Martin:2009iq, Lai:2010vv, 
Aaron:2009aa, Ball:2010de, Ball:2012cx}, which in practice restrict
the range of $x$ that can be accessed to $x \lesssim 0.7$.
For spin-dependent PDFs, the scarcity of high-energy data forces
most global analyses to use less restrictive cuts, of the order
$Q^2 > 1$~GeV$^2$ and $W^2 \gtrsim 4$~GeV$^2$.
On the other hand, there are a number of important reasons for needing
to know the large-$x$ behavior of PDFs, and several recent unpolarized
PDF analyses \cite{Alekhin:2012ig, Alekhin:2009ni, CJ10, CJ11, CJ12}
have been performed with relaxed cuts of $Q^2 \gtrsim (1.3\ \rm GeV)^2$
and $W^2 > 3$~GeV$^2$ which have allowed an expanded reach into large
$x$.  The improvement in the large-$x$ kinematic coverage amounts to
about 1300 more data points for proton and deuteron targets, representing
some $\sim 50\%$ increase in the total number of DIS data points compared
with the more restrictive cuts.
Most importantly, the resulting leading twist PDFs fits have proved
to be very stable even when such low cuts have been applied \cite{CJ10}.

With the inclusion of the kinematic regions in which subleading
$1/Q^2$ effects play a non-negligible role, it is obviously crucial
to account for the power corrections that might otherwise obfuscate
the leading twist PDFs.  Among the different categories of $1/Q^2$
effects, the simplest are the target mass corrections (TMCs), which
are formally associated with matrix elements of leading twist operators
\cite{Georgi:1976ve, Matsuda:1979ad, Piccione:1997zh, Blumlein:1998nv}
and are of kinematical origin.  Others include genuine higher twist
corrections, which arise from dynamical, multi-parton correlations,
as well as higher order perturbative QCD corrections, which can also
resemble power suppressed contributions at low $Q^2$.

The standard method to compute TMCs in DIS is based on the operator
product expansion, and was first formulated by Georgi and Politzer
\cite{Georgi:1976ve}, and expressions for all unpolarized and
polarized structure functions now exist in both $x$ and Mellin
space \cite{Matsuda:1979ad, Piccione:1997zh, Blumlein:1998nv,
Kretzer:2003iu, Steffens:2012jx}.
An alternative method based on collinear factorization (CF) in
momentum space was developed by Ellis, Furmanski and Petronzio
\cite{Ellis:1982cd}, and extended by various authors
\cite{Kretzer:2003iu, Accardi:2008ne, Accardi:2009md}
(see also the recent reviews of TMCs in
Refs.~\cite{Schienbein:2007gr, Brady:2011uy}).
The advantage of the OPE method is that TMCs can be calculated
to all orders in $1/Q^2$ in DIS, whereas TMCs in the CF approach
have only been computed to ${\cal O}(1/Q^2)$.  On the other hand,
the OPE is limited to inclusive DIS, while the CF framework can
be applied to computing TMCs also in other processes
\cite{Accardi:2009md}.  Since typically the non-DIS data are
taken at very high $Q^2$ where TMC effects are very small,
the OPE method is usually adopted.

All of the TMC methods also suffer to some extent from the threshold
problem, whereby the target mass corrected structure function remains
nonzero as $x \to 1$ \cite{Steffens:2012jx, Bitar:1978cj,
Steffens:2006ds}.
For the purposes of global fits, however, the region where the
threshold effects become problematic is $W < 2$~GeV 
\cite{Steffens:2012jx}, which is mostly outside of where even the
most liberal cuts in $W$ and $Q^2$ are made \cite{Alekhin:2009ni,
CJ10, CJ11}.
From the seminal work of De~R\'ujula {\it et al.} \cite{DeRujula:1976ih,
DeRujula:1976tz}, the appearance of the threshold problem in the
analysis of TMCs is attributed to the neglect of dynamical higher
twist corrections, both of which scale as powers in $1/Q^2$.
(In fact, higher order perturbative QCD corrections can also resemble
power suppressed contributions at low $Q^2$.)

Regardless of their origin, the various power suppressed corrections
that are not included in a leading twist calculation can be absorbed
into phenomenological functions; for example, for an unpolarized
structure function $F_i$,
\begin{equation}
F_i(x,Q^2)
= F_i^{\rm LT}(x,Q^2) + \frac{h_i(x,Q^2)}{Q^2}
  + \frac{h_i^\prime(x,Q^2)}{Q^4} + \ldots,
\end{equation}
where $F_i^{\rm LT}$ denotes the leading twist contribution including
TMCs.  The higher twist corrections are sometimes assumed to be
multiplicative, with the functions $h_i, h_i^\prime$ proportional to the
leading twist contribution, $h_i(x,Q^2) = F_i^{\rm LT}(x,Q^2)\, c(x)$.
Possible additional $Q^2$ dependence of the higher twist contributions,
from radiative $\alpha_s(Q^2)$ corrections, is usually neglected.
The leading twist PDFs were found in Ref.~\cite{CJ10} to be
essentially independent of the TMC prescription adopted, with the
HT parameters able to compensate for the variations due to the
different TMC formulations.
The isospin dependence of the HT corrections was also studied in
Refs.~\cite{Virchaux:1991jc, Alekhin:2003qq, Blumlein:2008kz, JR13}.
Existing data do not allow for an accurate determination of the
${\cal O}(1/Q^4)$ corrections $h_i^\prime$, and attempts to include
them in global fits produce anomalously small values of $\chi^2$
and $\alpha_s(M_Z^2)$ in conjuction with large compensating twist--4
and twist--6 contributions \cite{JR13, AKP}, possibly due to
overfitting and blurring of the scaling violations.

In the polarized case, the greater scarcity of data means that
typically one cannot afford the luxury of $Q^2$ and $W$ cuts as
stringent as those applied in some of the unpolarized PDF analyses.
Most global analyses of spin-dependent PDFs therefore include
structure function measurements down to $Q^2 = 1$~GeV$^2$,
where higher twist corrections are believed to be important.
The higher twist contributions to $g_1$ and $g_2$ may be treated
in an analogous way to the unpolarized $F_i$ structure functions,
with the important difference that the twist--3 contributions to
$g_2$ are not $Q^2$-supressed.  Phenomenological PDF analyses
exist which include HT contributions to $g_1$ \cite{Leader:2010rb,
Blumlein:2010rn}, $g_2$ \cite{Blumlein:2012se, Accardi:2009au},
and to both functions simultaneously \cite{JAM13}.
While the focus in the present work is on the leading twist
PDFs, the higher twist contributions to $g_1$ and $g_2$ are also
of intrinsic interest in themselves, containing information,
for example, about the correlations of color electric and magnetic
fields in the nucleon with the nucleon's spin \cite{Ji:1993sv,
Stein:1995si, Meziani:2004ne, Deur:2004ti, Osipenko:2004xg}.

\subsection{Nuclear corrections}
\label{ssec:nuclear}

Since nucleons bound in a nucleus are not free, the parton distributions
$f_i^A$ in a nucleus $A$ deviate from a simple sum of PDFs in the free
proton and neutron, $f_i^A \neq Z f_i^p + (A-Z) f_i^n$, where $Z$ is
the number of protons.  This phenomenon is especially relevant at small
values of $x$, where nuclear shadowing (or screening) effects suppress
the nuclear to free isoscalar nucleon ($N$) ratio, $f_i^A/(A f_i^N)<1$,
and at large $x$, where the effects of Fermi motion, nuclear binding,
and nucleon off-shellness give rise to the ``nuclear EMC effect''
\cite{Aubert:1983rq, Geesaman:1995yd, Norton:2003cb}.
In addition, for spin-dependent PDFs, the different polarizations
of the bound nucleons and nuclei need to be taken into account.

In the nuclear impulse approximation, where scattering is assumed
to take place incoherently from partons inside individual nucleons,
the PDF in a nucleus can be expressed as a convolution of the PDF in
a bound nucleon and a momentum distribution function $\varphi_{N/A}$
of nucleons in the nucleus \cite{Melnitchouk:1993nk, Kulagin:1994fz,
Kulagin:2004ie}.
Coherent rescattering effects involving partons in two or more nucleons
give rise to nuclear shadowing corrections to the impulse approximation.
In general, at large $Q^2$ the PDFs in a nucleus and in a nucleon are
related by
\begin{eqnarray}
f_i^A(x,Q^2)\
&=& \sum_{N=p,n}
    \int {dz \over z}\, \varphi_{N/A}(z)\, f_i^N (x/z,Q^2) 
\nonumber\\
& &
 +\ \delta^{(\rm off)} f_i^A(x,Q^2)\
 +\ \delta^{(\rm shad)} f_i^A(x,Q^2),
\end{eqnarray}
where the additive term $\delta^{(\rm off)} f_i^A(x,Q^2)$
represents nucleon off-shell or relativistic corrections,
and $\delta^{(\rm shad)} f_i^A(x,Q^2)$ parametrizes the
shadowing corrections.
A similar expression can be written for spin-dependent PDFs.

The momentum distribution, or ``smearing function'', $\varphi_{N/A}$,
can be computed from nuclear wave functions, incorporating nuclear
binding and Fermi motion effects.
At $Q^2 \to \infty$ the smearing function has a simple probabilistic
interpretation in terms of the light-cone momentum fraction
\mbox{$z = (M_A/M)(p \cdot q / P_A \cdot q)   
   \approx (M_A/M)(p^+/P_A^+)$}
of the nucleus carried by the struck nucleon, where $p$ and $P_A$
are the four-momenta of the nucleon and nucleus, respectively,
and $M_A$ is the nuclear mass.  In this case the smearing function
is normalized to unity, $\int dz\, \varphi_{N/A}(z) = 1$.
At finite $Q^2$, however, the smearing function depends in addition
on the parameter
   $\gamma^2 = {\bm q}^2/\nu^2$ \mbox{ $ = 1 + 4x^2 M^2/Q^2$},
where $\nu$ and ${\bm q}$ are the energy and three-momentum transfer,
respectively, which characterizes the deviation from the Bjorken limit
\cite{Kulagin:2004ie, Kahn:2008nq}.
Typically, the function $\varphi_{N/A}$ is steeply peaked around
$z \approx 1$, becoming broader with increasing mass number $A$
as the effects of binding and Fermi motion become more important.
In the limit of zero binding, $\varphi_{N/A}(z) \to \delta(1-z)$,
and one recovers the free-nucleon case.  This is the usual assumption
made in most global PDF analyses.

Recently, several analyses \cite{Alekhin:2012ig, Alekhin:2009ni,
CJ10, CJ11, CJ12} have accounted for the nuclear effects by
explicitly calculating the corrections from microscopic nuclear models,
or attempted to constrain them phenomenologically \cite{Martin:2012da}.
The most straightforward calculation is for the simplest nucleus ---
the deuteron, for which both nonrelativistic and relativistic wave
functions are available \cite{AV18, CDBonn, WJC}, constrained by
high-precision nucleon-nucleon scattering data.
Experiments with deuterium targets play a vital role, in fact,
in determining the flavor decomposition of the proton PDFs.
Traditionally, the standard method for disentangling the $u$ and $d$
PDFs has been through charged lepton DIS, which for a proton target
is sensitive at large $x$ to the combination $4u + d$.  DIS from a
neutron would constrain $4d + u$; however, the absence of free neutron
targets has necessitated the use of deuterium as effective neutron
targets.  While the nuclear corrections in the deuteron are typically
a few percent, in some regions of kinematics, most notably at large $x$,
they can give rise to large uncertainties in the extracted $d$ quark
PDF in particular.

An alternative, purely phenomenological approach \cite{Ball:2013gsa}
has attempted to constrain the nuclear corrections in deuterium
directly from the data.  However, without an independent, high-precision
measurement of the $d$ quark PDF from processes other than inclusive
DIS, it is difficult to unambiguously separate the effects of the
nuclear corrections from uncertainties in the $d$ quark PDF,
especially at high values $x$.
Until future experiments (see Ref.~\cite{CJ11} for a discussion)
are able to determine the $d$ quark PDF independent of uncertainties
in nuclear models, a more practical approach adopted by the CJ
Collaboration \cite{CJ12} has been to produce a set of global PDFs
for a range of nuclear models, corresponding to mild (CJ12min),
medium (CJ12mid), and strong (CJ12max) nuclear corrections in the 
deuteron.

The current uncertainties in PDFs from nuclear corrections can have
important consequences far beyond the fixed-target experiments where
they are encountered most directly.
In fact, the effects of the nuclear smearing corrections are not
suppressed at large $Q^2$, and must be considered at all scales
wherever data at $x > 0.5$ are used \cite{CJ10, Arrington:2008zh,
Arrington:2011qt}.
Through $Q^2$ evolution, PDFs at large $x$ and small $Q^2$ evolve
to lower $x$ and higher $Q^2$, so that large-$x$ uncertainties in
fixed-target experiments can have significant consequences for
collider measurements \cite{Kuhlmann:1999sf}, examples of which
were discussed for selected observables at the Tevatron and LHC
in Ref.~\cite{Brady:2011hb}.

For neutrino scattering, in order to increase the relatively low rates
and obtain sufficient statistics, experiments have often resorted to
using heavier nuclear targets, such as iron or lead.  This is
particularly relevant for determinations of the strange quark PDF,
which is typically extracted from opposite-sign dimuon events in
$\nu$ and $\bar \nu$ charm production,
  $W^+ + s \to c$ or $W^- + \bar s \to \bar c$
\cite{Bazarko:1994tt, Mason:2007zz}.
Such extractions are complicated by the presence of nuclear corrections
in neutrino structure functions \cite{Kulagin:2004ie, Kulagin:2007ju},
as well as effects of the nuclear medium on the charm quark propagation
in the final state \cite{Accardi:2009qv}.
Uncertainties in the strange quark PDF can have significant impact
on $W$ and $Z$ boson measurements at the LHC, for example, so that
understanding of the nuclear effects will have impact far beyond
lepton--nucleus DIS \cite{Schienbein:2009kk}.

In order to systematically study the nuclear dependence of PDFs,
without assuming specific relations between PDFs in nuclei and
nucleons, several groups have parametrized nuclear PDFs directly
\cite{Schienbein:2009kk, Hirai:2007sx, Eskola:2009uj, Kovarik:2010uv}.
One approach has been to parametrize the $A$ dependence of the initial
proton PDF parameters, and then perform a global fit of the available
hard scattering data on nuclear targets.  This implicitly assumes a
smooth $A$ dependence for the PDFs, which is reasonable for large $A$,
but may break down for light nuclei such as deuterium.
An alternative method is to fit to one type of data set and examine
how the results differ from those on proton targets.  This can be
useful for testing whether the nuclear corrections are consistent
with those obtained from various data sets.  Such comparisons have
revealed, for instance, a controversy about whether or not the $A$
dependence of the neutrino-nucleus data from the NuTeV collaboration
is compatible with that observed in charged lepton DIS
\cite{Kovarik:2010uv, Paukkunen:2013grz, deFlorian:2011fp}.
Differences between neutrino and electromagnetic nuclear interactions
can arise, for instance, from the presence of the parity-odd
$F_3$ structure function in $\nu/\bar\nu$ scattering, which does
not contribute to charged-lepton scattering \cite{Kulagin:2007ju}.

For spin-dependent scattering, the scarcity of data and their larger
uncertainties at very small $x$ and at high $x$, where nuclear corrections
are most prominent, has meant that almost all global analyses have thus
far relied exclusively on the effective polarization {\it ansatz},
in which the polarized PDF in the nucleus $\Delta f_i^A$ is related
to the polarized PDFs in the proton and neutron as
$\Delta f_i^A \approx \langle \sigma \rangle^p\, \Delta f_i^p
		    + \langle \sigma \rangle^n\, \Delta f_i^n$,
with $\langle \sigma \rangle^{p (n)}$ the average polarizations of
the proton (neutron) in the nucleus.
In practice, only polarized deuterium and $^3$He nuclei have been
used in DIS experiments, in addition to protons.
The new global analysis of helicity PDFs by the JAM collaboration
\cite{JAM13} is the first systematic attempt to incorporate the effects of
nuclear smearing in DIS structure functions of the deuteron and $^3$He.
This will be important as future spin-dependent scattering experiments
probe the nucleon spin structure at increasingly large values
of~$x$~\cite{E12-06-109, E12-06-110, E12-06-122, Accardi:2012hwp}.

\subsection{PDF parametrizations and sum rule constraints}
\label{ssec:param}

The concept of global fitting for PDFs relies on the formalism outlined
in the previous sections.  Partonic cross sections for various hard
scattering processes are convoluted with scale-dependent PDFs to
generate results for physical observables which can then be compared
to data.  The PDFs are parametrized at some convenient input scale
$Q_0$ and then evolved using the appropriate evolution equations to
the scales needed for each calculation.
The values of the input parameters are estimated using a $\chi^2$
minimization technique.  All of the global PDF fitting groups use
some variation of this technique.
A typical parametrization at the input scale $Q_0$ for a generic
polarized or unpolarized parton (quark, antiquark or gluon)
distribution function $f$ is
\begin{equation}
xf(x,Q_0^2) = a_0\, x^{a_1} (1-x)^{a_2}\, P(x),
\label{eq:param}
\end{equation}
where $P(x)$ represents some smoothly varying function usually chosen
as a polynomial in $x$ or $\sqrt{x}$, although exponential functions
are used as well.  In addition, there are parametrizations inspired by
a statistical model of the nucleon \cite{Bourrely:2001du}, as well as
those based on neural networks \cite{Forte:2002fg} and self-organizing
maps \cite{Honkanen:2008mb}.
%
%

Some of the parameters in the input distributions can be determined
from physical constraints.  For example, in the unpolarized case valence
quark number is conserved,
\begin{equation}
\int_0^1 \, dx\, \left( q(x,Q_0^2) - \bar q(x,Q_0^2) \right)
= \left\{
  \begin{array}{ll}
    2   & q = u,    \\
    1   & q = d,    \\
    0   &\mbox{otherwise,}
  \end{array}\right.
\label{eq:sum_rules}
\end{equation}
while the momentum sum rule requires
\begin{equation}
\int_0^1\, dx\, x
\left[ \sum_q^{n_f} q^+(x,Q_0^2) + g(x,Q_0^2) \right] = 1,
\label{eq:mom_sum_rule}
\end{equation}
where we use the notation $q^+ \equiv q + \bar q$,
and the number of flavors at the input scale $Q_0^2$ is usually
taken to be $n_f=3$ (see Sec.~\ref{ssec:HQ}).
In the polarized case the first moments of the charge-conjugation
even (or $C$-even) distributions can be related to octet baryon weak
decay constants.  For the isovector combination, corresponding to the
Bjorken sum rule,
\begin{equation}
\int_0^1 \, dx\,
\left( \Delta u^+(x,Q_0^2) - \Delta d^+(x,Q_0^2) \right) = g_A,
\label{eq:su2}
\end{equation}
where $g_A = 1.270 \pm 0.003$ is the nucleon axial charge,
while for the SU(3) octet one has
\begin{equation}
\int_0^1 \, dx\,
\left( \Delta u^+(x,Q_0^2) + \Delta d^+(x,Q_0^2) - 2 \Delta s^+(x,Q_0^2) 
\right)
= a_8,
\label{eq:hyperon}
\end{equation}
where the octet axial charge $a_8 = 0.58 \pm 0.03$ is extracted
from hyperon $\beta$-decays assuming SU(3) flavor symmetry
\cite{Close:1993mv}.
Note that the sum rules (\ref{eq:sum_rules})--(\ref{eq:hyperon})
are preserved under $Q^2$ evolution.

While most of the PDF groups use similar procedures and data, one
can nevertheless obtain PDF results that can be rather different.
Some of the reasons for this include the following:

\begin{itemize}

\item
Differences in the specific parametrizations and input scale $Q_0^2$.

\item
Differences in data selection, choices of data sets, kinematic cuts,
and the specific treatment of the correlated errors of the data.
This typically limits the amount of data available at large values
of $x$.

\item
Differences in the theoretical framework used, including the particular
form of the solutions of the RGE employed, the treatment of heavy quarks,
inclusion of higher twist contributions, target mass effects, and nuclear
corrections.

\end{itemize}

These differences should be born in mind when comparing the results
from different PDF groups.

\subsection{PDF errors}
\label{ssec:errors}

PDFs are an essential ingredient for producing predictions for processes
in high energy experiments.  The uncertainties in these predictions
depend, in part, on how well determined the PDFs are themselves.
It is important to bear in mind that the predictions that are compared
to data are convolutions of PDFs with partonic hard scattering cross
sections.  There are thus three main sources of PDF uncertainties:
the fitted data, the partonic cross sections, and the parametrizations
used to describe the PDFs.  The following list describes the main sources
of uncertainty in the determination of PDFs.

\begin{itemize}

\item The experimental errors on the fitted data can be directly 
propagated to the fitted PDFs. Standard techniques include the Hessian 
\cite{Pumplin:2001ct}, Lagrange \cite{Stump:2001gu}, and Monte Carlo 
\cite{DelDebbio:2007ee} methods.

\item Uncertainties due to the use of perturbation theory: These can be 
estimated to some extent by doing LO, NLO, and NNLO fits, although not
all processes are known to NNLO accuracy.

\item Scale dependence: The perturbative predictions depend to some
extent on the choices made for the renormalization and factorization
scales for each process.  These choices will change the results for
each process and the fitted PDFs must compensate these changes
(see Sec.~III of Ref.~\cite{Stump:2003yu} for an example of the effects
of choosing different scales).

\item Choice of the value of the running coupling $\alpha_s(M_Z)$: some 
PDF determinations fit $\alpha_s(M_Z)$ while others use the global
average value.  This is mostly a question of philosophy.  On the one
hand, since the strong coupling is a parameter of QCD, then there is no
freedom to choose a different value for each process.  Thus, information
from data types not included in the global fits (such as data from
$e^+e^-$ processes) provides valid constraints on the value of the
coupling and are included in determining the global average.  On the
other hand, fitting the value of the strong coupling and comparing it
to the global average provides an interesting consistency check on the
description of the data provided by QCD.

\item Choice of data sets and kinematic cuts: these choices can affect
the fitted PDFs and the user should be aware of these differences.

\item Treatment of heavy quarks: the various schemes that are currently 
used differ at higher orders and such differences can affect the fitted 
PDFs.

\item Parametrization dependence: fitted PDFs can differ simply through 
the choice of the initial parametrizations.  Extensive efforts are made
to choose flexible parametrizations which are well-constrained by data,
but there is no control over the behavior outside the kinematic region
covered by the data.  This can lead to different extrapolations at very
large or very small values of $x$.  A method for estimating this
remaining uncertainty has been suggested in
Ref.~\cite{JimenezDelgado:2012zx}.

\end{itemize}

Of all these sources of error, the easiest to treat is the first ---
the propagation of the experimental errors.

\subsubsection{Hessian method.}
\label{sssec:Hessian}

The Hessian method is described in detail in Ref.~\cite{Pumplin:2001ct}.
The elements of the Hessian matrix are given by
\begin{equation}
H_{ij}
= \frac{1}{2} \frac{\partial^2 \chi^2}{\partial a_i \, \partial a_j}
\end{equation}
where $a_i$ denotes the $i^{\rm th}$ PDF parameter.  The Hessian matrix
is generated during the actual minimization procedure and its inverse
is the error matrix.  The eigenvectors of the error matrix can then be
used to define eigenvector parameter sets which, in turn, can be used
to calculate error bands for the PDFs or for specific processes.
One particular subtlety is that the error bands generally depend on
a $\chi^2$ tolerance.  Mathematically, one expects the $1\sigma$
parameter errors to correspond to an increase of $\chi^2$ by one
unit from the minimum value.  However, it has been suggested
\cite{Pumplin:2002vw} that inconsistencies between different data sets
may require a larger value to be used.  This ``$\chi^2$ tolerance''
varies between groups and allowance must be made for this when comparing
the resulting error bands.

\subsubsection{Lagrange multiplier method.}
\label{sssec:Lagrange}

The Lagrange multiplier \cite{Stump:2001gu} method is useful when one
wants to determine the PDF error on a specific observable $X$ such as
the cross section for $W$ or $Z$ production.  Let $\chi^2_{\rm global}$
denote the $\chi^2$ for the global data set.
Then one minimizes the function
\begin{equation}
\Psi(a, \lambda) = \chi^2_{\rm global} + \lambda X(a),
\label{eq:psi}
\end{equation}
where $a$ denotes the set of PDF parameters and minimization is done
for a range of values of $\lambda$.  The end result is a relation
between $\chi^2_{\rm global}$ and the value $X$ of the chosen observable.
Once one specifies the $\chi^2$ tolerance, {\it i.e.}, the range of
allowable $\chi^2$ values, this maps out a range of values
\begin{equation*}
X_0 - \Delta X \le X \le X_0 + \Delta X
\end{equation*}
where $X_0$ is the value of $X$ at the global minimum.

\subsubsection{Monte Carlo method.}

An alternative to the usual linear propagation of errors (Hessian method)
which can be useful for minima that are not well behaved or defined,
is the so--called Monte Carlo method.  In order to propagate the
experimental errors a number of `replica' data sets are generated by
using random numbers and the original errors to generate new data
points \cite{DelDebbio:2007ee}.
These replica data sets are then fitted and the resulting replica
PDF sets are treated using standard statistics; the central values
are given by the average over replicas, and the uncertainties by the
envelope of predictions.

\subsection{Data types}

The main motivation for pursuing global fitting as a technique for
determining PDFs is that the use of a wide range of data types
with different kinematic coverage places many constraints simultaneously
on the PDFs. Each type of observable depends on a particular linear
combination of PDFs or products of PDFs. Obtaining the best fit to
all of the observables simultaneously has proven to be an efficient
method for extracting PDFs. Nevertheless, it remains true that specific
observables may be sensitive to a specific PDF or combination of PDFs.
In this section some examples using simplified LO kinematics will be
presented since these can be useful in understanding how PDF errors can
be reduced by future measurements.

\subsubsection{Unpolarized experiments.}
\label{sssec:unpol}

{\bf Deep-inelastic scattering} experiments provide direct information
on PDFs since both the structure functions and the cross sections depend
linearly on the PDFs.  The following list summarizes the main dependences.
\begin{enumerate}

\item
Charged lepton neutral current measurements on a proton target
constrain the combination
$$
4 u^+ + d^+ + s^+,
$$
where we have assumed that only single photon exchange contributes
($Z$ boson exchange involves a different linear combination of PDFs).
At large values of $x$ the antiquark PDFs become negligible leaving the   
combination $4 u_v + d_v$.  If one had a neutron target, then the linear   
combination would be
$$
4 d^+ + u^+ + s^+,
$$
which becomes $4d_v + u_v$ at large values of $x$.  Of course, neutron
targets are not available, so deuterium is often used for this purpose
requiring that nuclear corrections be made in order to extract the 
neutron target information as discussed in Sec.~\ref{ssec:nuclear}.

\item
Charged current neutrino interactions constrains the combinations
$$
\sum_i (q_i \pm \bar q_i),
$$
where the plus (minus) sign corresponds to $F_{1,2}$ ($xF_3$).
The actual linear combinations depend on the type of target used;
high statistics neutrino experiments employ different types of heavy
targets for which model-dependent nuclear corrections must be made.
Note that one can, in principle, isolate specific combinations of
$q$ or $\bar q$ PDFs, depending on the type of target used.
A special case is provided by charm production in neutrino and
antineutrino scattering as these are proportional to the $s$ and
$\bar s$ PDFs, respectively.  The experimental signal is the
production of opposite sign muon pairs requiring one to take
into account the charm fragmentation and decay as well as nuclear
corrections, depending on the type of target employed.

\item
The gluon PDF enters in DIS at ${\cal O}(\alpha_s)$.
Thus, it is mainly constrained through the $Q^2$ dependence of
the structure functions and Rosenbluth (or longitudinal/transverse)
separated $F_L$ data (see Sec.~\ref{ssec:gluon}).

\end{enumerate}

{\bf Vector boson production} (lepton pairs, $W^{\pm}$, and $Z^0$)
in LO proceed through $q \bar q$ annihilation.  A few key examples are:
\begin{enumerate}

\item
Lepton pair production in proton--proton and proton--neutron
collisions depend on the combinations
\begin{eqnarray*}
\sigma^{pp} &\sim& 4 u(x_a)\bar u(x_b) + d(x_a) \bar d(x_b)
		+ (x_a \leftrightarrow x_b) + \cdots	\\
\sigma^{pn} &\sim& 4 d(x_a)\bar d(x_b) + u(x_a) \bar u(x_b)
		+ (x_a \leftrightarrow x_b) + \cdots
\end{eqnarray*}
where $x_{a,b} = (M_B/\sqrt{s})\, e^{\pm y}$, and the\ $\cdots$\
indicate contributions from the $s$, $c$ and $b$ quarks.
Of course, the $pn$ cross section is obtained from data from
a deuterium target.  These cross sections can be used to
constrain the $\bar d/\bar u$ ratio, for example.

\item
$W^{\pm}$ production constrains products of the form $q \bar q'$
with specific weights given by the appropriate CKM matrix elements,
whereas $Z^0$ production constrains $q \bar q$.  For $p \bar p$
collisions at the Tevatron at large values of rapidity, one has
approximately,
\begin{eqnarray*}
\sigma^{W^+} &\sim& u(x_a) d(x_b) + \bar d(x_a) \bar u(x_b) + \cdots\\
\sigma^{W^-} &\sim& d(x_a) u(x_b) + \bar u(x_a) \bar d(x_b) + \cdots
\end{eqnarray*}
where $a\, (b)$ denotes the proton (antiproton), and the $\cdots$
represents contributions from heavier quarks.
Note that these results are written in terms of proton PDFs.
If the rapidity is large and positive, then $x_a > x_b$ and one can
neglect the antiquark terms so that these cross sections directly
constrain the $u {\rm \ and\ } d$ PDFs.  Due to the missing
neutrino from the $W$ decays, one cannot directly reconstruct the
rapidity distributions.  What is usually presented is the charged
lepton rapidity asymmetry for $W^{\pm}$ production.  In this case the
decay process means that the constraints on the PDFs are less direct,
but such measurements still provide useful constraints on the $d/u$
ratio at moderate values of $x$.  Of particular interest is the model
dependent determination of the $W$ rapidity asymmetry by the CDF
Collaboration \cite{Aaltonen:2009ta}, which directly constrains the
$d/u$ ratio at large values of $x$ since the asymmetry is for the $W$
and not the charged lepton from the decay.

\item
Recently the ATLAS Collaboration has presented an analysis
\cite{Aad:2012sb} using $W$ decay lepton and $Z^0$ rapidity
distribution data which shows potential for constraining the
strange quark PDF.
This analysis makes use of the fact that the $Z^0$ cross section
receives a significant contribution from $s \bar s$ annihilation
while the $W$ cross sections help constrain the $\bar u$ and
$\bar d$ PDFs.

\end{enumerate}

The other major class of observables includes inclusive
{\bf jet or photon, dijet}, and {\bf photon + jet production}.
Each of these constrain products of PDFs summed over all flavors.
Nevertheless, there are certain PDFs that can be constrained by
these observables.  This follows since the $u$ and $d$ PDFs are well
constrained by the DIS data and the vector boson and lepton pair
production data provide strong constraints on the $\bar u$ and
$\bar d$ PDFs.  Hence, these data have the greatest impact on
the gluon distribution.

\begin{enumerate}

\item
Direct photon production in LO has two subprocesses:
	$qg \to \gamma q$
and
	$q \bar q \to \gamma g$.
Hence, this process was though to be a good candidate for constraining
the gluon, especially at fixed target energies where the available
$x_T$ range extended out to about 0.6.  However, photons can also be
created by bremsstrahlung from the charged quarks -- often referred
to as the fragmentation process.  Indeed, one can define photon
fragmentation functions to help describe this production component.
It turns out that at fixed target energies this component receives
very large soft gluon corrections, requiring threshold resummation
techniques \cite{deFlorian:2005wf}.  This, combined with some apparent
disagreements between experimental data sets \cite{Aurenche:1998gv,
Aurenche:2006vj} meant that photon production has not fulfilled the
original promise of constraining the gluon PDF.
Recently, however, an analysis \cite{d'Enterria:2012yj} suggests that 
the use of {\it isolated} photon collider data may well help constrain
the gluon PDF.  The use of isolation cuts reduces the fragmentation
contribution and the use of higher energy collider reduces the need
for threshold resummation.

\item
High-$p_T$ jet production has played a significant role in
constraining the gluon PDF.  Even though the quark and antiquark
PDFs are well constrained by other types of data, there are still
significant contributions from $qg$ and $gg$ subprocesses 
\cite{Stump:2003yu}.

\item
Dijet production triple differential cross sections yield more
information than single jet cross sections because the rapidity of
the second jet is also constrained, thereby helping to constrain the
momentum fractions of the PDFs.  By tuning the rapidity ranges for the
two jets one can explore different regions for these momentum fractions.
An example is given in Ref.~\cite{Stump:2003yu}.

\item
Photon + jet production offers similar constraints, but now the 
subprocesses are weighted by the squared charge of the parton to
which the photon couples. 

\end{enumerate}

\subsubsection{Polarized experiments.}
\label{sssec:pol}

As with unpolarized measurements, historically most constraints on
spin-dependent PDFs have come from {\bf polarized charged-lepton DIS}
experiments.

\begin{enumerate}

\item
In the one-photon exchange approximation,
the difference of inclusive DIS cross sections for
leptons polarized longitudinally with spin parallel and antiparallel
to the target hadron polarization measures $C$-even quark combinations
$\Delta q^+$.  For proton targets, one has the combination
$$
4 \Delta u^+ + \Delta d^+ + \Delta s^+,
$$
while for the neutron the combination would be 
$$
4 \Delta d^+ + \Delta u^+ + \Delta s^+,
$$
with contributions from heavy quark polarization expected to be
negligible.  In practice, polarized $^3$He targets are usually
used as effective sources of polarized neutron, since the neutron
carries almost 90\% of the spin of $^3$He, while polarized deuterons,
to which the proton and neutron spins contribute equally, are used
to provide the isoscalar combination
$$
  5 (\Delta u^+ + \Delta d^+) + 2 \Delta s^+.
$$
Because of the greater sensitivity of the proton $g_1$ structure
function measurements to the polarized $u$ quark PDF, and
uncertainties associated with nuclear corrections when extracting
the neutron from $^3$He or deuterium data, the uncertainty on
the $\Delta d$ PDF is significantly larger at high $x$ than on
$\Delta u$.

\item
At NLO, the polarized gluon distribution $\Delta g$ also enters
in the $g_1$ structure function.  The $Q^2$ evolution of the
flavor singlet contribution to $g_1$ can then be used to constrain
$\Delta g$, as for the unpolarized gluon PDF discussed above,
although the constraints are weaker in the polarized because
$\Delta g/g \ll 1$ at small $x$.

\end{enumerate}

{\bf Semi-inclusive DIS} provides additional independent combinations
of spin-dependent PDFs that can be used to reconstruct individual
quark and antiquark flavors.

\begin{enumerate}

\item
Semi-inclusive production of hadrons $h$ in the current fragmentation
region, primarily pions or kaons, is proportional to products of
spin-dependent PDFs and quark $\to$ hadron fragmentation functions,
$$
\sum_q e_q^2\, \Delta q(x)\, D_q^h(z),
$$
where $z$ is the fraction of the quark's energy carried by the hadron
$h$.  The fragmentation functions $D_q^h$ can be determined from other
reactions, such as inclusive hadron production in $e^+ e^-$ annihilation.
One can weight particular quark or antiquark flavors by selecting
favored (such as $D_u^{\pi^+}$ or $D_{\bar d}^{\pi^+}$) or unfavored
($D_d^{\pi^+}$ or $D_{\bar u}^{\pi^+}$) fragmentation functions for
specific hadrons (in this case $h=\pi^+$).  Information on the
polarized strange quark PDF $\Delta s$ in particular can be obtained
from data on $K$ production.

\item
The polarized gluon distribution $\Delta g$ can be constrained by
semi-inclusive DIS data on charmed or high-$p_T$ hadron production
through the photon--gluon fusion process.  Recently the analysis of
this process has been performed at NLO, extending earlier LO
extractions of $\Delta g$.

\end{enumerate}

Inclusive particle production in {\bf polarized proton--proton
collisions} provides an additional method of determining
spin-dependent sea quark and gluon distributions.

\begin{enumerate}

\item
$W^\pm$ production cross sections for scattering longitudinally
polarized protons from unpolarized protons,
	$\vec p\, p \to W^\pm\, X$,
depend on products of spin-dependent and spin-averaged PDFs,
for example,
\begin{eqnarray*}
\Delta\sigma^{W^+}
&\sim& \Delta \bar d(x_a) u(x_b) - \Delta u(x_a) \bar d(x_b)	\\
\Delta\sigma^{W^-}
&\sim& \Delta \bar u(x_a) d(x_b) - \Delta d(x_a) \bar u(x_b).
\end{eqnarray*}
At large positive or negative rapidities, $x_a \gg x_b$ or $x_a \ll x_b$,
the cross sections (or asymmetries) are dominated by a single flavor,
while at mid-rapidities both $u$ and $d$ flavors contribute.

\item
Inclusive jet or $\pi^0$ production in double-polarized proton--proton
scattering,
	$\vec p\, \vec p \to {\rm jet}/\pi^0 + X$,
is sensitive to the polarized gluon PDF.  The first evidence for a small,
but nonzero $\Delta g$ was recently observed by the STAR Collaboration
at RHIC in jet data at $\sqrt{s}=200$~GeV.  Direct photon production in
polarized $pp$ scattering has also been suggested \cite{Berger:1988ke}
as a means of probing $\Delta g$.

\end{enumerate}

\section{Unpolarized parton distributions}
\label{sec:uno}

Using the technology outlined in Sec.~\ref{sec:QCD}, a number of
global QCD analyses of the world's high-energy scattering data have
produced sets of proton PDFs, up to next-to-leading order (NLO)
or next-to-next-to-leading order (NNLO) accuracy.
The efforts have indeed been global, with groups in Europe and
the US in the forefront of the data analyses.
The PDFs sets include parametrizations from the MSTW
\cite{Martin:2009iq} and ABM \cite{Alekhin:2012ig} groups,
both of which use standard global fitting methodology;
the NNPDF \cite{Ball:2012cx} collaboration, which uses a newer
approach based on neural networks;
the JR \cite{JimenezDelgado:2009tv} PDFs, which are dynamically
generated through $Q^2$ evolution from a low $Q^2$ input scale;
and the HERAPDF \cite{Aaron:2009aa} group, which includes
only data from the H1 and ZEUS experiments at HERA.
The US-based efforts have been centered around the CTEQ Collaboration
\cite{CTEQweb}, which at present involves two derivative analyses of
nucleon PDFs, by the CT \cite{Lai:2010vv} and CJ (CTEQ-Jefferson Lab)
\cite{CJ12} groups, as well as the nCTEQ \cite{Schienbein:2009kk,
Kovarik:2010uv} analysis of nuclear PDFs.  The CJ Collaboration in
particular has focused on developing the methodologies needed for
describing data over a broad energy range including the low-$Q^2$
and $W$ domain \cite{CJ10}.

In this section we summarize the results for unpolarized PDFs
from the various PDF fitting groups, discussing their similarities
and contrasting their differences.
We will focus mostly on the physics issues, rather than on technical
aspects of PDF fitting.  Of course with new data arriving or soon
anticipated at the LHC, Jefferson Lab, and other facilities,
the global fitting efforts are constantly evolving, so that the
information presented here can only be viewed as a snapshot of the
field at the present time.

\begin{figure}[bt]
\begin{center}
\includegraphics[width=8cm]{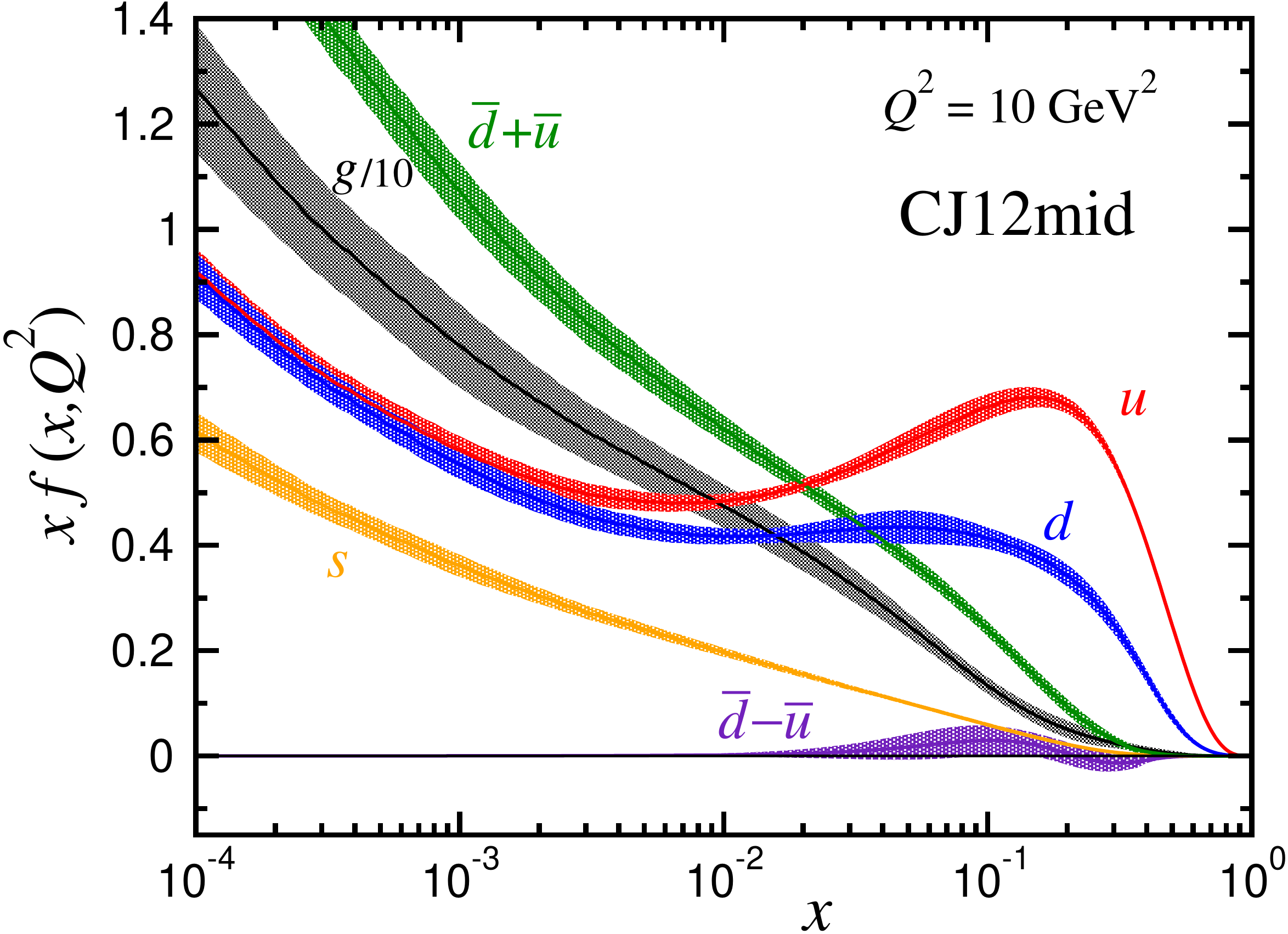}
\includegraphics[width=7.2cm]{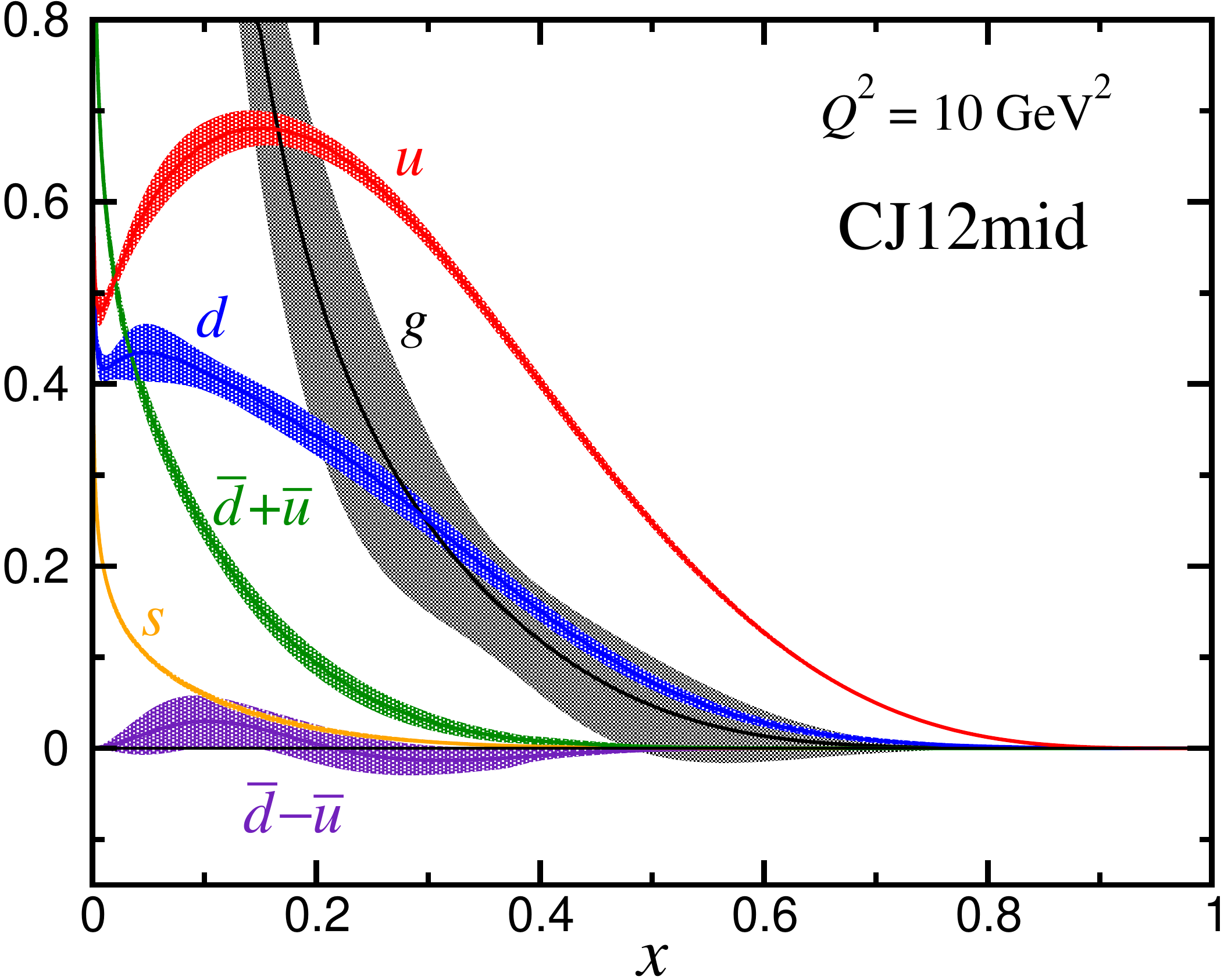}
\caption{Uncertainty bands for the $xu$ (red), $xd$ (blue),
	$x(\bar d+\bar u)$ (green), $x(\bar d-\bar u)$ (violet),
	$xs$ (orange) and $xg$ (black) PDFs for the CJ12mid
	fit \cite{CJ12} at $Q^2=10$~GeV$^2$, shown on
	logarithmic {\it (left)} and linear {\it (right)}
	scales in $x$.  Note that in the left panel the gluon
	is scaled by a factor 1/10.}
\label{fig:allpdfs}
\end{center}
\end{figure}

As a typical example of modern PDFs and their uncertainties,
Fig.~\ref{fig:allpdfs} shows the $xu$, $xd$, $x(\bar d+\bar u)$,
$x(\bar d-\bar u)$, $xs$ and $xg$ PDFs (with the gluon scaled by a
factor 1/10) for the CJ12 fit \cite{CJ12} (for the case of moderate
nuclear corrections, CJ12mid) at a scale of $Q^2=10$~GeV$^2$.
The general behavior of the PDFs is similar for all the
parametrizations \cite{JimenezDelgado:2009tv, Alekhin:2012ig,
Martin:2009iq, Lai:2010vv, Aaron:2009aa, Ball:2012cx, CJ12},
particularly where sufficient data exist to constrain the distributions
in specific regions of $x$ and $Q^2$.  In regions where data are
scarce, such as at very low $x$ ($x \lesssim 10^{-4}$) or high $x$
($x \gtrsim 0.4-0.6$) for certain observables, or where data on
specific PDFs (such as the strange quark) are subject to large
experimental or theoretical uncertainties, the details of the PDFs
can depend strongly on the assumptions adopted in the extrapolations.

\begin{figure}[t]
\begin{center}
\includegraphics[width=12cm]{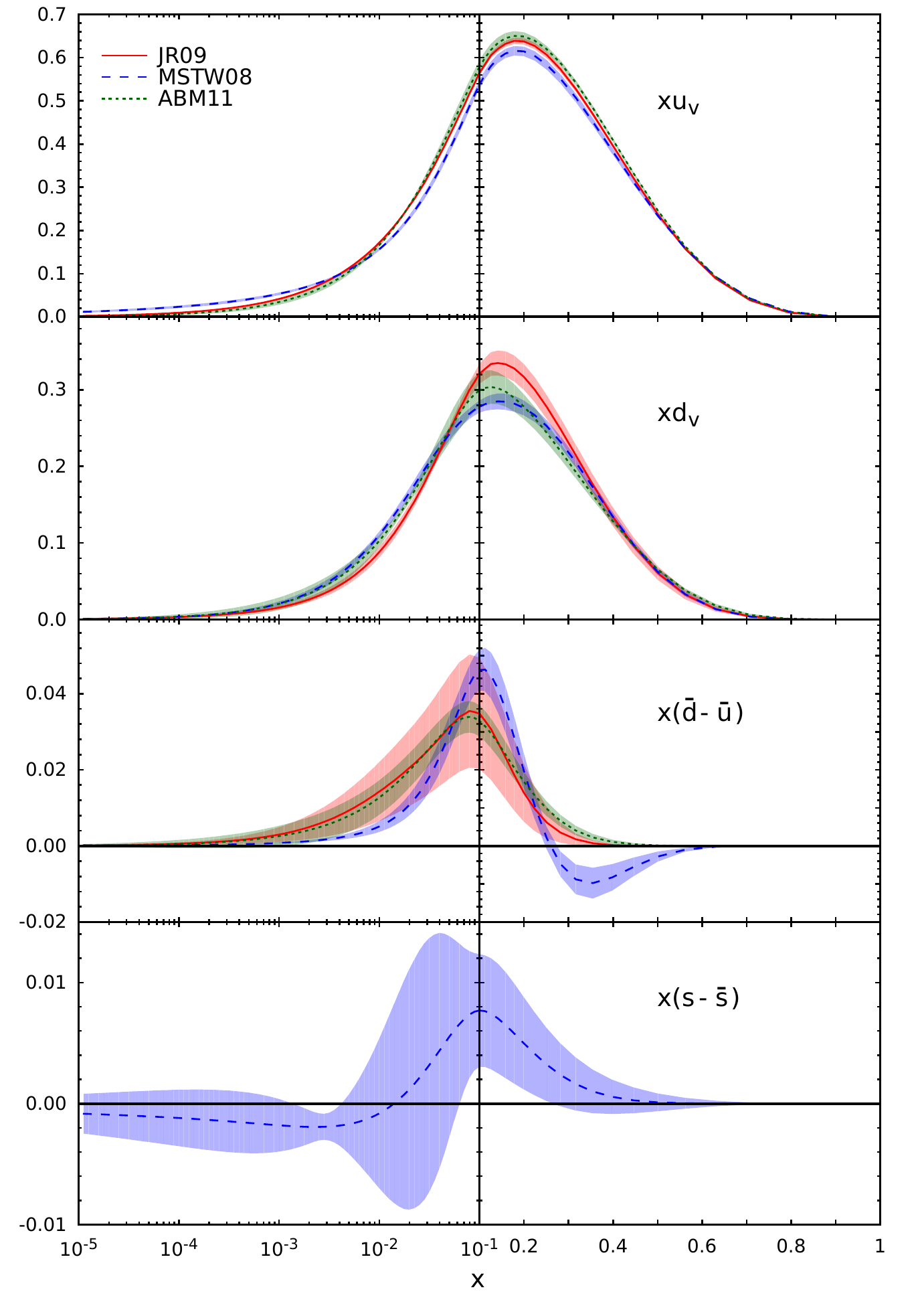}
\caption{Comparison of nonsinglet PDF combinations, including
	the valence $xu_v$ and $xd_v$, and the $x(\bar d-\bar u)$
	and $x(s-\bar s)$ sea distributions, for the available
	3-flavor NNLO PDF sets from JR09 \cite{JimenezDelgado:2009tv}
	(red solid line), MSTW08 \cite{Martin:2009iq} (blue dashed)
	and ABM11 \cite{Alekhin:2012ig} (green dotted),
	at $Q^2=10$~GeV$^2$.}
\label{fig:NSpdf}
\end{center}
\end{figure}
\begin{figure}[t]
\begin{center}
\includegraphics[width=12cm]{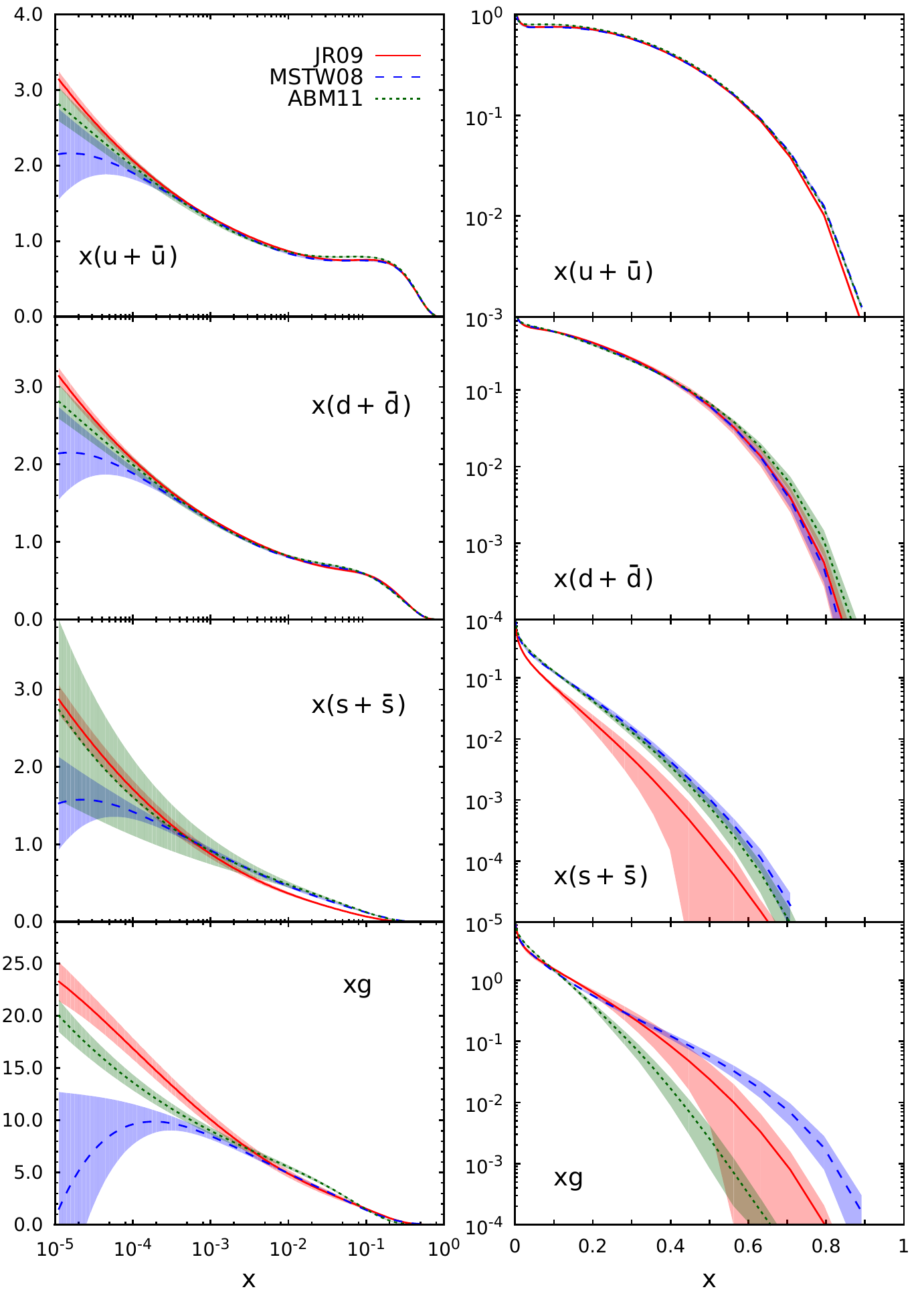}
\caption{As in Fig.~\ref{fig:NSpdf}, but for the singlet-sector
	distributions $x(u+\bar u)$, $x(d+\bar d)$ and $x(s+\bar s)$
	and the gluon $xg$ PDF.}
\label{fig:Spdf}
\end{center}
\end{figure}

Some of these differences are evident in Figs.~\ref{fig:NSpdf} and
\ref{fig:Spdf}, which illustrate the flavor nonsinglet combinations
$\{ x(u-\bar u),\, x(d-\bar d),\, x(\bar d-\bar u),\, x(s-\bar s) \}$,
and the singlet-sector distributions
$\{ x(u+\bar u),\, x(d+\bar d),\, x(s+\bar s),\, xg \}$,
respectively, for the NNLO MSTW08 \cite{Martin:2009iq},
ABM11 \cite{Alekhin:2012ig}, NNPDF \cite{Ball:2012cx}
and JR09 \cite{JimenezDelgado:2009tv} PDF sets.
While the valence $u_v$ and $d_v$ PDFs are fairly well constrained
in the intermediate-$x$ region, at large $x$ there are significant
uncertainties on the $d_v$ distribution in particular.
The SU(2) nonsinglet distribution $\bar d-\bar u$ is well
determined by data from the Drell-Yan reaction over the range
$x \approx 0.05-0.25$, but is not constrained at higher $x$.
The strange nonsinglet distribution is the most difficult to
determine, requiring a combination of neutrino and antineutrino
scattering data, which are subject to large experimental and
nuclear uncertainties.

For the $C$-even distributions in Fig.~\ref{fig:Spdf}, the $u+\bar u$
and $d+\bar d$ are the dominant quark distributions at large $x$, but
the strange quark PDF becomes increasing more important at small $x$.
Numerically, the gluon distribution dominates all other PDFs at small
$x$, but has sizable uncertainties at large $x$ values.
In the following sections we discuss features of each of the PDFs in
more detail.

\subsection{Valence quarks at large $x$}
\label{ssec:valence}

Valence quarks give the global properties of the nucleon, such as its
charge and baryon number.  Knowledge of their momentum distributions
is important for many reasons, especially at high values of $x$ where
a single quark carries most of the nucleon's momentum.  The large-$x$
region is in fact a unique laboratory for studying the nonperturbative
flavor and spin dynamics of quarks \cite{Feyn72, Close:1973xw,
Melnitchouk:1995fc, Holt:2010vj}, as well as testing predictions from
perturbative QCD for the behavior of PDFs in the limit $x \to 1$
\cite{Farrar:1975yb, Blankenbecler:1974tm, Gunion:1973nm,
BrodskyLepage79}.

Reliable determination of PDFs at large $x$ is also important for
searches for new physics beyond the Standard Model in collider
experiments at the LHC.  Through perturbative QCD evolution,
uncertainties in PDFs from fixed-target experiments at high $x$
and low $Q^2$ can propagate to larger $Q^2$ to affect cross sections
at smaller $x$ values \cite{Kuhlmann:1999sf, Brady:2011hb}.
This is especially true for the forward production of particles of
mass $m$ at large rapidities $y$, whose cross sections are given
by products of PDFs with one evaluated at small
	$x \approx (m/\sqrt{s})\, e^{-y}$
and the other at large
	$x \approx (m/\sqrt{s})\, e^y$.
The production of heavy $W'$ and $Z'$ bosons, for example, is sensitive
to $d$ quark PDF uncertainties at high rapidities, exceeding 100\% in
the $W'^-$ channel, which places limits on the accuracy of cross section
measurements for masses near the kinematic thresholds \cite{Brady:2011hb}.
Furthermore, understanding PDFs at large $x$ is vital for the analysis
of neutrino oscillation experiments \cite{Itow:2001ee, Ayres:2004js,
Raby:2008pd}, where one of the most significant uncertainties comes
from neutrino--nucleus cross sections at the interface of the DIS and
resonance regions.

The growing need to better understand large-$x$ PDFs and their
uncertainties has been reflected in the greater attention being paid
recently to the physics of the large-$x$ region, with dedicated
experiments planned at Jefferson Lab following its 12~GeV upgrade
\cite{BONUS12, MARATHON, SOLID}, as well as at proposed new
facilities such as the LHeC \cite{AbelleiraFernandez:2012ty}
and the Electron-Ion Collider \cite{Accardi:2012hwp}.
It has also been a catalyst for the recent concerted theoretical
efforts by the CJ Collaboration \cite{CJweb} in their global QCD
analyses of PDFs extending into the lower $Q^2$ and $W^2$ regions,
with the aim of providing better constraints on PDFs at large
values of $x$ \cite{CJ10, CJ11, CJ12}.

\begin{center} 
\begin{figure}[ht]
\includegraphics[width=7.5cm]{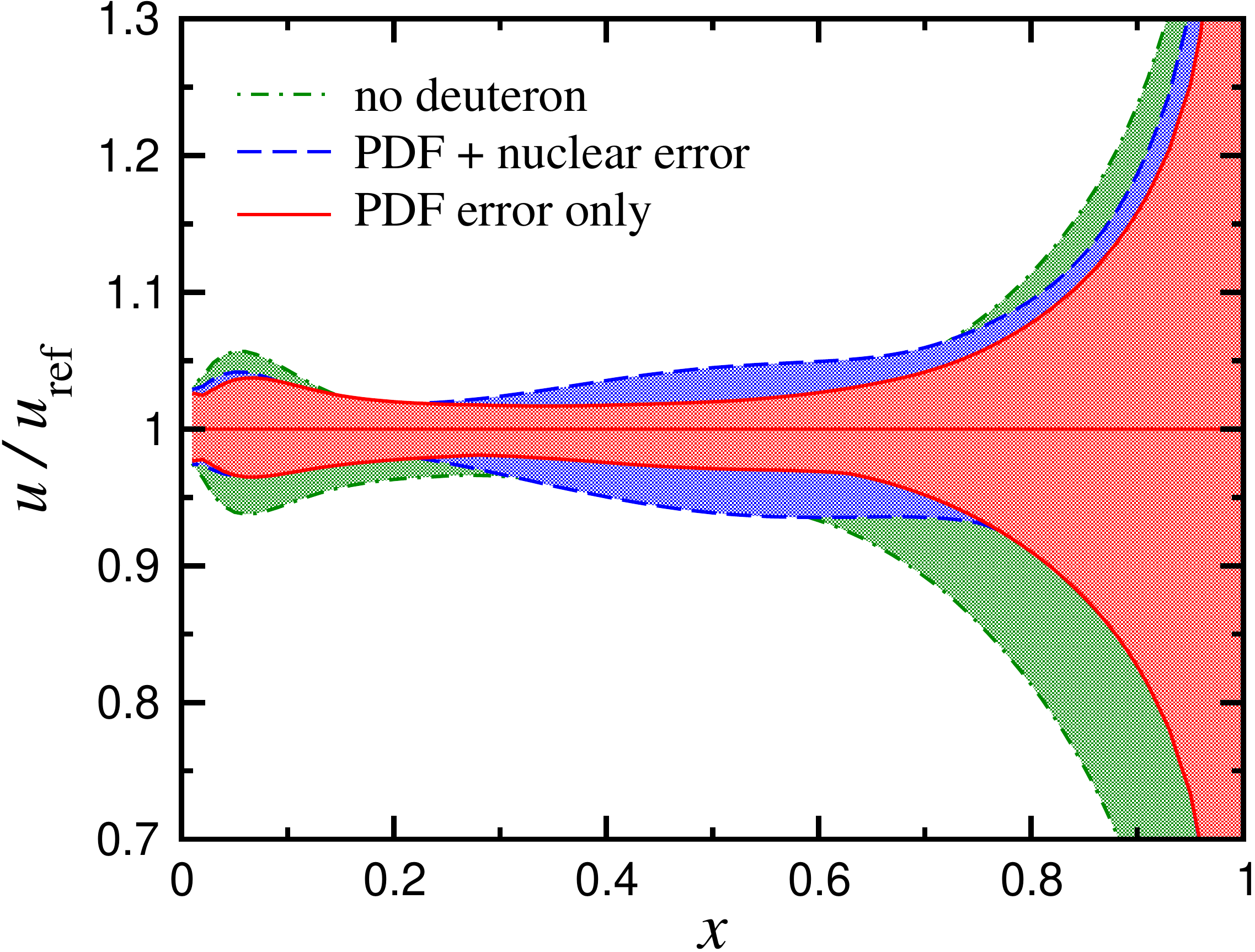}\ \ \ \
\includegraphics[width=7.5cm]{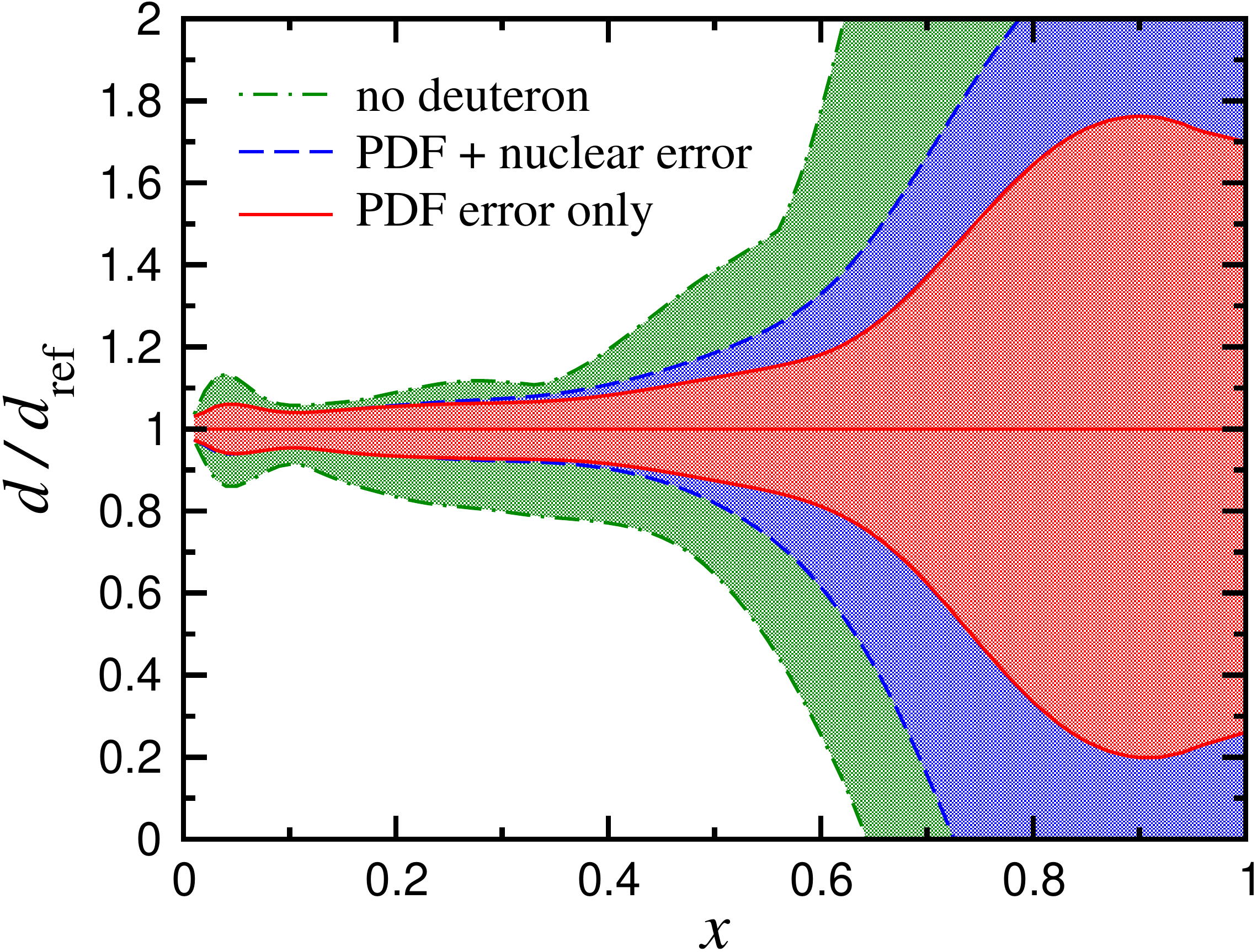}
\caption{$u$ and $d$ quark distributions from the CJ12mid PDFs
	\cite{CJ12}, with PDF errors only (solid, red shaded) with
	combined PDF and nuclear uncertainties (dashed, blue shaded),
	and with a fit excluding all deuterium data (dot-dashed,
	green shaded), relative to the reference CJ12mid set at
	an arbitrary scale $Q^2=100$~GeV$^2$.  Note the different
	vertical scales for the $u$ and $d$ quark PDFs.}
\label{fig:duNOD}
\end{figure}
\end{center}

The current constraints on the $u$ and $d$ quark distributions are
illustrated in Fig.~\ref{fig:duNOD} for the CJ12mid PDF set \cite{CJ12},
where the uncertainties arising from nuclear corrections and PDF
(experimental) errors are indicated separately.
Since the $u$ quark PDF is relatively well constrained by the proton
DIS data at large $x$, the effect of the nuclear uncertainties is
minor, increasing the total uncertainty at intermediate $x$ by a
few percent.  For the $d$ quark distribution, on the other hand,
whose determination at present requires both proton and deuterium
DIS data, the nuclear correction uncertainties in the deuteron
significantly increase the overall error for $x \gtrsim 0.6$.
However, even though the use of deuterium data introduces the need
to confront the problem of nuclear effects and their uncertainties,
without these data the error bands on the $d$ quark PDF would be even
larger over most of the $x$ range, as Fig.~\ref{fig:duNOD} demonstrates.

In addition to accounting for the finite-$Q^2$ (TMC and higher twist)
and nuclear effects which become increasingly important as $x$ tends
to 1 at fixed $Q^2$, the behavior of PDFs in the $x \to 1$ limit also
depends critically on the functional form used for the $x$ dependence.
The conventional parametrizations utilized in most global PDF analyses
assume a $\sim (1-x)^{a_2}$ dependence for both the $u$ and $d$ quark
PDFs, as in Eq.~(\ref{eq:param}) above, so that in the limit as
$x \to 1$, the $d/u$ ratio tends either to zero or infinity.
Accardi {\it et al.} \cite{CJ11} allowed for a more flexible
parametrization of the valence $d$ quark PDF, in which the $d_v$
distribution receives a small admixture from the $u_v$ PDF,
\begin{equation}
d_v \to d_v + b\, x^c u_v,
\label{eq:du}
\end{equation}
with $b$ and $c$ as free parameters.  Provided the $d$ quark has
a softer momentum dependence (larger exponent $a_2$) than the $u$
quark, which agrees with phenomenology, the ratio of the distributions
will approach a constant value, $d_v/u_v \to b$, as $x \to 1$.
A~nonzero, finite value in this limit is in fact expected in
several different nonperturbative models of nucleon structure
\cite{Melnitchouk:1995fc, Holt:2010vj, Farrar:1975yb}.

\begin{figure}[ht]
\begin{center}
\includegraphics[width=7.5cm]{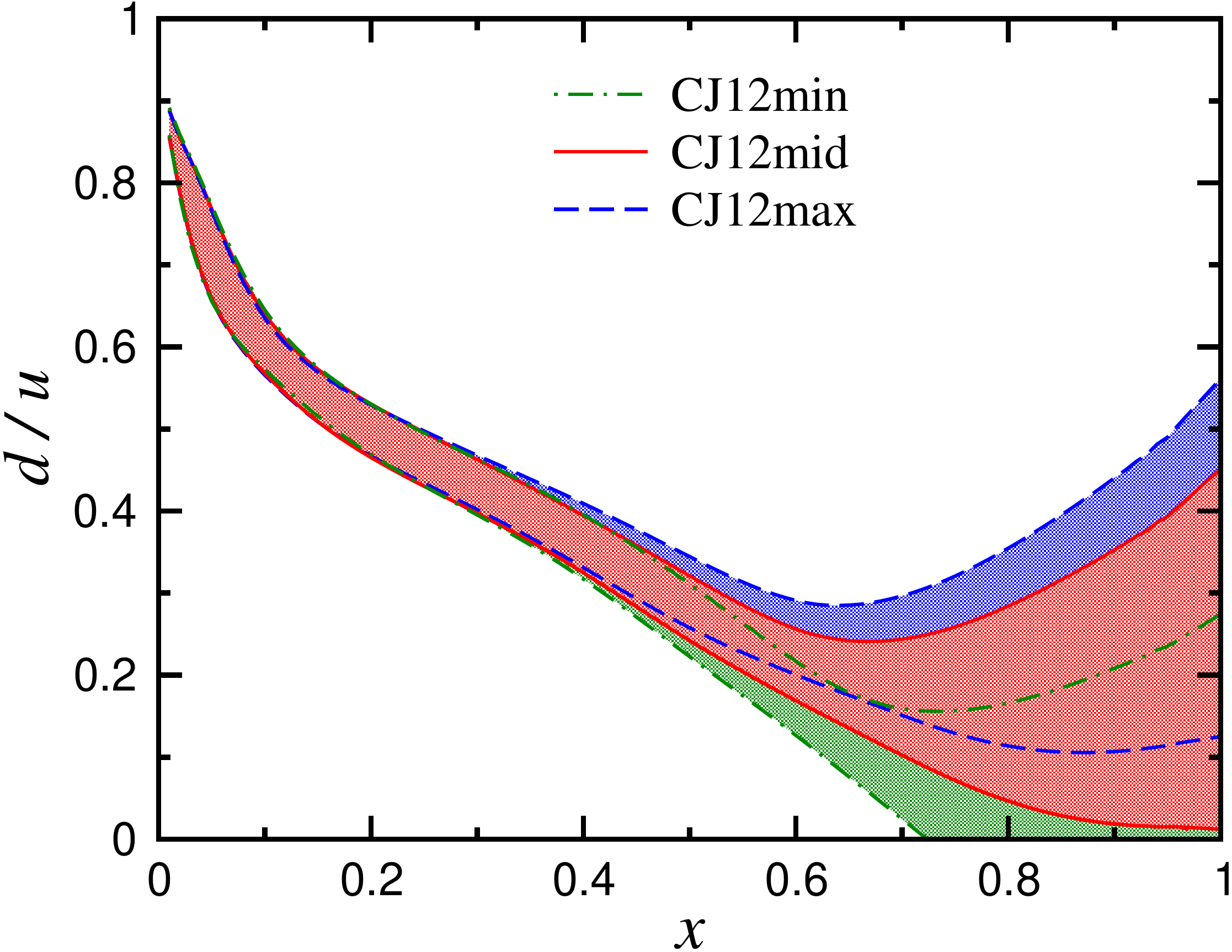}\ \ \ \
\includegraphics[width=7.5cm]{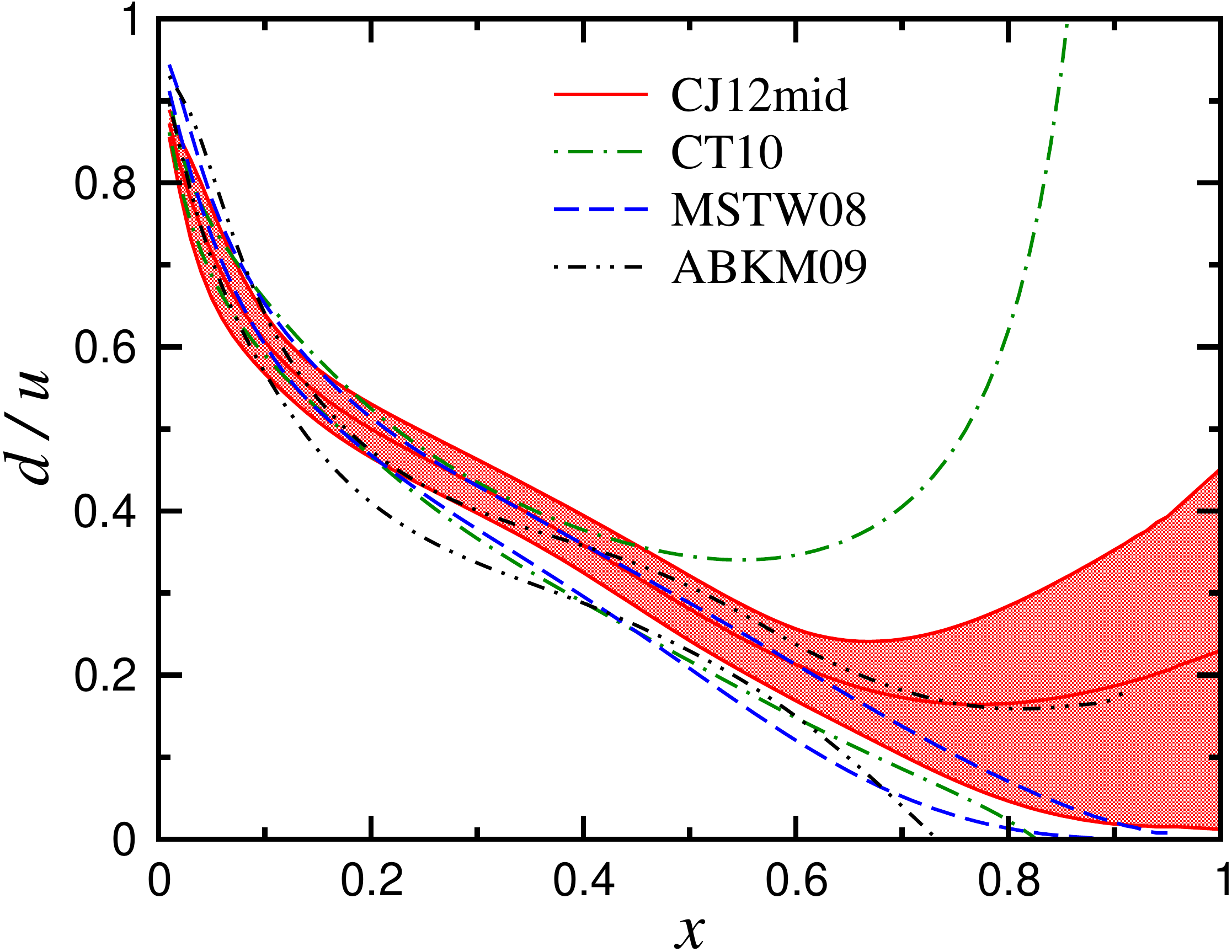}
\caption{$d/u$ ratio for the 
	{\it (left)} CJ12min (dot-dashed), CJ12mid (solid) and
	CJ12max (dashed) PDFs \cite{CJ12}, and
	{\it (right)} CJ12mid (solid) compared with the CT10
	\cite{Lai:2010vv} (dot-dashed), MSTW08 \cite{Martin:2009iq}
	(dashed) and ABKM09 \cite{Alekhin:2009ni} (dot-dot-dashed)
	parametrizations at $Q^2=100$~GeV$^2$.}
\label{fig:duPDF}
\end{center}
\end{figure}

The ratio of the $d$ to $u$ quark distributions is illustrated in
Fig.~\ref{fig:duPDF} for several modern NLO PDF sets, including the
CJ12 \cite{CJ12}, CT10 \cite{Lai:2010vv}, MSTW08 \cite{Martin:2009iq}
(dashed) and ABKM09 \cite{Alekhin:2009ni} distributions.
As expected, the total uncertainty on the $d/u$ ratio at large $x$
increases when different nuclear correction models are considered,
with the central values extrapolated to $x=1$ increasing from
$d/u \approx 0$ for the CJ12min PDFs, with the minimum nuclear
correction model, to $d/u \approx 0.3$ for the CJ12max set with
the maximum nuclear correction model from Ref.~\cite{CJ12}.
Combining all uncertainties, the CJ12 analysis found
$d/u \to 0.22 \pm 0.20 \, \rm{[{\small PDF}]} \pm 0.10 \,\rm{[nucl]}$
as $x \to 1$, where the first error is from the PDF fits and the
second is from the nuclear correction models.

In contrast, using the more restrictive, conventional parametrizations,
the $d/u$ ratios for the MSTW08 and ABKM09 distributions tend to zero
in the $x \to 1$ limit, while that for CT10 tends to infinity.
This behavior distorts somewhat the error bands, making the MSTW08 ratio
appear to have an anomalously small uncertainty, while the CT10 result
has a much larger one.  The relative errors, however, are similar for
the MSTW08 and CT10 PDFs, while the CJ12 and ABKM09 uncertainties are
reduced at larger $x$ because of the additional high-$x$ data used in
those analyses.
Unfortunately, the extrapolated values of $d/u$ at $x=1$ span most
of the range of available predictions \cite{Feyn72, Close:1973xw,
Melnitchouk:1995fc, Holt:2010vj, Farrar:1975yb}, so that discrimination
between the physical mechanisms that lead to the different $x \to 1$
behaviors will only be possible with constraints from new experiments
that do not involve deuterium data and the nuclear model uncertainties
\cite{BONUS12, MARATHON, SOLID}.

\subsection{Light quark sea}
\label{ssec:lightsea}

Inclusive electromagnetic DIS probes $C$-even combinations of PDFs,
$q + \bar q$, weighted by their squared charges, or in terms of
valence and sea quark (or antiquark) distributions, $q_v + 2 \bar q$.
The shapes of the quark valence and sea distributions differ markedly
as a function of $x$.  As we have seen in the preceding section, the
valence quark PDFs dominate at intermediate and large values of $x$,
but vanish as $x q_v \to 0$ for $x \to 0$.  The sea, in contrast,
dominates at small $x$, but is strongly suppressed as $x \to 1$.
For the light antiquarks $\bar u$ and $\bar d$, structure function
data with proton targets are sensitive at small $x$ to the combination
$4 \bar u + \bar d$, while deuterium targets (nuclear corrections aside)
probe the isosinglet combination $\bar u + \bar d$.
Under $Q^2$ evolution, the singlet quark distribution mixes with the
gluon, which leads to the rapid growth in the antiquark PDFs with
decreasing $x$.  With increasing $Q^2$, the uncertainties on
$\bar u + \bar d$ also decrease.

Inclusive DIS measurements can therefore provide important constraints
on the behavior of both the $q_v$ and $\bar q$ distributions in the
regions where their contributions are dominant.
However, in the intermediate-$x$ region, $x \sim 0.01-0.1$, where
the magnitude of valence and sea quark PDFs is comparable, additional
information is required to uniquely determine the distributions
individually.
Here data from neutrino scattering can allow access to the
parity-violating $F_3$ structure function, which measures $C$-odd
combinations of PDFs, $q-\bar q$.
Combined with the $C$-even PDFs from the parity-conserving $F_1$
and $F_2$ structure functions, one can then uniquely determine the
$q$ and $\bar q$ distributions individually.
Unfortunately, neutrino scattering cross sections are usually
statistics limited compared with electromagnetic data, and with
few exceptions, have necessitated the use of nuclear targets such
as iron or lead as a means of increasing the rates.
This subsequently brings with it the inherent complications associated
with accounting for nuclear corrections in relating the nuclear
structure functions with those in a free nucleon \cite{Kulagin:2004ie,
Kulagin:2007ju}.

More direct constraints on $\bar q$ distributions can be obtained
from lepton pair production in inclusive hadron--hadron scattering,
which involves products of PDFs evaluated at different values of
beam ($x_a$) and target ($x_b$) momentum fractions.  As discussed
in Sec.~\ref{sssec:h-h}, the Drell-Yan process or $W^\pm$-boson
production in $pp$ scattering at nonzero rapidity, for example,
allows the antiquark distribution at small $x_a$ to be extracted
with knowledge of the corresponding quark distribution at $x_b$.

\begin{figure}[t]
\begin{center}
\includegraphics[width=12cm]{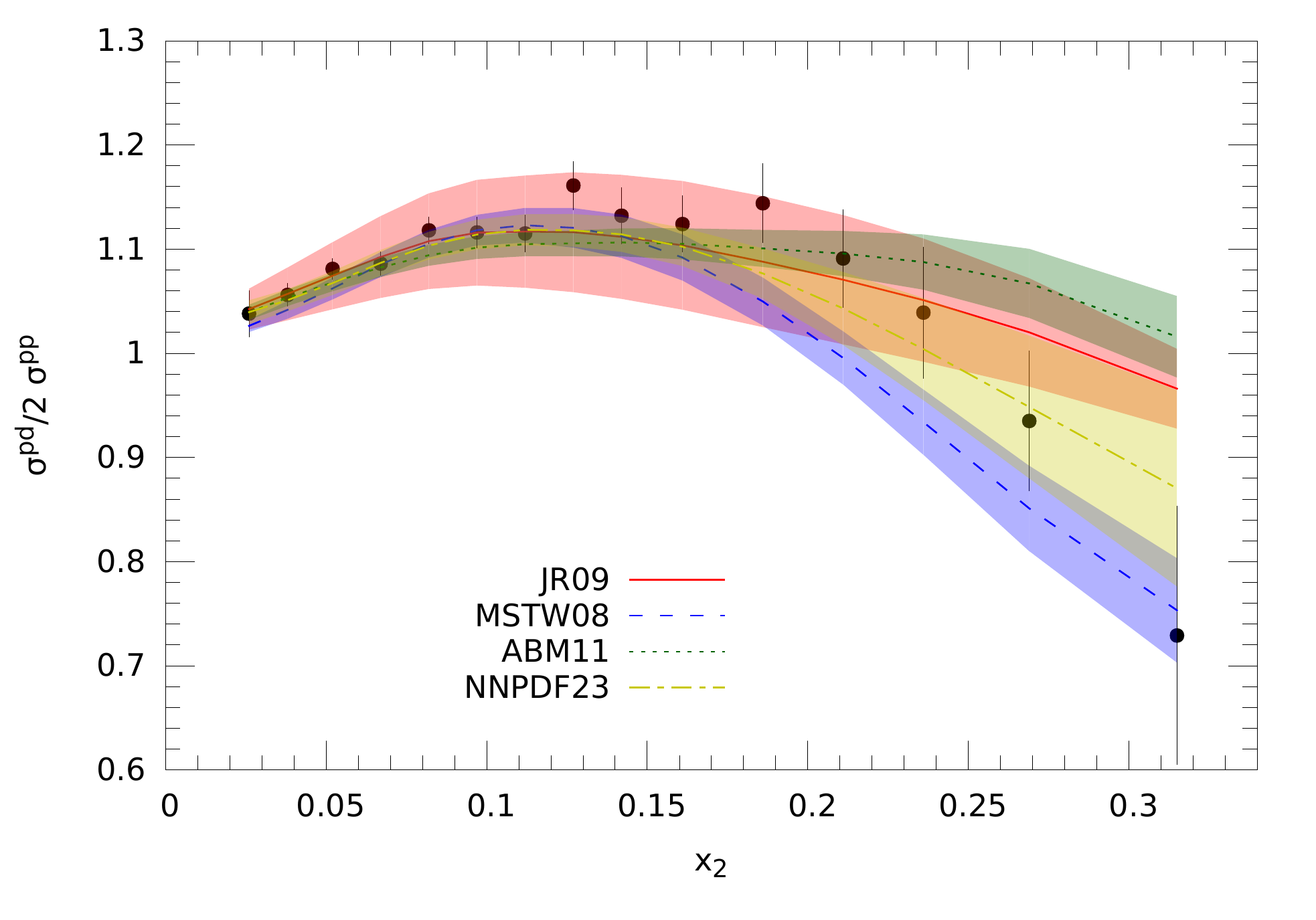}
\vspace*{-0.5cm}
\caption{Ratio of lepton pair production cross sections in
	$pd$ and $pp$ scattering as a function of the fractional
	momentum $x_b$ of the target parton.  The data points from
	the Fermilab E866/NuSea Collaboration \cite{Towell:2001nh}
	are compared with the cross section ratio calculated using
	the JR09 \cite{JimenezDelgado:2009tv} (red solid line),
	MSTW08 \cite{Martin:2009iq} (blue dashed), and ABM11
	\cite{Alekhin:2012ig} (green dotted), and NNPDF23
	\cite{Ball:2012cx} (yellow dot-dashed) PDF parametrizations.}
\label{fig:dbar_ubar}
\end{center}
\end{figure}

In particular, the Drell-Yan reaction \cite{Drell:1970wh} has been
used by the NA51 Collaboration at CERN \cite{NA51} and the E866/NuSea
Collaboration at Fermilab \cite{Hawker:1998ty, Towell:2001nh} to
determine the ratio of $\bar d$ to $\bar u$ distributions in the proton.
Naive expectations from perturbative QCD based on $g \to q\bar q$
splitting suggest that since the masses of the $u$ and $d$ quarks
are very similar, the production of $u\bar u$ and $d \bar d$ pairs
should be nearly identical, even with the inclusion of higher-order
radiative corrections \cite{Ross:1978xk}.
The first indication of a significant asymmetry in the proton sea was
obtained by the New Muon Collaboration (NMC) \cite{Amaudruz:1991at,
Arneodo:1994sh} at CERN, who performed an accurate measurement of the
ratio of DIS cross sections for hydrogen and deuterium, from which
the Gottfried sum $S_G$ was determined~\cite{Gottfried:1967kk},
\begin{eqnarray}
S_G &=& \int_0^1 {dx \over x}\, \left( F_2^p - F_2^n \right)\
     =\ {1 \over 3}\ +\ {2 \over 3} \int_0^1 dx\, (\bar u - \bar d),
\end{eqnarray}
where charge symmetry is assumed (see Sec.~\ref{ssec:csv} below).
For a flavor symmetric SU(2) proton sea, the prediction for the
Gottfried sum would be $S_G = 1/3$.  In contrast, the NMC
measurement found $S_G = 0.235 \pm 0.026$ \cite{Arneodo:1994sh},
indicating a strong violation of flavor symmetry in the light
antiquark sea,
\begin{eqnarray}
\int_0^1 dx\, (\bar d - \bar u) &=& 0.148 \pm 0.039.
\end{eqnarray}
Following the suggestion by Ellis and Stirling \cite{Ellis:1990ti}
that the light antiquark sea could be more directly probed in
the Drell-Yan process, the NA51 experiment used the ratio of
dimuon cross sections for $pd$ and $pp$ scattering to extract
the ratio $\bar d/\bar u = 1.96 \pm 0.15 \pm 0.19$ at an average
$\langle x \rangle = 0.18$ at a Feynman-$x$ of
$x_F = x_a - x_b \approx 0$.
Subsequently, the E866/NuSea experiment at Fermilab measured
the $\sigma^{pd}/\sigma^{pp}$ ratio at large $x_F$, where at
leading order
\begin{eqnarray}
{\sigma^{pd} \over 2\sigma^{pp}}
&\approx& {1\over 2}
	  \left[ 1 + {\bar d(x_b) \over \bar u(x_b)} \right],\ \ \ \ \
x_a \gg x_b.
\label{eq:sig_pd_pp}
\end{eqnarray}
The results, illustrated in Fig.~\ref{fig:dbar_ubar}, confirmed a
significant deviation of the cross section ratio from the naive
perturbative QCD expectation, indicating a large asymmetry between
$\bar d$ and $\bar u$.

The earliest efforts to explain an enhancement of $\bar d$ quarks
over $\bar u$ in the proton invoked the effects of Pauli blocking,
which arise from the presence of different numbers of valence $u$
and $d$ quarks in the proton \cite{Field:1976ve}.  Estimates of this
effect are very difficult to quantify reliably \cite{Schreiber:1991tc,
Signal:1991ug}, with some calculations \cite{Steffens:1996bc} even
suggesting that it produces a small excess of $\bar u$ over $\bar d$.
Alternative explanations have focused on the role of chiral symmetry
breaking and the pion cloud of the nucleon \cite{Thomas:1983fh}.
Because of the Heisenberg uncertainty principle, a component of the
proton's wave function involves its dissociation into a nucleon and
a virtual pion state.  The preferential coupling of a proton to a
$n \pi^+$, with the $\pi^+$ containing a valence $u \bar d$ pair,
implies a natural mechanism for generating an excess of $\bar d$
over $\bar u$.  While many models have been proposed to compute this
effect quantitatively (see Refs.~\cite{Speth:1996pz, Kumano:1997cy}
for reviews), the underlying physics principles are model independent
and simply the chiral symmetry properties of QCD \cite{Thomas:2000ny,
Burkardt:2012hk}.

While most of the modern PDF parametrizations are able to accommodate
an enhanced $\bar d$ sea over the range $0.02 \lesssim x \lesssim 0.25$,
there is some uncertainty in the trend of the $\bar d/\bar u$ ratio
at larger $x$.  The new Drell-Yan experiment E906/SeaQuest at Fermilab
\cite{SeaQuest} will extend the $x$ coverage up to $x \approx 0.45$,
and a proposal at the J-PARC facility in Japan \cite{J-PARC-P04,
Kumano:2010qm} would extend the range even further.

\subsection{Charge symmetry violation in PDFs}
\label{ssec:csv}

Most PDF analyses have been performed under the assumption of charge
symmetry, or independence of interactions under rotations in isospin
space.  Charge symmetry implies that the $u$ quark distribution in
the proton is identical to the $d$ quark distribution in the neutron,
and {\it vice versa}, and is believed to be accurate in nature at the
$\lesssim 1\%$ level.  This symmetry is broken explicitly by light
quark mass differences, $m_u \not= m_d$, as well as by electromagnetic
corrections.  Such effects lead to nonzero values of the ``majority''
and ``minority'' charge symmetry violating asymmetries in the nucleon,
defined as
\begin{eqnarray}
\delta u(x,Q^2) &=&  u^p(x,Q^2) - d^n(x,Q^2),
\label{eq:csv_u}					\\
\delta d(x,Q^2) &=&  d^p(x,Q^2) - u^n(x,Q^2),
\label{eq:csv_d}
\end{eqnarray}
respectively, and similarly for the antiquarks $\delta\bar{q}$.

Although there is currently no direct evidence for charge symmetry
violation (CSV) in PDFs, it has been predicted in several
nonperturbative models for both valence and sea quark PDFs
\cite{Sather:1991je, Rodionov:1994cg, Londergan:2003pq,
Londergan:2009kj}.  In addition, electromagnetic radiative effects
can be included in the $Q^2$ evolution equations
(Sec.~\ref{ssec:Q2evol}), in analogy with usual gluon radiation,
by adding a term that accounts for the emission of photons by
quarks \cite{Gluck:2005xh, Martin:2004dh},
\begin{equation}
Q^2 \frac{\partial f_i(x,Q^2)}{\partial Q^2}
= \sum_j \left( P_{ij}^{\rm (QCD)} \otimes f_j
	      + P_{ij}^{\rm (QED)} \otimes f_j
	 \right).
\label{eq:QEDDGLAP}
\end{equation}
Here $P_{ij}^{\rm (QED)}$ are the new QED splitting functions
(expanded in terms of the electromagnetic coupling $\alpha$), and
the sum over partons includes also the photon PDF $\gamma(x,Q^2)$.
Even with charge symmetric PDFs at the input scale, the different
electric couplings of the photon to $u$ and $d$ quarks in the
modified evolution equations will induce nonzero CSV distributions
(\ref{eq:csv_u}) and (\ref{eq:csv_d}) at a higher scale.

\begin{figure}[t]
\begin{center}
\includegraphics[width=6.7cm]{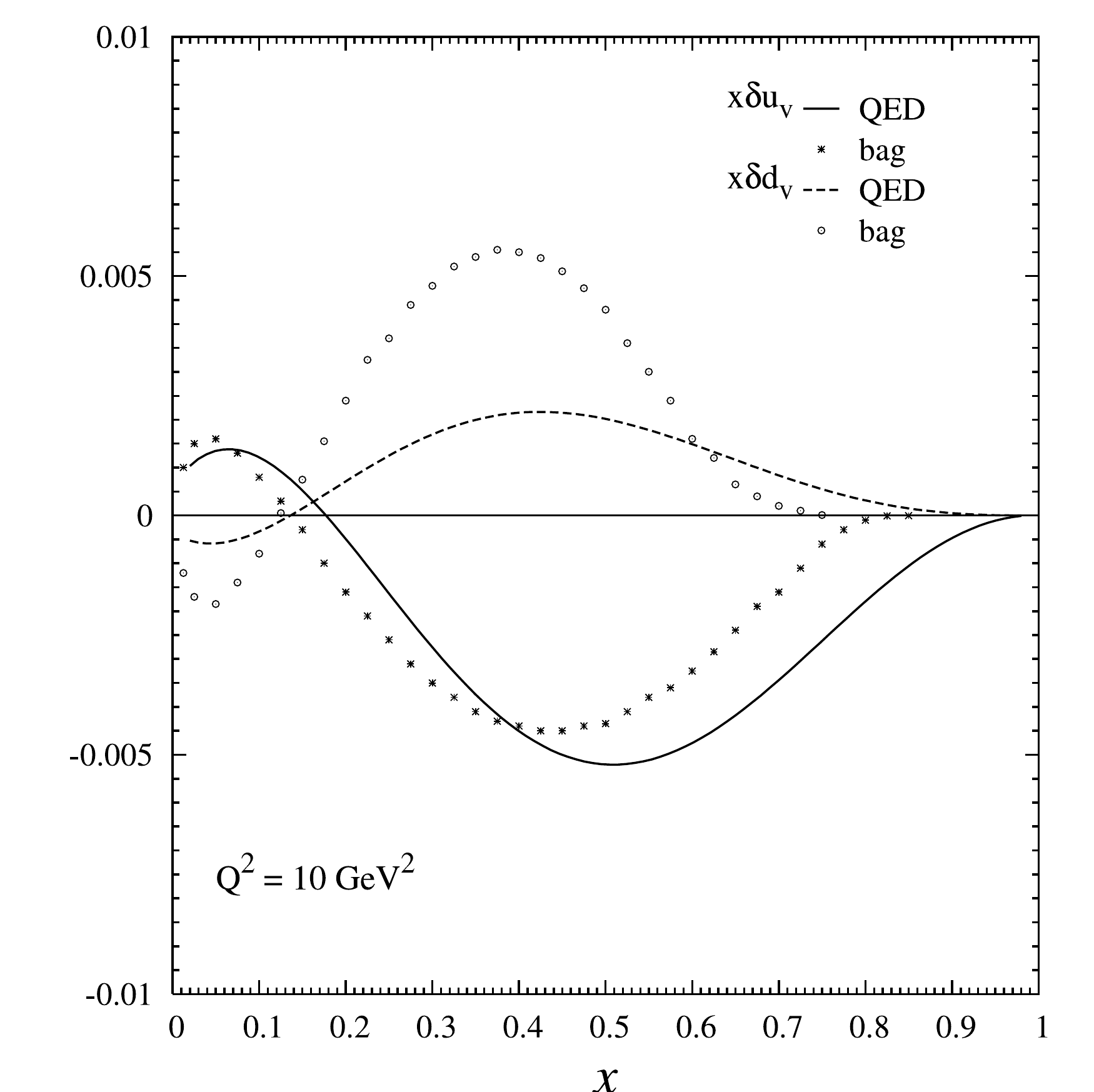}\hspace*{-0.3cm}
\includegraphics[width=9.65cm]{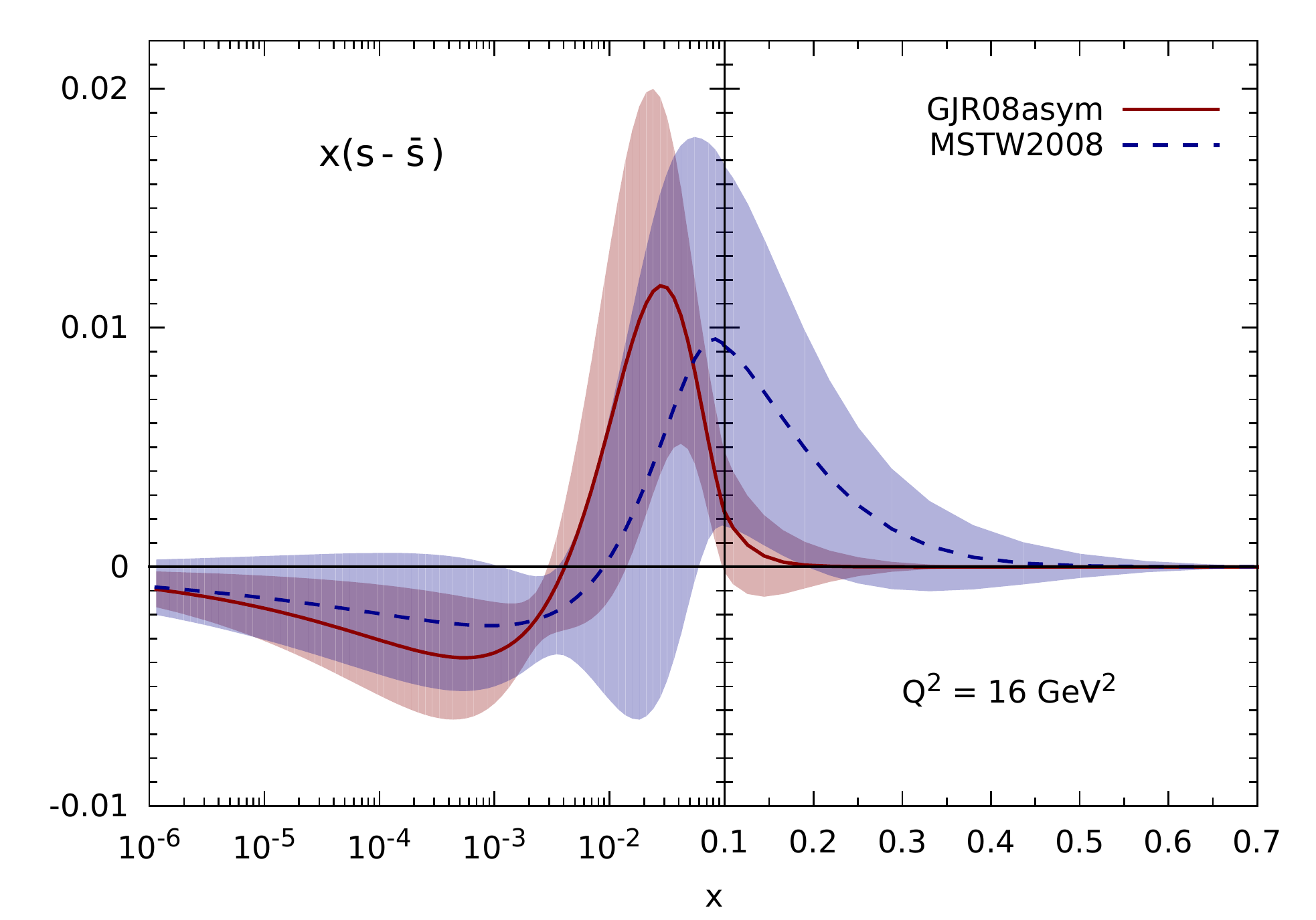}
\caption{{\it (Left)}
	The CSV ``majority'' $\delta u_v$ and ``minority'' $\delta d_v$
	valence quark distributions for the radiative QED fit of
	Ref.~\cite{Gluck:2005xh} at $Q^2 = 10$~GeV$^2$, compared
	with the bag model estimates have been taken from
	Ref.~\cite{Londergan:2003pq}.
	{\it (Right)}
	Strange--antistrange asymmetry at $Q^2 = 16$~GeV$^2$
	(appropriate for the NuTeV experiment) for the asymmetric
	GJR08 \cite{JimenezDelgado:2010pc} (red solid) and the
	MSTW2008 \cite{Martin:2009iq} (blue dashed) PDFs.}
\label{fig:csv_ssbar}
\end{center}
\end{figure}

The valence quark asymmetries resulting from the QED-modified $Q^2$
evolution (\ref{eq:QEDDGLAP}) are illustrated in Fig.~\ref{fig:csv_ssbar}
at a scale $Q^2 = 10$ GeV$^2$.  The perturbatively generated CSV
distributions from Gl\"uck {\it et al.} \cite{Gluck:2005xh} are
calculated with an average isoscalar quark mass $m_q = 10$~MeV,
with the QED radiative corrections computed after fixing the QCD
evolution effects.  The majority valence $\delta u_v$ distribution
is negative at intermediate $x$ values, and is of opposite sign to
the minority valence distribution $\delta d_v$, which is smaller
in magnitude.
Qualitatively, these features are similar to the nonperturbative
CSV valence asymmetries computed in the bag model \cite{Sather:1991je,
Rodionov:1994cg, Londergan:2003pq}, which arise from differences
between the spectator $uu$ and $dd$ diquark masses, as well as
the proton--neutron mass difference.

The MRSTQED \cite{Martin:2004dh} analysis of CSV was similar to that
in Ref.~\cite{Gluck:2005xh}, but involved unequal photon PDFs in the
proton and neutron resulting from QED radiation with quark masses
$m_{u(d)} = 6 (10)$~MeV.  Since the photon now also contributes to
the momentum sum rule, Eq.~(\ref{eq:mom_sum_rule}), the CSV photon
distributions induce charge symmetry violation in the valence quark
PDFs, assuming the sea and gluon are charge symmetric.
Despite the different assumptions, the resulting distributions turn
out to be similar to those in Fig.~\ref{fig:csv_ssbar}.

While the overall magnitude of the possible CSV effects is rather
small, they can nevertheless play an important role in understanding
the discrepancy between the NuTeV measurement of $\sin^2\theta_W$
\cite{Zeller:2001hh} and the Standard Model value \cite{PDG12}.
The NuTeV Collaboration extracted the Paschos-Wolfenstein (PW)
ratio \cite{Paschos:1972kj} of total neutral to charged current
cross sections, involving the difference between $\nu$ and $\bar\nu$
cross sections.  In addition to assuming small uncertainties on the
nuclear corrections necessary for translating the measured iron target
data to an isoscalar nucleon target, the analysis also neglected any
CSV effects.  Including CSV corrections, the experimental PW ratio
would receive a contribution proportional to
	$\int_0^1 dx\, x\, (\delta d_v - \delta u_v)$
\cite{Londergan:2009kj}.
Depending on the sign of the CSV distributions, the asymmetry can
therefore either remove some of the discrepancy with the NuTeV
result, or exacerbate it.  The signs and magnitude of the predicted
perturbative QED and nonperturbative quark model asymmetries in
Fig.~\ref{fig:csv_ssbar} would in fact remove approximately 2/3
of the discrepancy.

An additional correction to the PW ratio can arise from differences
between strange and antistrange distributions in the nucleon,
$\int_0^1 dx\, x\, (s-\bar s)$, which we discuss in the next section.

\subsection{Strange quarks}
\label{ssec:strange}

Traditionally the strange quark distribution has been more
difficult to determine experimentally than the non-strange sea.
In practice one has often assumed that the $s$ and $\bar s$\, PDFs
(which are often assumed to be identical) are proportional to the
$\bar u + \bar d$ distributions,
\begin{eqnarray}
\kappa &=& {s + \bar s \over \bar u + \bar d},
\label{eq:kappa}
\end{eqnarray}
with $\kappa$ ranging between $\sim 0.2$ and 0.5.
The $s$ and $\bar s$ distributions can be directly determined,
however, through the measurement of dimuon pairs in neutrino DIS,
which are produced through the charged current by scattering from
strange quarks with subsequent generation of charm.
In particular, while neutrino scattering produces charmed hadrons
($W^+ s \to c$), antineutrino scattering results in anticharmed
hadrons ($W^- \bar s \to \bar c$).
The semileptonic decay of the charmed hadrons (such as $D$ and
$\bar D$ mesons) then yields $\mu^+ \mu^-$ pairs along with the
associated hadronic debris.
Because of the low rates involved in neutrino scattering experiments,
in order to increase the statistics these have typically made use of
heavy nuclear targets, such as iron.  Since the nuclear effects
for neutrino and charged lepton scattering are in general different
\cite{Kulagin:2007ju}, the nuclear corrections introduce an additional
source of uncertainty in extractions of the strange quark PDF.

A number of nonperturbative models have in fact predicted an
asymmetry between the $s$ and $\bar s$ distributions in the
nucleon, which can be probed by taking the difference between
the $\nu$ and $\bar\nu$ cross sections. 
For example, in analogy with the pion cloud models invoked
to explain the $\bar d$ excess over $\bar u$ in the proton
in Sec.~\ref{ssec:lightsea}, the dissociation of a nucleon
to a hyperon and a virtual kaon naturally produces unequal
$s$ and $\bar s$ distributions, although the magnitude and
even the sign is difficult to determine unambiguously
\cite{Signal:1987gz, Burkardt:1991di, Melnitchouk:1996fj,
Alwall:2004rd}.
In addition, an asymmetry can also be generated through
perturbative QCD evolution at NNLO because of unequal valence
$u$ and $d$ distributions \cite{Catani:2004nc}.

The CCFR \cite{Bazarko:1994tt} and NuTeV \cite{Mason:2007zz}
collaborations at Fermilab have collected the highest statistics
data on inclusive charm production in $\nu$ and $\bar\nu$
scattering on an iron target, enabling the most direct constraints
to be placed on the strange quark PDFs as a function of $x$.
The latter in particular was the first analysis using a complete
NLO description of charm production.
The cross section for (anti)neutrino charm production was computed
to NLO in Refs.~\cite{Gottschalk:1980rv, Gluck:1996ve} within the
FFNS, in which, apart from the gluon, only the light quark flavors
($u, d, s$) are included as massless partons (see Secs.~\ref{ssec:HQ}
and \ref{ssec:HQresults}).  The fully differential NLO calculation,
including acceptance corrections, was performed in
Ref.~\cite{Kretzer:2001tc}.
(Note, however, that the use of the expressions in \cite{Gluck:1996ve}
together with VFNS distributions, as implemented in the NuTeV
analysis \cite{Mason:2007zz}, is strictly not consistent.)

Comparison of the $\nu$ and $\bar\nu$ events found a preferred fit for
the $s(x) - \bar s(x)$ distribution that peaked at $x \sim 0.05-0.1$,
with the compensating negative contribution required by the sum rule
$\int_0^1 dx\, (s - \bar s) = 0$ to be restricted to the unmeasured
region at $x \lesssim 0.004$ \cite{Mason:2007zz}.
For the first moment, the value obtained at $Q^2 = 16$~GeV$^2$ was
$S^-\ \equiv\ \int_0^1 dx\, x (s - \bar s)
  = \left( 1.96 \pm 0.46 \pm 0.45\ {}^{+1.48}_{-1.07} \right)
    \times 10^{-3}$,
where the errors are statistical, systematic, and external.
The latter includes a dominant contribution from the charm
semileptonic branching ratio uncertainty, as well as from nuclear
corrections \cite{Mason:2007zz}.

The strangeness asymmetry in the nucleon is shown in
Fig.~\ref{fig:csv_ssbar} at $Q^2 = 16$~GeV$^2$, as relevant for
the NuTeV experiment, for several different PDF parametrizations.
The MSTW08 \cite{Martin:2009iq} strange asymmetry is fitted to
the CCFR and NuTeV data assuming the form
\begin{eqnarray}
(s - \bar s) &=& a_0\, x^{a_1} (1-x)^{a_2} (1-x/x_0),
\label{eq:sminus}
\end{eqnarray}
where the factor $(1-x/x_0)$ ensures zero net strangeness
in the nucleon.
The strange asymmetry in Ref.~\cite{JimenezDelgado:2010pc}
is generated radiatively from the boundary condition
$(s + \bar s)(x,Q_0^2) = 0$ at $Q_0^2 = 0.5$~GeV$^2$, but
allowing $s-\bar s$ to have the form in Eq.~(\ref{eq:sminus}).
The resulting $s-\bar s$ asymmetry at the experimental scale
is similar to the MSTW08 fit, and to the phenomenological
extraction in Ref.~\cite{Mason:2007zz}, within the currently
sizable uncertainties.

The values for the momentum weighted asymmetry are the order
$S^- \approx 2 \times 10^{-3}$ for the MSWT08 fit \cite{Martin:2009iq}
and $S^- \approx 0.8 \times 10^{-3}$ for the dynamically generated PDFs
\cite{JimenezDelgado:2010pc} at scales $Q^2 \approx 10-20$~GeV$^2$.
The CTEQ Collaboration found a strange asymmetry in the range
$-0.001 < S^- < 0.004$ \cite{Olness:2003wz}.
These values have the correct sign and similar magnitude to that
required to reconcile the NuTeV $\sin^2\theta_W$ measurement with
the Standard Model value \cite{Zeller:2001hh}, once the charge
symmetry violating effects in Sec.~\ref{ssec:csv} are also included
\cite{Gluck:2005xh}.

Finally, a recent measurement by the ATLAS Collaboration
\cite{Aad:2012sb} of inclusive $W^\pm$ and $Z$ boson production
in $pp$ collisions at the LHC found a significantly larger value
of the strange to non-strange sea ratio in Eq.~(\ref{eq:kappa}),
	$\kappa = 1.00\, {}^{+0.25}_{-0.28}$
at $x = 0.023$ for $Q^2 = 1.9$~GeV$^2$, than in previous analyses.
A reanalysis of the ATLAS and HERA data by the NNPDF group
\cite{Ball:2012cx} found, however, that the uncertainty on the
extracted $\kappa$ was significantly underestimated, and that
a more complete global analysis gives $\kappa$ values consistent
with the traditional values.

\subsection{Heavy quarks}
\label{ssec:HQresults}

Heavy quark production contributes considerably to inclusive DIS;
for example, the charm (bottom) contributions to the dominant $F_2$
structure function constitute up to 30\% (3\%) of the total in the
small--$x$ region.  Besides these contributions, heavy quark
electroproduction is directly accessed experimentally in semi--inclusive
DIS, typically by detecting charm (bottom) mesons in the final state.
Because the main production mechanism is photon--gluon fusion,
data on this process provide valuable constraints on the gluon PDF
in the nucleon.  Furthermore, they provide an appropriate context for
testing the different approaches to these calculations that have been
adopted in various PDF analyses (see Sec.~\ref{ssec:HQ}).

\begin{figure}[t]
\begin{center}
\includegraphics[width=7.7cm]{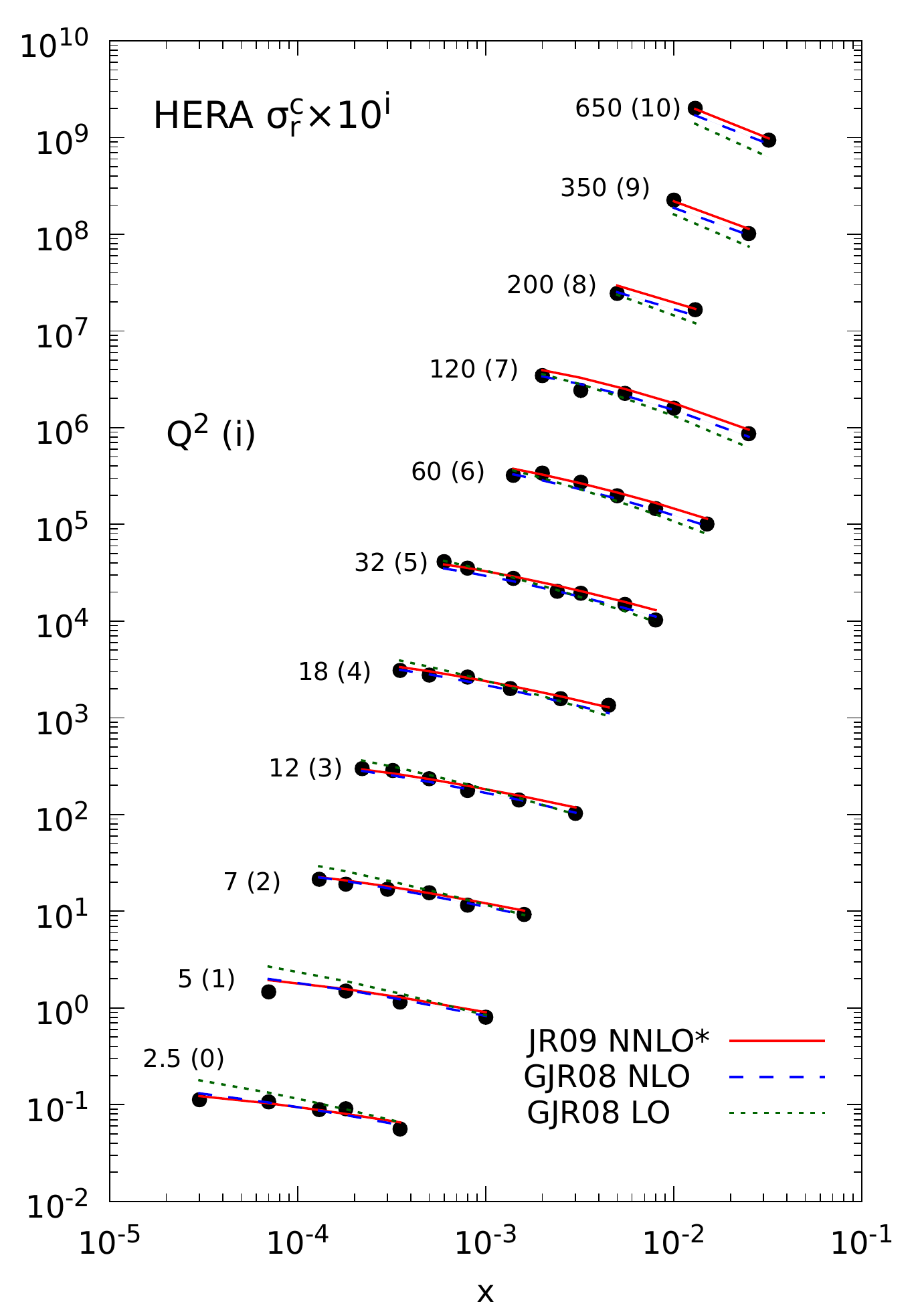}
\includegraphics[width=7.7cm]{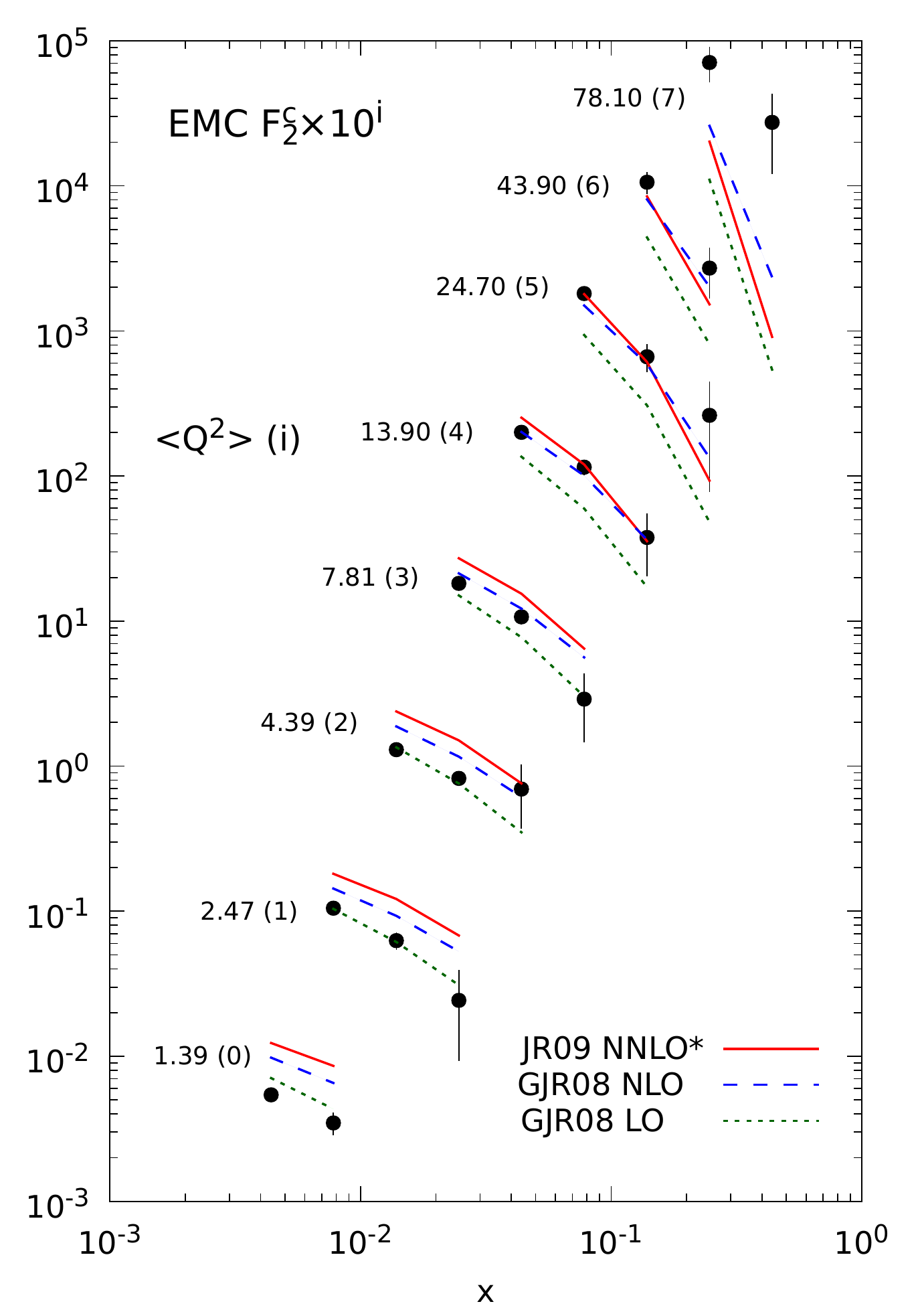}
\caption{Reduced charm production cross section data from HERA
	\cite{Abramowicz:1900rp} {\it (left)}
	and charm structure function data from EMC
	\cite{Aubert:1981ix, Aubert:1982tt} {\it (right)},
	compared with global PDF fits at LO, NLO and NNLO*
	\cite{JimenezDelgado:2009tv, Gluck:2007ck}.
	Note that for the EMC iron data the nuclear corrections
	of Ref.~\cite{deFlorian:2003qf} were used.}
\label{fig:charm}
\end{center}
\end{figure}

Since it is by far the most relevant, we focus the discussion on the
charm flavor, although similar considerations often also apply to
bottom (top contributions to this process, on the other hand, are
negligible at accessible kinematics).  The most relevant data on
heavy quark electroproduction are the (H1 + ZEUS) combination of
HERA measurements \cite{Abramowicz:1900rp}.  A comparison of
these data with FFNS calculations at LO, NLO and NNLO*
(see Sec.~\ref{ssec:HQ}) in Fig.~\ref{fig:charm}~(left) shows
excellent agreement between the FFNS calculations.
In fact, the experimental analysis of the HERA data is based
primarily on the FFNS fully differential (exclusive) calculations
of Ref.~\cite{Harris:1997zq}.  Despite this fact, the constructs
in the GM-VFNSs are often motivated by arguing that
pseudo-mass-divergences (non-collinear logarithms of the form 
$\ln Q^2/m^2$) which appear in the Wilson coefficients of the
massive calculations need to be resummed.  However, the FFNS results
are notably stable even for very large values of $Q^2 \gg m^2$
\cite{Gluck:1993dpa}, in particular at the largest $Q^2$ values
accessible at HERA, suggesting little need to resum these supposedly
``large logarithms'' in the context of global PDF analyses.

Finally, in addition to the perturbative generation of charm,
it is also possible for charm to be produced through nonperturbative
mechanisms, such as those discussed in Refs.~\cite{Steffens:1999hx, 
Brodsky:1980pb, Vogt:1995dn, Navarra:1995rq, Vogt:2000sk,
Pumplin:2005yf} (and references therein).  First postulated to
account for some early charm production data in hadronic reactions,
these ``intrinsic'' charm contributions share several characteristic
features, even though their details depend somewhat on the models.
In particular, they are typically valence-like, with significantly
harder $x$ distributions than those generated perturbatively, and can
produce large asymmetries between the $c$ and $\bar c$ distributions.

Measurements of the charm structure function $F_2^c$ by EMC
\cite{Aubert:1981ix, Aubert:1982tt}, especially the two data points
at the highest $Q^2$ in Fig.~\ref{fig:charm}~(right), have been
interpreted as indicating a nonzero intrinsic charm contribution
at large $x$.  The evidence is not conclusive, however, in view of
the disagreement evident in Fig.~\ref{fig:charm}~(right) between
the EMC data and theoretical predictions at smaller $x$, where the
predictions are in agreement with the more recent HERA results in
Fig.~\ref{fig:charm}~(left).

The current limits on the normalization of the nonperturbative charm
quark distribution range from $\sim 0.5\%$ to 2\%, depending on which
data sets are fitted and which models of intrinsic charm considered
\cite{Steffens:1999hx, Martin:2009iq, Brodsky:1980pb, Vogt:2000sk,
Pumplin:2005yf, Harris:1995jx, Hobbs:2013}.  Although current global
PDF analyses do not require an intrinsic charm component, the existence
of nonperturbative charm remains an intriguing issue which can be
addressed by high-precision measurements of charm production at
future facilities such as
J-PARC \cite{Kumano:2010qm, J-PARC13},
GSI-FAIR \cite{Riedl:2007sv},
Jefferson Lab at 12~GeV \cite{PR12-07-106, Brambilla:2010cs},
AFTER@CERN \cite{Brodsky:2012vg}, or an
Electron-Ion Collider \cite{Accardi:2012hwp}.

\subsection{Gluons}
\label{ssec:gluon}

\begin{figure}[t]
\begin{center}
\includegraphics[width=10cm]{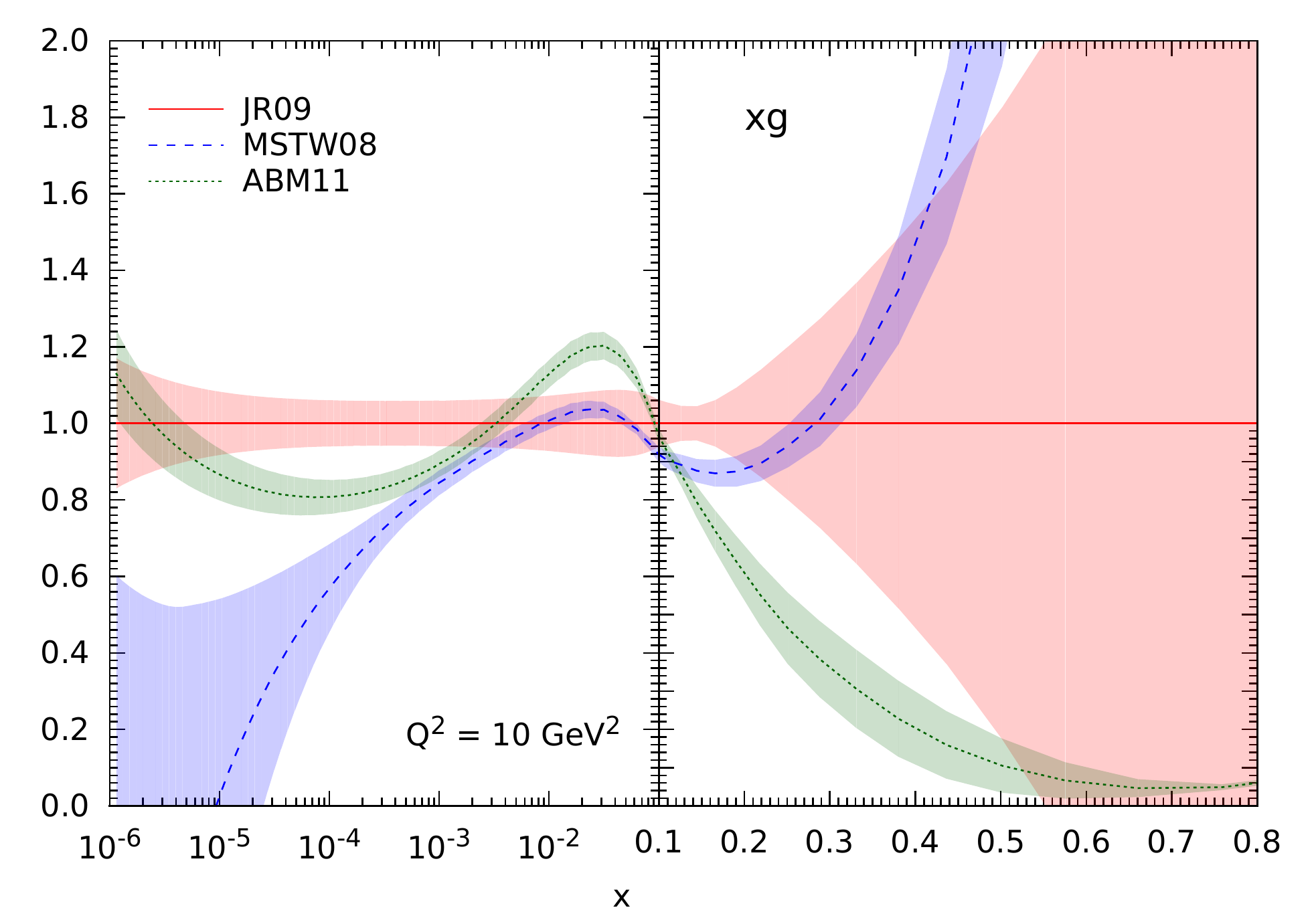}
\caption{Gluon distributions for the 3-flavor NNLO PDF sets from
	JR09 \cite{JimenezDelgado:2009tv} (red solid line),
	MSTW08 \cite{Martin:2009iq} (blue dashed) and
	ABM11 \cite{Alekhin:2012ig} (green dotted), relative
	to the JR09 gluon PDF, at $Q^2=10$~GeV$^2$.}
\label{fig:gluon}
\end{center}
\end{figure}

In addition to the heavy quark production discussed above, there are
three primary sources of information on the gluon distribution:
the $Q^2$ dependence of the DIS structure function $F_2$ at low values
of $x$; the longitudinal structure function $F_L$ at all $x$ values;
and jet (both at HERA and at hadron colliders) and photon
production cross sections at moderate to high values of $x$.
Furthermore, the momentum sum rule (\ref{eq:mom_sum_rule}) provides
an additional constraint on the gluon PDF.  Since the normalizations
of the valence PDFs are fixed by the flavor sum rules in
Eq.~(\ref{eq:sum_rules}) and the contributions of the light
antiquarks are relatively well determined by DIS and lepton pair
production data, the momentum sum rule effectively constrains the
gluon contribution.  The effect is an anticorrelation between the
low- and high-$x$ behavior of the gluon PDF: a decrease (increase)
of the gluon distribution at low $x$ leads to an increase (decrease)
at high values of $x$.  Interestingly, this can result in more
constrained gluon distributions in the medium-$x$ region
(see Fig.~\ref{fig:gluon}).

To understand the relationship between the scale dependence of
$F_2$ and the gluon PDF, consider Eq.~(\ref{eq:RGE}) for the scale
dependence of the quark PDF.  Multiply both sides by $x$ and by the
squared parton charge and sum over the active flavors.  In this way
one derives (using the LO expression for $F_2$),
\begin{equation}
\hspace{-2cm}
Q^2\, \frac{\partial F_2(x,Q^2)}{\partial Q^2} 
= \int_x^1 dy
  \left[ P_{qq}(y,Q^2) F_2\left(\frac{x}{y},Q^2\right)
  + \sum_i e_i^2 P_{qg}(y,Q^2)
    \frac{x}{y} g\left(\frac{x}{y},Q^2\right)
  \right]
\label{eq:F2Q2}
\end{equation}
where $P_{qq}$ and $P_{qg}$ are the quark--quark and quark--gluon
splitting functions.
At small values of $x$ the gluon term dominates the integrand so that
the scale dependence of $F_2$ places a constraint on the gluon PDF.

A second constraint on the gluon from DIS comes from the longitudinal   
structure function $F_L$, which is defined as
\begin{eqnarray}
& &
\hspace{-2.2cm}
F_L(x,Q^2)
= \left(1 + \frac{4 M^2 x^2}{Q^2}\right) F_2(x,Q^2)
- 2 x F_1(x,Q^2)
\stackrel{Q^2 \rightarrow \infty}{\longrightarrow}
F_2(x,Q^2) - 2 x F_1(x,Q^2).	\nonumber\\
& &
\end{eqnarray}
In lowest order, where only quarks contribute to $F_1$ and $F_2$,
one has $F_2 = 2 x F_1$ so that $F_L = 0$.  The first nonzero
contribution to $F_L$ starts at ${\cal O}(\alpha_s)$ where, at
small $x$, there is a significant contribution from the subprocess
$\gamma^* g \to q \bar q$, which again imposes strong constraints
on the gluon PDF at small $x$.
At larger $x$ values the valence quark contributions increase, the
gluon PDF decreases, and the subprocess $\gamma^* q \to qg$ dominates.
However, since the same combinations of quark distributions are well
determined from $F_2$, where they enter with one power of $\alpha_s$
less and the gluon does not contribute, precision measurements of the
longitudinal structure function provide valuable information on the
gluon PDF at large $x$ \cite{Monaghan:2012et}.

Note also that the gluon PDF is accompanied by the same powers of
$\alpha_s$ in all DIS structure functions, starting with $\alpha_s\, g$.
This results in a correlation between the value of $\alpha_s$ obtained
in global PDF analysis and the size and shape of the gluon distribution,
with larger $\alpha_s$ leading to a smaller gluon PDF in the small-$x$
region and (via the momentum sum rule correlation) a larger gluon PDF
at large $x$.  This can be seen in Fig.~\ref{fig:gluon}, where the
JR09 \cite{JimenezDelgado:2009tv} and ABM11 \cite{Alekhin:2012ig}
analyses obtain $\alpha_s$ values and small-$x$ gluon PDFs of
similar size, while the MSTW08 fit \cite{Martin:2009iq} finds a
larger $\alpha_s$ value and a smaller gluon PDF in this region
(which even turns negative).  Both possibilities describe well the
inclusive cross sections measurements at HERA.

This ambiguity can be reduced by using processes with a different
leading $\alpha_s$ power, typically jet production at the Tevatron,
which, as noted in Sec.~\ref{sssec:unpol}, can provide constraints
on the gluon PDF.  To understand this, recall that the calculation
of the jet cross section involves a sum over $qq$, $qg$ and $gg$
induced subprocesses where ``$q$'' stands for any flavor of quark
or antiquark.  At low values of jet $x_T$, the $gg$ subprocesses
dominate, at intermediate values all three contribute, and eventually
as $x_T$ approaches one only the $qq$ terms survive -- see Fig.~3 in
Ref.~\cite{Stump:2003yu}, for example.  Jet data at the Tevatron
extend to $x_T$ values of approximately 0.5, so these data provide
some constraints on the gluon PDF, at least in the mid-$x$ range
where the $qg$ subprocesses make substantial contributions.

As discussed in Ref.~\cite{d'Enterria:2012yj}, isolated photon production
at collider energies has the potential to tighten the error bands
in the $x$ range below about 0.2.  It remains a challenge to find
processes that will constrain the gluon beyond $x \approx 0.5$.
The fundamental problem is that the valence PDFs here are larger
than the gluon PDF.  Most hard processes that receive contributions
from $qg$ subprocesses also receive contributions from $qq$ processes,
which reduce the sensitivity to the gluon PDF.  One counterexample
is provided by direct photon production in $pp$ collisions.
The two dominant subprocesses are initiated by $q \bar q$ and $qg$
initial states.  At large values of $x_T$ the fragmentation process
is suppressed; it can be further reduced by photon isolation cuts.
The single photon cross section involves an integration over the
awayside jet rapidity which tends to smear out the $x$ values of the
contributing PDFs.  There are regions where one $x$ is small and one
is large --- one might think that this could constrain the large-$x$
gluon.  However, the favored configuration will be where the gluon or
$\bar q$ is at small values of $x$ while the valence quark is at the
larger values.
Instead, consider the photon + jet cross section with equal and
opposite photon and jet rapidities, in which case both momentum
fractions will be equal to $x_T \cosh y_{\gamma}$ using LO kinematics.
Going to large values of $x_T$ and $y_{\gamma} = -y_{jet}$ would
then yield information on the gluon at large $x$, since the $\bar q$
contribution is much smaller there than that of the gluon.
This would also be the case for lepton pair \cite{Berger98:lpp}
and vector boson production in $pp$ collisions, provided that one
considers $p_T \gtrsim M_B/2$ in order for the parton momentum
fraction to be large.  The analysis in Ref.~\cite{Carminati:2013}
suggests that photon + jet data with increased statistics from
the LHC may help constrain the gluon and light quark PDFs.

Another source of information on the gluon PDF is top quark pair
production, for which the lowest order subprocess is $gg \to t\bar t$.
In Ref.~\cite{Czakon:2013} the constraints on the gluon PDF provided
by $t \bar t$ data from the Tevatron and LHC were examined, with the
results indicating that these data place strong constraints on the
large-$x$ gluon PDF.
Also of note is the analysis of ratios and double ratios of cross
sections measured at different LHC energies \cite{Mangano:2013er}.
Such ratios have reduced systematic errors and have the potential
to provide strong constraints on large-$x$ PDFs.

In addition to the issues discussed above, an obstacle for the
determination of the gluon distribution, and the highly correlated
$\alpha_s$ values from PDF analysis, is that the calculations for
the relevant processes are not available beyond NLO, with the
important exceptions of the (light quark contributions to the)
inclusive DIS structure functions and Drell-Yan cross sections.
For example, the relative uncertainties of the JR09 gluon PDF in
Fig.~\ref{fig:gluon} are below about 10\% for $x \lesssim 0.1$,
and increase as one proceeds to higher values of $x$, where the
constraints of the $F_2$ structure function from fixed target
experiments used in that fit are weaker.
The uncertainty at large $x$ could in principle be reduced by including
jet data.  However, this would also affect the central values obtained
for the gluon PDF, changing them in the relevant region by an amount
proportional to the missing NNLO corrections, which in the case of
jet production are expected to be large \cite{Ridder:2013mf}.
Global NNLO fits which include jet data, such as the MSTW08
\cite{Martin:2009iq} parametrization in Fig.~\ref{fig:gluon},
generally find larger gluon PDFs in the high-$x$ region and
correspondingly also larger values of $\alpha_s$.

\begin{figure}[h]
\begin{center}
\includegraphics[width=10cm]{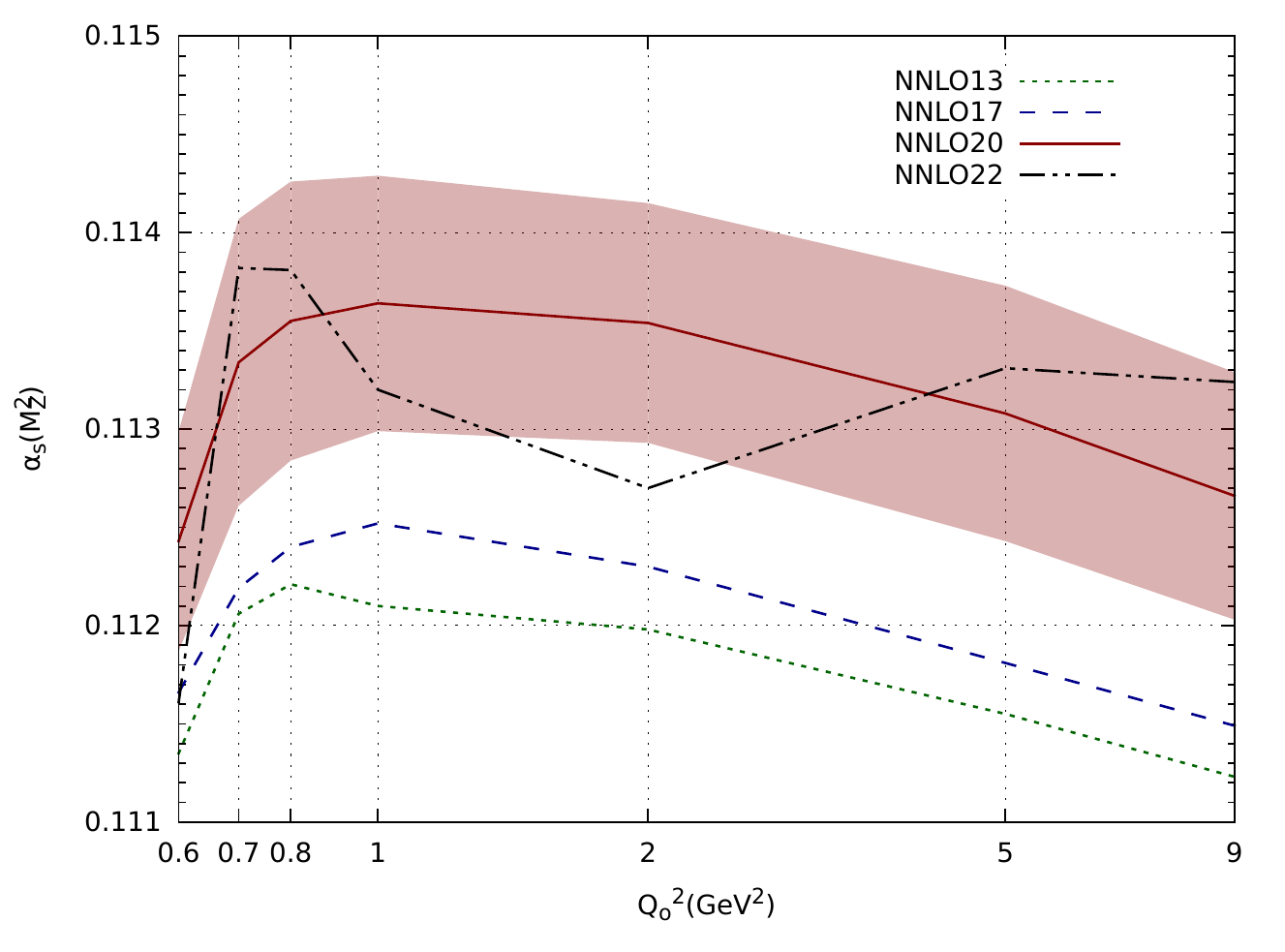}
\caption{Input scale dependence of $\alpha_s(M_Z^2)$ values obtained for
	different parametrizations in Ref.~\cite{JimenezDelgado:2012zx}.
	The band indicates the uncertainties due to propagation of the
	experimental errors for a particular parametrization (NNLO20).}
\label{fig:alphas}
\end{center}
\end{figure}

Yet another factor which can alter the PDFs and the value of $\alpha_s$
obtained in a particular analysis are the kinematic cuts applied;
of particular relevance are the cuts on the DIS structure function
data and the related treatment of contributions beyond leading twist
\cite{Alekhin:2012ig, JimenezDelgado:2012zx}.
Although data selection (data sets and cuts) bring about certain
arbitrariness in global PDF analysis, it cannot account for all the
existing differences between the various PDF sets, which are also
affected by theoretical issues such as heavy quark schemes
(Sec.~\ref{ssec:HQ}) and the particular solutions of the
RGE equations used.  Even within a given framework there are
uncertainties due to the inability of the estimation procedure
to find the optimal solution, for example, shortcomings of the
parametrization to reproduce the optimal shape of the distributions,
and the statistical estimation procedure.
These effects, referred in general as `procedural bias', induce a
dependence of the results on the choice of the scale at which the
input distributions are parametrized, which in principle should not
depend on these choices.  For particularly sensitive quantities such
as the gluon distribution and $\alpha_s$, this uncertainty can be
comparable to the experimental uncertainty \cite{JimenezDelgado:2012zx},
as can be seen in Fig.~\ref{fig:alphas}.

Another example of this is the differences between ``standard'' 
distributions, generated from input PDFs at values for the input
scale $Q_0^2 \geq 1$~GeV$^2$, and the so--called dynamical
distributions \cite{Gluck:1994uf, Gluck:1998xa, Gluck:2007ck,
JimenezDelgado:2008hf}, for which input scale is chosen to be
$Q_0^2 < 1$~GeV$^2$.  At these low scales the input distributions
naturally tend to valence--like (positive definite) shapes
\cite{JimenezDelgado:2012zx} that vanish in the small--$x$ limit, not
only for valence but also for the sea quark and gluon input densities.
It was shown in Refs.~\cite{Gluck:2007ck, JimenezDelgado:2008hf},
where also ``standard'' distributions were generated, that this
implies that the dynamical distributions at small $x$ are more
restricted and have smaller uncertainties than their standard
counterparts.  Furthermore, this procedure typically results in
smaller $\alpha_s$ values and steeper gluons in the small--$x$
region, although both approaches provide comparable descriptions of
data, in particular, of the inclusive DIS cross sections from HERA.

An alternative approach to fitting $\alpha_s$ as part of the global
fitting is to use the world average value for $\alpha_s(M_Z)$,
thereby avoiding the correlations during the fitting process.
Since $\alpha_s$ is a parameter of the QCD Lagrangian, it should
be the same for all processes, not just those in the global fit.
Of course, interesting information could be obtained by determining the
PDFs both with a fixed value of $\alpha_s(M_Z)$ and with a fitted value.
Comparing the results could indicate which processes in the global fit
are chiefly responsible for any differences.

\subsection{Lattice PDF moments}
\label{ssec:lattice}

The moments of leading twist PDFs can be evaluated from first principles
using lattice QCD.  Within the operator product expansion, the moments
of PDFs are related to matrix elements of local operators between hadron
states, which can be computed numerically on the lattice.
The lattice PDF moments can be related to the quark and antiquark
distributions in the nucleon as
\begin{eqnarray}
\langle x^n \rangle_q
&=& \int_0^1 dx\, x^n\ [q(x)-(-1)^n \bar q(x)],
\label{eq:mom_unp}                                 \\
\langle x^n \rangle_{\Delta q}
&=&\int_0^1 dx\, x^n\ [\Delta q(x)+(-1)^n \Delta \bar  q(x)],
\label{eq:mom:_pol}
\end{eqnarray}
for the unpolarized and helicity distributions
(see Sec.~\ref{sec:spin}), respectively.
The lattice moments alternate between the total (or $C$-even)
$q+\bar q$ and valence ($C$-odd) $q-\bar q$ distributions,
depending on whether $n$ is even or odd.
While reconstructing the $x$ dependence of the PDFs from a finite
number of calculated moments is challenging \cite{Detmold:2001dv},
comparison of the lattice moments with the phenomenological PDF
moments can provide important tests of lattice techniques and
provide constraints on specific PDFs which may be difficult to
measure experimentally.

Because of the discretization of space-time on the lattice, the 
rotational symmetry of the continuum is broken to the hyper-cubic 
subgroup, which introduces mixing with lower dimensional operators 
under renormalization.  As a result, only the lowest few moments,
corresponding to $n \leq 3$, can at present be reliably computed.
As well as the usual approximations of working at a finite lattice
spacing $a$, in a finite lattice volume $V$, and at unphysically
large values of the quark mass, simulations usually include 
contributions only from ``connected'' diagrams, representing
operator insertions on quark lines connected to the source.         
Meaningful comparisons of lattice data can therefore only be made 
with nonsinglet combinations, such as $u-d$, where the 
``disconnected'' contributions cancel.

Significant progress has been made in recent years in the computation
of matrix elements of twist-two operators on the lattice, using
dynamical (unquenched) configurations and quark (or pion, by the 
Gell-Mann--Oaks--Renner relation) masses approaching the physical
limit.  Unfortunately, the results for the lowest nontrivial
($n=1$) nonsinglet moment of the unpolarized distributions,
$\langle x \rangle_{u-d}$, lie systematically in the range
$\approx 0.23-0.25$ at a scale $\mu^2 = 4$~GeV$^2$, which is
some 40\% above the phenomenological PDF moments of
$\langle x \rangle_{u-d} \approx 0.17$ \cite{Blumlein:2012bf}.
A smaller, but persistent 10\% overestimate of the lowest
($n=0$) moment of the nonsinglet spin-dependent distribution,
$\langle 1 \rangle_{\Delta q} = g_A$, has posed a serious challenge
to lattice simulations \cite{Bratt:2010jn, Pleiter:2011gw}.

It has been suggested \cite{Detmold:2001jb, Procura:2006gq,
Shanahan:2013xw} that the discrepancy between the calculated and
experimental PDF moments could be resolved by the nontrivial
chiral behavior of the moments, when extrapolating the results
from the lowest $m_\pi$ values at which lattice data exist to
the physical mass.
Verification of this behavior will require lattice simulations
at very small quark masses, as well as at large enough volumes
to fully include the physics of the pion cloud \cite{Detmold:2002nf}
(see also Sec.~\ref{ssec:lightsea}).
While it may take some time before the lattice parameters can
become sufficiently close to those in the physical realm, such
comparisons will continue to reflect the important synergy
between lattice QCD and PDF phenomenology.

\section{Spin-dependent parton distributions}
\label{sec:spin}

Considerable progress has been made in understanding the spin structure
of the nucleon since the first precision polarized DIS experiments at
CERN in the late 1980s \cite{Ashman:1989ig} indicated an anomalously
small fraction of the proton spin carried by quarks.
A rich program of spin-dependent inclusive and semi-inclusive DIS,
as well as polarized proton--proton scattering experiments has
followed, vastly improving our knowledge of spin-dependent PDFs
of the nucleon over the last two decades (for a recent review,
see Ref.~\cite{Aidala:2012mv} and references therein).

While the spin-dependent data have not been as abundant as those
available for constraining spin-averaged PDFs, the existing empirical
information from polarized lepton and hadron facilities at
CERN \cite{Ashman:1989ig, SMC, COMPASS07, COMPASS10,
	   Adeva:1997qz, Alekseev:2010ub},
SLAC \cite{SLAC:E80/E130, SLAC:E142, SLAC:E154, SLAC:E143,
	   SLAC:E155, SLAC:E155x},
DESY \cite{HERMES97, HERMES07, HERMES12, Airapetian:2004zf,
	   Airapetian:2008qf},
Jefferson Lab \cite{E99-117, E97-103, E01-012, EG1a, EG1b}
and
RHIC \cite{Adams:1994bg, Adare:2008aa, Manion:2011zz,
	   Aggarwal:2010vc, Adamczyk:2012qj, collaboration:2011fga}
has enabled several dedicated global QCD analyses of spin-dependent
parton distributions to be performed.
These include the NLO analyses from the more established
DSSV \cite{deFlorian:2009vb} group, the European-based
LSS \cite{Leader:2010rb} and
BB \cite{Blumlein:2010rn} groups, as well as the Japanese
Asymmetry Analysis Collaboration (AAC) \cite{Hirai:2008aj},
using the standard global fitting methodology outlined in
Sec.~\ref{sec:QCD}.
More recent efforts have been made by the NNPDF Collaboration
\cite{Ball:2013lla}, extending the neural network approach to
the polarized sector, and the Jefferson Lab Angular Momentum (JAM)
Collaboration \cite{JAM13}, which has focused on utilizing data over
a large kinematic range that includes the large-$x$ and low-$Q^2$
regions, with a simultaneous fit of unpolarized PDFs.

\begin{figure}[t]
\begin{center}
\includegraphics[width=14.5cm]{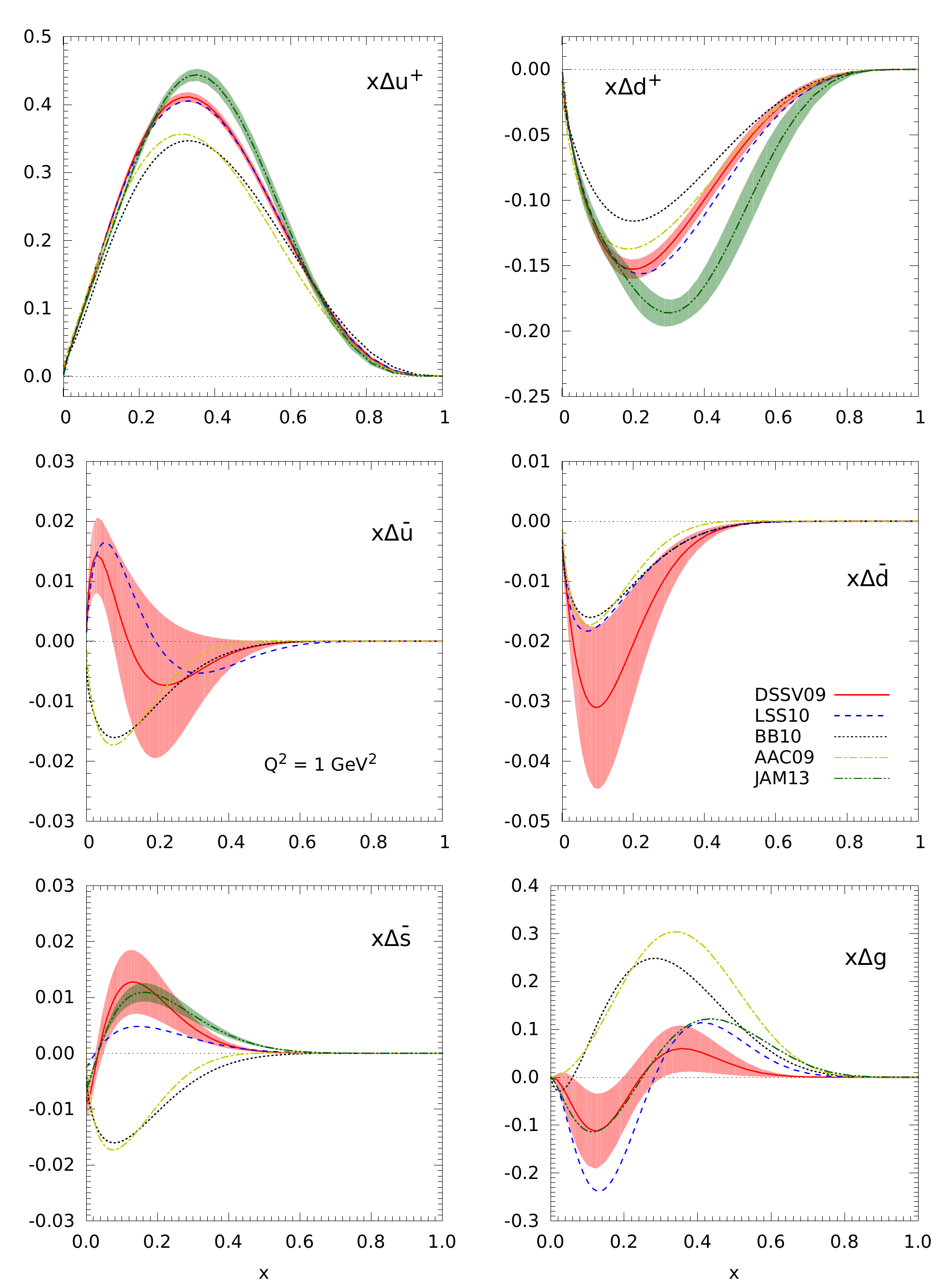}
\caption{Comparison of the spin-dependent PDFs for the
	total $x \Delta u^+ = x (\Delta u + \Delta \bar u)$
	and $x \Delta d^+ = x (\Delta d + \Delta \bar d)$,
	the antiquark $x \Delta \bar u$, $x \Delta \bar d$
	and $x \Delta \bar s$, and polarized gluon $x \Delta g$
	distributions at $Q^2 = 1$~GeV$^2$, for the
	DSSV09 \cite{deFlorian:2009vb},
	LSS10 \cite{Leader:2010rb},
	BB10 \cite{Blumlein:2010rn},
	AAC09 \cite{Hirai:2008aj},
	and JAM13 \cite{JAM13} PDF sets.}
\label{fig:spinPDFs}
\end{center}
\end{figure}

Representative examples of the spin-dependent PDFs are shown
in Fig.~\ref{fig:spinPDFs}, illustrating the total (or $C$-even)
$\Delta u^+$ and $\Delta d^+$ PDFs, as well as the sea quark
$\Delta \bar u$, $\Delta \bar d$ and $\Delta \bar s$
and polarized gluon $\Delta g$ distributions,
for the DSSV09 \cite{deFlorian:2009vb},
LSS10 \cite{Leader:2010rb}, BB10 \cite{Blumlein:2010rn},
AAC09 \cite{Hirai:2008aj}, and JAM13 \cite{JAM13} PDF sets
at $Q^2 = 1$~GeV$^2$.
The BB10 \cite{Blumlein:2010rn} and JAM13 \cite{JAM13} analyses are
based on inclusive DIS data only, while the LSS10 \cite{Leader:2010rb}
fit includes also semi-inclusive DIS data.  The DSSV09 PDFs are
constrained in addition by polarized $pp$ scattering data.
The latest AAC analysis \cite{Hirai:2008aj} is based on inclusive
DIS data and a $K$-factor approximation for the NLO corrections
for the $pp$ data.

As for the unpolarized case in Fig.~\ref{fig:NSpdf},
the $\Delta u^+$ distribution is the best constrained
polarized PDF, mostly by measurements of the proton $g_1$
structure function over a relatively broad range of $x$.
The corresponding $\Delta d^+$ distribution, which has the
opposite sign, is smaller in magnitude compared with the
$\Delta u^+$ with somewhat larger uncertainties, especially
at larger $x$ values.
The uncertainty band for the DSSV09 parametrization of
$\Delta u^+$ and $\Delta d^+$ is smaller than the variation
between the different PDF sets, which reflects the fact that
the systematic uncertainties associated with the choice of data
sets and parametrization assumptions, as well as other theoretical
inputs, are currently larger than the experimental errors.

The polarization of the sea is considerably smaller,
and more strongly dependent upon assumptions about the
flavor dependence in the analysis of semi-inclusive DIS data.
At present there is no conclusive experimental evidence of a
nonzero light quark sea, $\Delta \bar u$ and $\Delta \bar d$,
although there is a slight trend towards a more negatively
polarized $\bar d$ distribution than $\bar u$, with a positive
$\Delta \bar u - \Delta \bar d$.
The polarization of the strange sea is also very small,
in contrast to suggestions from the early spin-dependent DIS
data analyses of a large negative $\Delta \bar s$.
The polarized gluon distribution is essentially unconstrained by
existing inclusive and semi-inclusive DIS data, and almost all
information on $\Delta g$ comes from measurements of $c \bar c$
production in semi-inclusive DIS \cite{Adolph:2012ca}, and
inclusive pion and jet production in polarized $pp$ scattering
\cite{Adams:1994bg, Adare:2008aa, Manion:2011zz, Adamczyk:2012qj,
collaboration:2011fga}.
The data are consistent with a small value of $\Delta g/g$,
consistent with zero, although new measurements from RHIC
have the promise of resolving a small nonzero distribution.

\subsection{Polarized valence quarks at large $x$}
\label{ssec:pol_val}

At large values of $x$, the spin-dependent PDFs are even more
sensitive to the quark-gluon dynamics responsible for spin-flavor
symmetry breaking than the spin-averaged valence quark PDFs discussed
in Sec.~\ref{ssec:valence}.  However, while considerable progress has
been made in determining spin-dependent PDFs over the last two decades,
especially in the small-$x$ region, relatively little attention has
been paid to structure function measurements at large $x$.

The dearth of data in the valence region is especially striking for
the neutron, where, with the exception of several data points from
Jefferson Lab Hall~A \cite{E99-117} for the polarization asymmetry
$A_1^n$ extending to $x \approx 0.6$, there are essentially no
constraints for $x \gtrsim 0.4$.  On the other hand, there are a
number of predictions for the behavior of polarized PDFs and ratios
of polarized to unpolarized PDFs in the limit as $x \to 1$ that
differ even in sign.

At large $x$ there are dramatically different predictions for the
valence $\Delta u$ and $\Delta d$ distributions from nonperturbative
and perturbative calculations.  In the simple SU(6) spin-flavor
symmetric quark model, the equal probabilities of the spin-0 and
spin-1 spectator diquark configurations lead to the simple relation
$\Delta d / \Delta u = - 4$, and corresponding polarized to unpolarized
ratios $\Delta u/u = 2/3$ and $\Delta d/d = -1/3$.  In this limit the
polarization asymmetry of the proton, which at LO can be written
\begin{eqnarray}
A_1^p\ &\approx&\ { 4 \Delta u^+ + \Delta d^+ \over 4 u^+ + d^+ },
\end{eqnarray}
is predicted to be $A_1^p = 5/9$, while that of the neutron,
\begin{eqnarray}
A_1^n\ &\approx&\ { \Delta u^+ + 4 \Delta d^+ \over u^+ + 4 d^+ },
\end{eqnarray}
would vanish, $A_1^n = 0$.

While spin-flavor symmetry is of course broken in nature (in some cases
badly), with the unpolarized $d$ quark distribution significantly softer
than the $u$, for example, the details of this breaking can affect the
spin-dependent PDFs in strikingly different ways.  The mechanism of
spin-1 diquark suppression, which leads to the vanishing of the $d/u$
ratio as $x \to 1$, also predicts that $\Delta u/u \to 1$ in this limit,
while $\Delta d/d$ remains unchanged from the SU(6) expectation.

Arguments based on helicity conservation in perturbative QCD, on the
other hand, predict that for valence quarks in the ground state of
the nucleon the scattering at large $x$ is predominantly from
configurations in which the quark is coupled to a diquark of zero
helicity \cite{Farrar:1975yb}.  In this case the PDFs associated
with a quark polarized in the same direction as the nucleon are
favored over the helicity-antialigned configurations,
$q^\uparrow(x) \gg q^\downarrow(x)$ as $x \to 1$.  This model leads
to the expectation that $\Delta u/u \to 1$ and $\Delta d/d \to 1$
in the $x \to 1$ limit, and consequently a rapid approach to unity
of both the proton and neutron $A_1$ asymmetries.

An additional feature of the helicity conservation model is that it
correlates the $x \to 1$ behavior of the PDFs with the large-$Q^2$
behavior of elastic form factors \cite{Blankenbecler:1974tm, 
Gunion:1973nm, BrodskyLepage79}, which can be measured in elastic
lepton--nucleon scattering experiments at Jefferson Lab and elsewhere
\cite{Arrington:2006zm}.  This is also closely related to the notion
of quark-hadron duality, which connects averages over structure
functions in the resonance region at low $W$ with those extrapolated
from the deep-inelastic continuum at higher $W$ (lower $x$)
\cite{Melnitchouk:2005zr, Melnitchouk:2001eh, Bloom:1970xb}.
Since the existing DIS data are for the most part consistent
with a vanishingly small neutron asymmetry at $x \lesssim 0.5$,
a very dramatic change in $A_1^n$ would be expected at larger $x$.
An even more rapid transition at high $x$ may occur if quark orbital
angular momentum plays an important role in nucleon structure
\cite{Avakian:2007xa}.

\begin{figure}[t]
\begin{center}
\includegraphics[width=8cm]{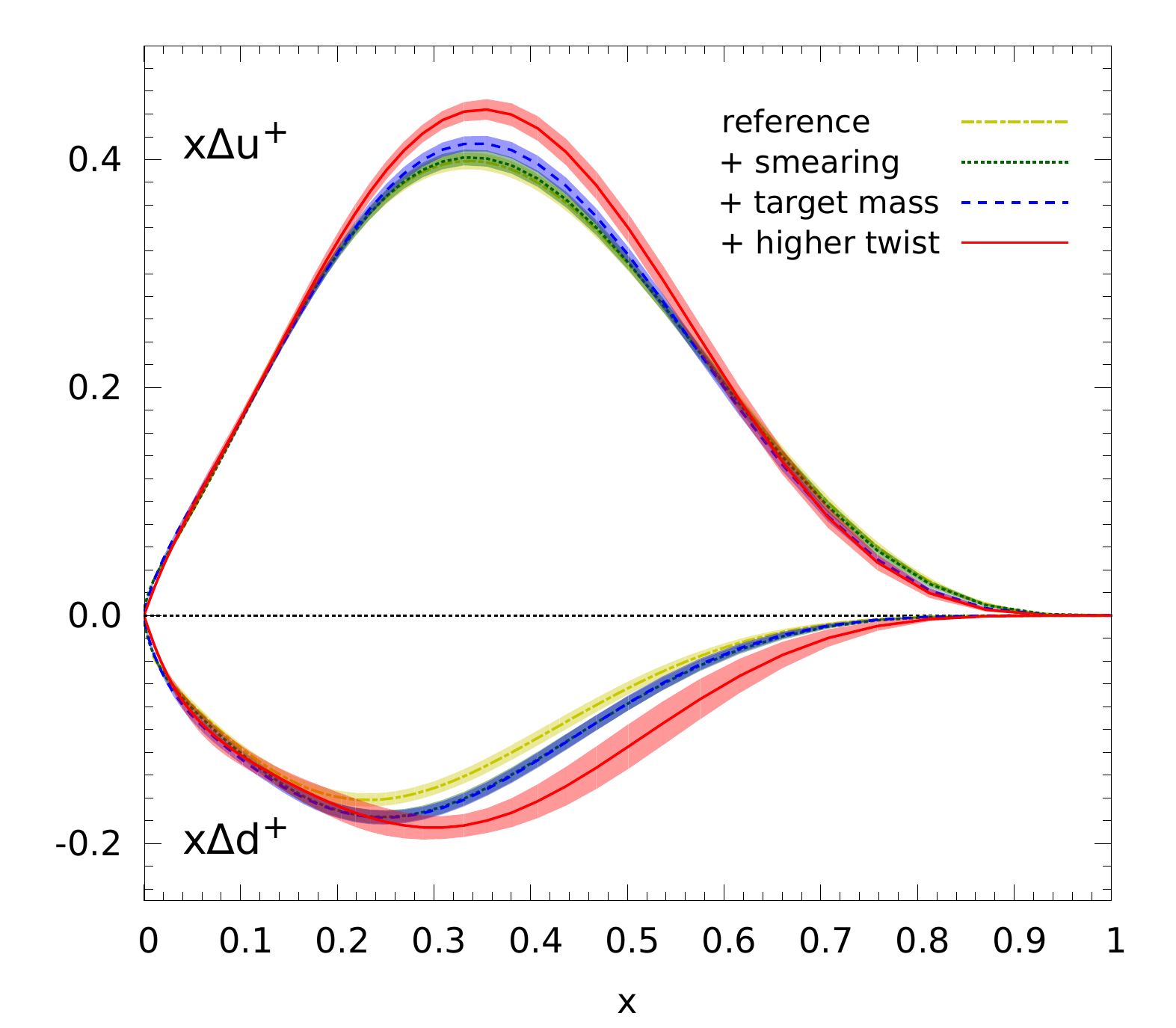}\hspace*{-0.4cm}
\includegraphics[width=8cm]{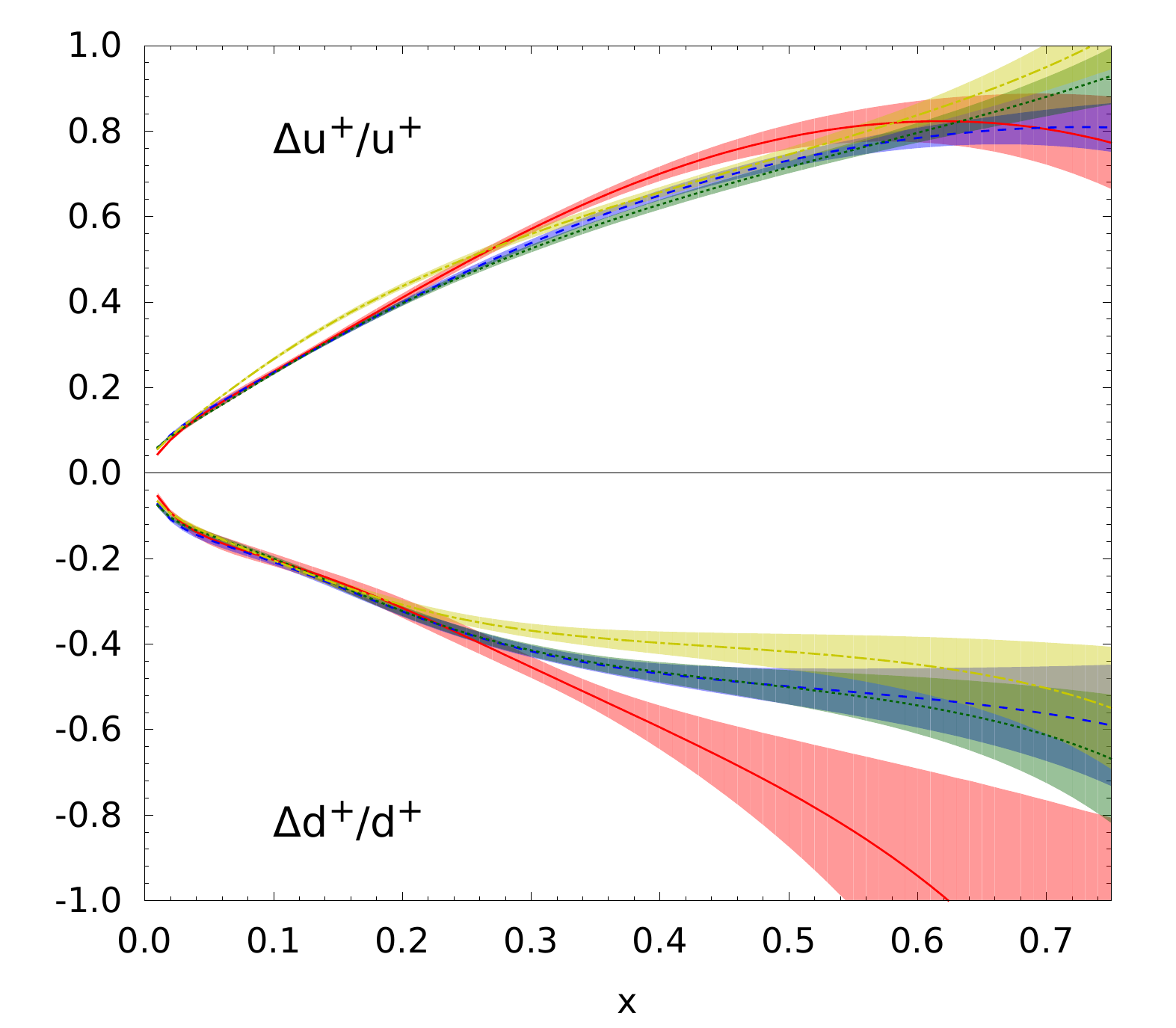}
\caption{{\it (Left)}
	Polarized $x \Delta u^+$ and $x \Delta d^+$ distributions
	and their uncertainties for the JAM13 PDF set \cite{JAM13}
	at $Q^2 = 1$~GeV$^2$, illustrating the effects of the
	nuclear smearing (green dotted), target mass (blue dashed)
	and higher twist (red solid) corrections, relative to the
	reference fit (yellow dot-dashed).  Note that the yellow
	and green bands overlap for $x \Delta u^+$ and the
	green and blue bands overlap for $x \Delta d^+$.
	{\it (Right)}
	Corresponding ratios of polarized to unpolarized
	$\Delta u^+/u^+$ and $\Delta d^+/d^+$ distributions.}
\label{fig:pol_val}
\end{center}
\end{figure}

Experimentally, the extraction of leading twist spin-dependent
PDFs at high $x$ faces similar challenges to those discussed in
Sec.~\ref{ssec:valence} for the unpolarized distributions.
In particular, information on the structure of the neutron is
typically obtained from experiments involving polarized deuterium or
$^3$He nuclei, requiring the nuclear corrections to be understood.
Analysis of high-$x$ data also necessitates careful treatment of
finite-$Q^2$ corrections such as target mass and higher twist
contributions.
The effects of these corrections are illustrated in
Fig.~\ref{fig:pol_val}, which shows the total $\Delta u^+$
and $\Delta d^+$ PDFs from the JAM13 analysis \cite{JAM13}
for various fits with or without the finite-$Q^2$ and nuclear
corrections.
At small values of $x$ their effects are negligible; however,
at $x \gtrsim 0.3$ they can give up $\sim 20\%$ corrections for
$\Delta u^+$ and more $\sim 50\%$ corrections for $\Delta d^+$
(the relative correction can be even larger at $x \gtrsim 0.8$,
although the PDFs are not constrained in this region).

The same effects are more clearly illustrated through the
polarization ratios $\Delta u^+/u^+$ and $\Delta d^+/d^+$ in
Fig.~\ref{fig:pol_val}, where the unpolarized distributions
are fitted simultaneously within the same analysis as the
polarized \cite{JAM13}.  This in principle eliminates any
bias arising from the use of spin-averaged PDFs taken from
analyses performed under different sets of assumptions.
In the intermediate-$x$ region the ratios for both $u$ and $d$
quarks are generally consistent with the symmetric quark model
expectations, with the $\Delta u^+/u^+$ increasing towards
unity at larger $x$.  The $\Delta d^+/d^+$ ratio, on the other
hand, remains negative for all $x$ where it is constrained,
and shows no indication of the upturn predicted by the
helicity conservation models.
The nuclear and finite-$Q^2$ effects can significantly impact
the limiting behavior as $x \to 1$, and clearly additional data
are needed in order to constrain their $x$ dependence at high $x$.
A dedicated program of large-$x$ structure function measurements
is planned at the 12~GeV energy upgraded Jefferson Lab facility
\cite{E12-06-109, E12-06-110, E12-06-122}.
More complete information on spin-dependent PDFs will also be
available once the $g_2$ structure function of the nucleon is
measured more accurately \cite{E12-06-121}.  This will provide
additional information on possible higher twist corrections
that enter in the extraction of leading twist PDFs from DIS
polarization asymmetries.

\subsection{Polarized sea quarks}
\label{ssec:pol_sea}

As for the unpolarized inclusive charged lepton DIS measurements,
inclusive $g_1$ structure function experiments measure $C$-even
combinations of PDFs, $\Delta q^+$.  The individual quark and
antiquark distributions can be separated with the help of 
semi-inclusive charged lepton scattering data, with coincident
measurement of fast pions or kaons in the final state.
Typically such experiments measure the semi-inclusive polarization
asymmetry, which at LO can be written
\begin{eqnarray}
A_1^h(x,z,Q^2)
&=& {\sum_q\, e_q^2\, \Delta q(x,Q^2)\, D_q^h(z,Q^2) \over
     \sum_q\, e_q^2\, q(x,Q^2)\, D_q^h(z,Q^2)},
\label{eq:A1sidis}
\end{eqnarray}
where $D_q^h(z,Q^2)$ is the fragmentation function for the
scattered quark or antiquark to produce a hadron $h$ ($h=\pi, K$)
in the current fragmentation region with a fraction $z = E_h/\nu$.
For large $z$, the produced hadron has a high probability of
containing the scattered parton, hence providing a tag on the
initial state distribution of quarks and antiquarks.
The fragmentation functions $D_q^h$ are extracted from independent
fits to single hadron production cross sections in $e^+ e^-$, $pp$
and other reactions.

\begin{figure}[t]
\begin{center}
\includegraphics[width=9cm]{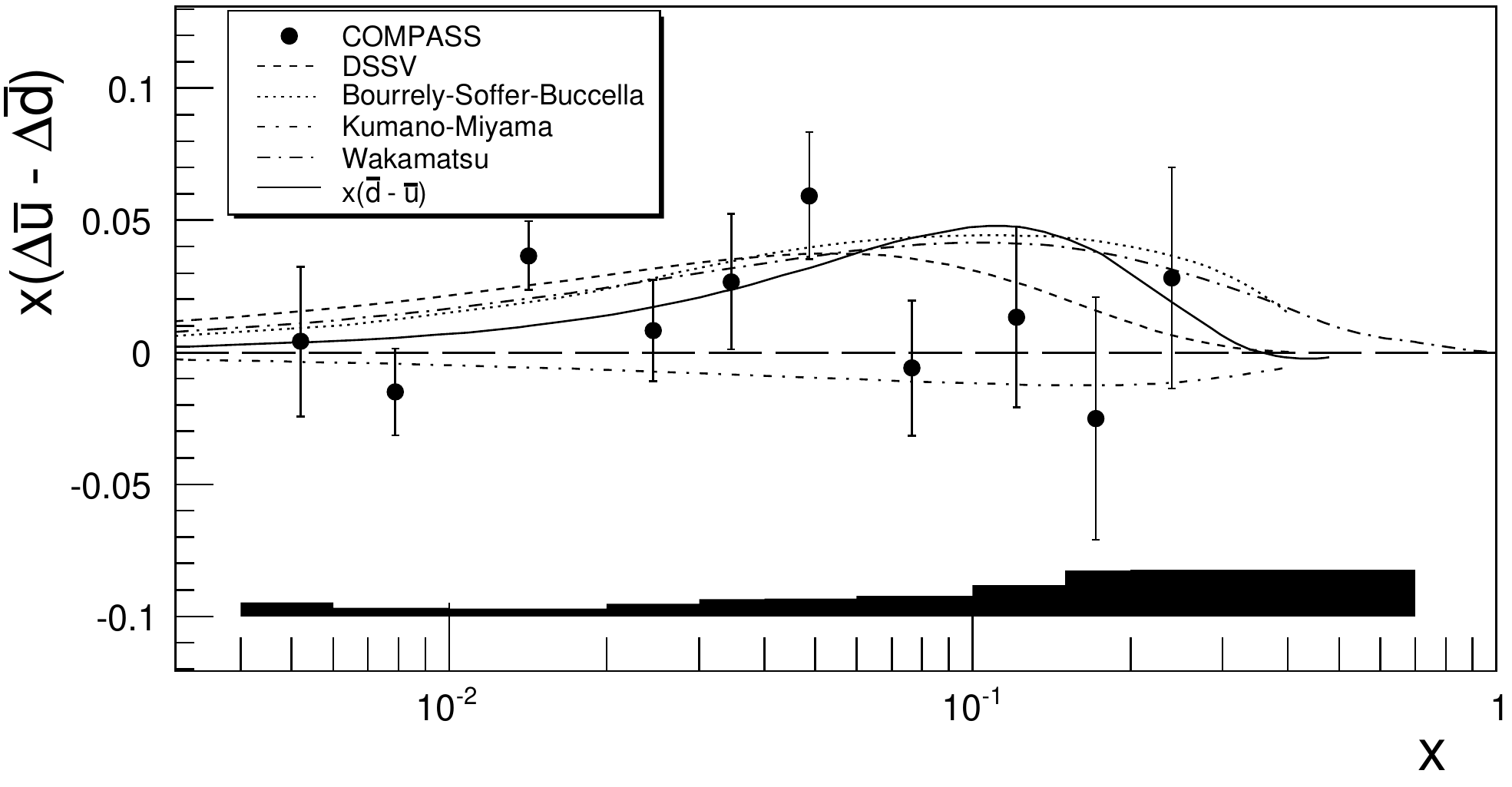}
\includegraphics[width=6cm]{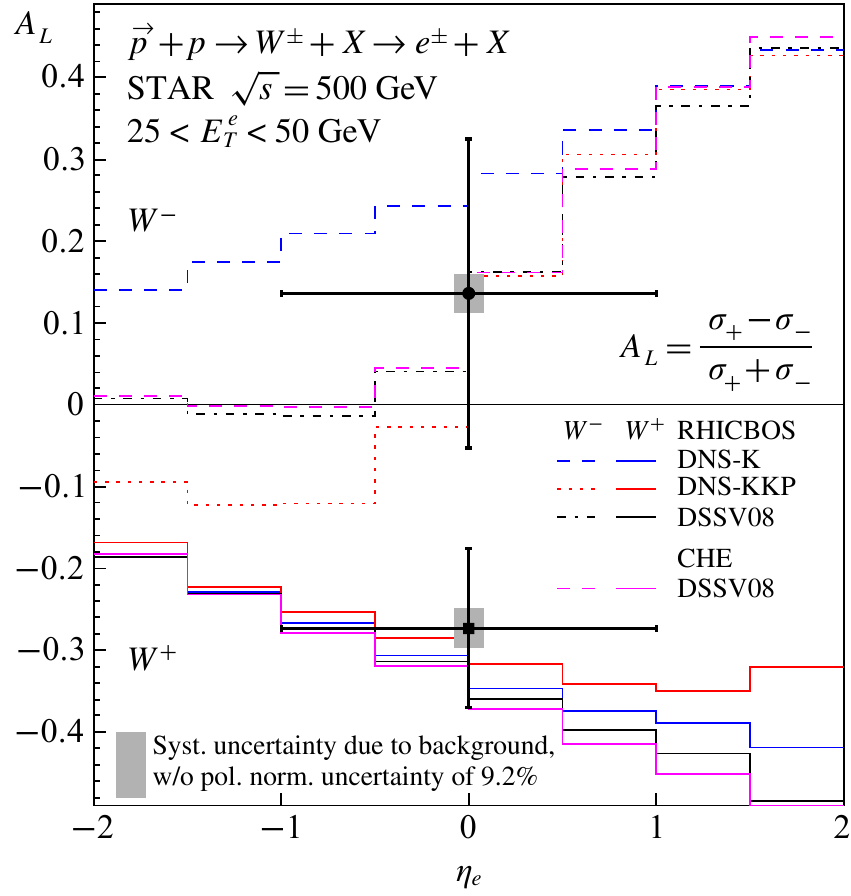}
\caption{{\it (Left)}
	Flavor asymmetry of the polarized sea,
	$x (\Delta \bar u - \Delta \bar d)$, from the COMPASS
	semi-inclusive DIS measurement at $Q^2 = 3$~GeV$^2$
	\cite{Alekseev:2010ub}, compared with several parametrizations
	and model calculations, and the unpolarized $x(\bar d-\bar u)$
	asymmetry (solid curve).
	(Figure from Ref.~\cite{Alekseev:2010ub}.)
	{\it (Right)}
	Single polarization asymmetry $A_L$ for inclusive $W^\pm$
	production in $\vec p\, p$ scattering from the STAR experiment
	at RHIC, at central electron pseudorapidity $\eta_e = 0$.
	(Figure from Ref.~\cite{Aggarwal:2010vc}.
	Copyright (2011) by the American Physical Society.)}
\label{fig:pol_sea}
\end{center}
\end{figure}

The program of semi-inclusive DIS measurements from proton and
deuteron targets at HERMES \cite{Airapetian:2004zf}, COMPASS
\cite{Alekseev:2010ub}, and earlier SMC \cite{Adeva:1997qz},
have produced intriguing glimpses into the flavor content of
the polarized sea.  The extracted sea quark and antiquark
asymmetries are generally rather small, and consistent with
zero within the current uncertainties.  For the light quark
polarized sea, there is a slight trend towards a positive
$\Delta \bar u$ (with an $x$-integrated value of
$+0.02 \pm 0.02\, {\rm (stat.)} \pm 0.01\, {\rm (syst.)}$
from the latest COMPASS data \cite{Alekseev:2010ub})
and a negative $\Delta \bar d$ (with an $x$-integrated value of
$-0.05 \pm 0.03\, {\rm (stat.)} \pm 0.02\, {\rm (syst.)}$),
so that the difference $\Delta \bar u - \Delta \bar d$ is
slightly positive, at the $1.5 \sigma$ level \cite{Alekseev:2010ub},
as Fig.~\ref{fig:pol_sea} indicates.
This is in contrast to the large negative values of the unpolarized
$\bar u - \bar d$ asymmetry (see Sec.~\ref{ssec:lightsea}), which
would disfavor models such as the chiral quark soliton model in the
large-$N_c$ limit \cite{Wakamatsu:1991yu, Pobylitsa:1998tk} that
predict large
	$\Delta \bar u - \Delta \bar d \gg \bar d - \bar u$.
Very small antiquark spin asymmetries would be expected in models
based on the pion cloud of the nucleon, such as those discussed in
Sec.~\ref{ssec:lightsea}.
One should caution, however, that the experimental distributions
in Fig.~\ref{fig:pol_sea} were extracted assuming LO expressions
for the semi-inclusive asymmetries, which, strictly speaking,
cannot be directly compared with NLO PDFs \cite{deFlorian:2009vb}.
The comparison should therefore be viewed as a qualitative one,
although the general indication of a small polarized non-strange
sea is unlikely to be qualitatively modified in an NLO treatment.

The semi-inclusive DIS measurements are also consistent with a
very small strange quark polarization, $\Delta s$.  The HERMES
data \cite{Airapetian:2008qf} give a slightly positive value
when integrated between $x = 0.02$ and 0.6, while the COMPASS
data \cite{Alekseev:2010ub} are consistent with zero.
This is in contrast with the analysis of inclusive DIS data on
protons, deuteron and $^3$He, which suggests a small negative value.
The extractions of $\Delta s$ from the inclusive and semi-inlcusive
data are somewhat dependent on the assumptions of SU(3) flavor
symmetry for the axial charges of hyperons, and on the specific
fragmentation functions used in the semi-inclusive analysis.
Future measurements of kaon fragmentation functions $D_q^K$
should improve the accuracy of the flavor decomposition of the
proton's helicity distributions.

A complementary method, discussed by Bourrely and Soffer
\cite{Bourrely:1993dd}, of constraining the $\Delta \bar u$ and
$\Delta \bar d$ distributions, which does not depend on knowledge
of fragmentation functions, is through longitudinal polarization
asymmetries in inclusive $W^\pm$ boson production from $pp$
scattering with one proton polarized and the other unpolarized,
$\vec p\, p \to W^\pm\, X \to e^\pm\, X$.
The asymmetry for $W^-$ production, for example, can then be
written at LO as \cite{deFlorian:2010aa}
\begin{eqnarray}
A_L^{W^-}
&=& { \Delta \bar u(x_a)\, d(x_b) (1-\cos\theta)^2
    - \Delta d(x_a)\, \bar u(x_b) (1+\cos\theta)^2 \over
      \bar u(x_a)\, d(x_b) (1-\cos\theta)^2
    + d(x_a)\, \bar u(x_b) (1+\cos\theta)^2 },
\label{eq:polW-}
\end{eqnarray}
where $\theta$ is the scattering angle of the electron in the
partonic center-of-mass frame, while the corresponding $W^+$
asymmetry at LO is
\begin{eqnarray}
A_L^{W^+}
&=& { \Delta \bar d(x_a)\, u(x_b) (1+\cos\theta)^2
    - \Delta u(x_a)\, \bar d(x_b) (1-\cos\theta)^2 \over
      \bar d(x_a)\, u(x_b) (1+\cos\theta)^2
    + u(x_a)\, \bar d(x_b) (1-\cos\theta)^2 }.
\label{eq:polW+}
\end{eqnarray}
At large negative rapidity, where $x_a \ll x_b$, the $W^-$
asymmetry then directly probes the $\Delta \bar u$ PDF,
$A_L^{W^-} \sim \Delta \bar u(x_a)/\bar u(x_a)$,
while at large positive rapidity one has $x_a \gg x_b$,
and the asymmetry is sensitive to the $\Delta d$ PDF at
large $x$, $A_L^{W^-} \sim -\Delta d(x_a)/d(x_a)$.
For $W^+$ production, the term proportional to $\Delta \bar d(x_a)$
will be similarly enhanced at small $x_a \ll x_b$; however, the
angular factor will produce a suppression of this contribution at
backward angles relative to the $\Delta u(x_a)$ term, so that both
terms will give competing contributions.

The first data from the STAR experiment (Run~9) at RHIC
\cite{Aggarwal:2010vc} are shown in Fig.~\ref{fig:pol_sea}
for zero electron rapidity, and are generally in agreement
with NLO calculations using existing PDF parametrizations.
Data from Run~12 at RHIC have subsequently been taken at both
forward and backward rapidities \cite{Aschenauer:2013woa}, and will
provide important constraints on the antiquark polarization.

\subsection{Gluon helicity}
\label{ssec:pol_gluon}

Information on the polarized gluon distribution $\Delta g$ can be
obtained in several complementary ways.  Firstly, from the $Q^2$
evolution equations, the evolution of the quark singlet contribution
to the $g_1$ structure function mixes with the gluon contribution,
so that in principle the study of scaling violations in $g_1$ can
constrain $\Delta g$.  In practice, however, the $Q^2$ range and
precision of the available inclusive DIS data are not sufficient to
provide meaningful constraints, and DIS-only fits of spin-dependent
PDFs effectively leave $\Delta g$ undetermined.

\begin{figure}[t]
\begin{center}
\includegraphics[width=8.5cm]{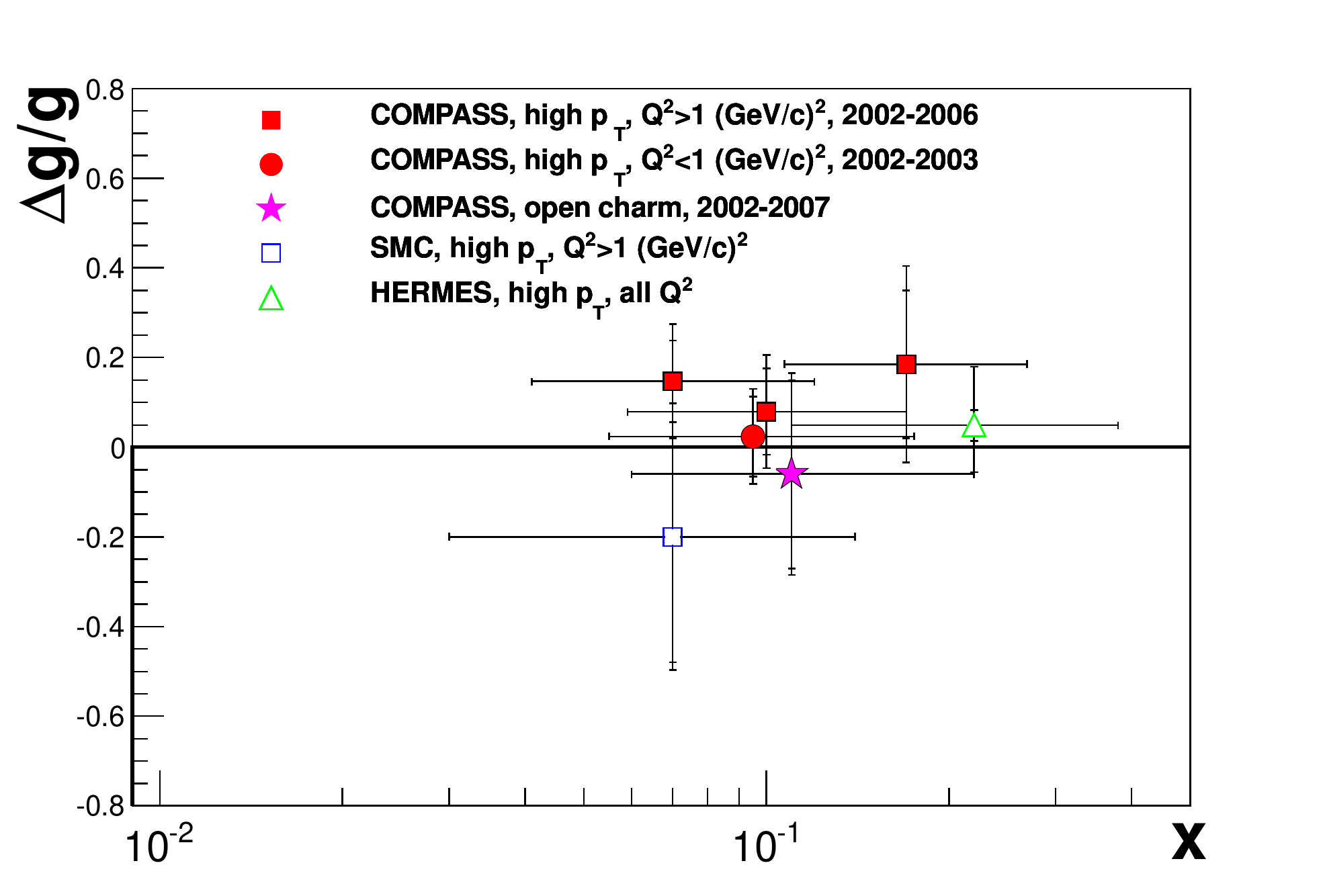}\hspace*{-0.5cm}
\includegraphics[width=8.5cm]{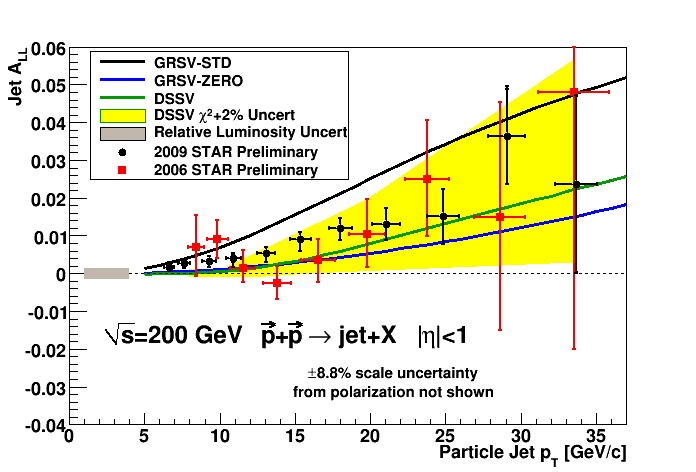}
\caption{{\it (Left)}
	Gluon polarization ratio $\Delta g/g$ values extracted
	at LO from charm and high-$p_T$ hadron production data
	from COMPASS, SMC and HERMES.
        (Figure from Ref.~\cite{Adolph:2012ca}.
	Copyright (2013) by the American Physical Society.)
	{\it (Right)}
	Longitudinal double-spin asymmetry $A_{LL}$ for inclusive
	jet production, as a function of the jet transverse momentum,
	in polarized $\vec p\, \vec p$ scattering from the STAR
	experiment at RHIC \cite{Adamczyk:2012qj}, compared with
	calculations based on several PDF parametrizations.
        (Figure from Ref.~\cite{collaboration:2011fga}.)}
\label{fig:pol_g}
\end{center}
\end{figure}

More direct information on gluon polarization has come from the
production of charm mesons and high-$p_T$ hadrons in semi-inclusive
DIS, through the process of photon--gluon fusion
(see Sec.~\ref{ssec:HQresults}).  Data on hadron production from
HERMES, SMC and COMPASS, and open charm data from COMPASS have
been analyzed over a range of $x$ values, averaging from
$\langle x \rangle = 0.07$ to $\approx 0.2$, and $Q^2$ between
$\approx 1$~GeV$^2$ and 13~GeV$^2$.
Most of the analyses have been performed at LO, finding values of
$\Delta g/g$ consistent with zero, albeit with large uncertainties,
as illustrated in Fig.~\ref{fig:pol_g}.  The recent COMPASS experiment
\cite{Adolph:2012ca} was the first to be analyzed at NLO, yielding
$\Delta g/g = -0.13 \pm 0.15\, {\rm (stat.)} \pm 0.15\, {\rm (syst.)}$
over the range $0.12 < x < 0.33$ with an average
$\langle x \rangle \approx 0.2$.

An alternative reaction that can be used to constrain $\Delta g$
is the inclusive production of jets or neutral pions in polarized
proton--proton scattering.
The longitudinal double-spin asymmetry for this process,
\begin{eqnarray}
A_{LL}
&=& { \sigma^{++} - \sigma^{+-} \over \sigma^{++} + \sigma^{+-}},
\label{eq:A_LL}
\end{eqnarray}
where $\sigma^{++} (\sigma^{+-})$ is the differential cross section
for longitudinally polarized protons with equal (opposite) helicities,
is then sensitive to the polarized quark--gluon scattering cross section
at the parton level.
Following an early measurement of $A_{LL}$ for multi-$\gamma$
pairs from two-jet type events by the Fermilab E581/704 Collaboration
\cite{Adams:1994bg}, which found a small $\Delta g/g$ at
$0.05 \lesssim x \lesssim 0.35$, most of the experimental effort
at measuring $\Delta g$ in $pp$ scattering has been by the
PHENIX and STAR Collaborations at RHIC.

The STAR experiment measures inclusive jet production in
$\vec p\, \vec p \to {\rm jet} + X$ at an invariant center of mass
energy $\sqrt{s}=200$~GeV.  The $A_{LL}$ data from the 2006 and
2009 runs \cite{Adamczyk:2012qj, collaboration:2011fga} shown in
Fig.~\ref{fig:pol_g} provide the clearest evidence to date for a
small but nonzero $\Delta g$.  A nonzero asymmetry is also observed
by STAR in preliminary data on dijet events at mid-rapidity
\cite{Walker:2011vs}, which may provide even better constraints
on the shape of $\Delta g$.

The PHENIX data on inclusive neutral pion production,
$\vec p\, \vec p \to \pi^0 + X$, from 2005 to 2009 suggest
a very small asymmetry up to $p_T \approx 10$~GeV
\cite{Adare:2008aa, Manion:2011zz}.
Taken together, the inclusive RHIC data from several different
channels suggest a small but non-vanishing gluon polarization
in the proton.  Though not as large as had been proposed in
some early explanations of the small contribution of quarks
to the proton spin, the precise role played by $\Delta g$ in
the proton spin budget will be intriguing to understand from
future measurements.

\section{Outlook}

In this topical review we have presented a summary of the current
knowledge of spin-averaged and (longitudinal) spin-dependent PDFs.
While great progress has been made in recent years, there are still
regions of Bjorken-$x$ where the PDF uncertainty bands are uncomfortably
large.  This is primarily due to either lack of data in kinematic
regions that would otherwise provide constraints on the PDFs, or to
the particular PDF being smaller in that region than those of other
flavors, or both.  One might naively conclude that if existing data
do not constrain particular PDFs in certain domains, then it may not
be too important to seek ways to constrain them there.  However, this
viewpoint overlooks two important points: (i) there is intrinsic
interest in the behavior of the different PDFs and relations to the
properties of the parent hadrons, and (ii) it is possible that new
phenomena may occur in extended kinematic regions which would require
precise knowledge of the PDFs there.

As an example, suppose one were searching for a massive state produced
at large rapidity in hadron--hadron collisions.  This would typically
require one parton to have a large value of $x$ and the other a small
value.  The errors at large $x$ may well dominate the uncertainty in
the resulting prediction.  The key here is to measure Standard Model
processes in expanded kinematic regions and include these processes in
the global fits.  The resulting reduction in the errors can then be
used to improve searches for new phenomena in these regions.  Moreover,
if one reduces the errors at large values of $x$, this will necessarily
lead to reductions in the errors at intermediate $x$.

Since deuterium targets in DIS are a prime source of information
on the $d$ quark PDF at large values of $x$, nuclear corrections
are a significant source of uncertainty in its determination.
Several planned experiments at the 12~GeV energy upgraded Jefferson Lab
\cite{MARATHON, BONUS12, SOLID} will use techniques that reduce the
effects of nuclear corrections on the extraction of $d/u$ at large $x$.
Constraints on the $d/u$ ratio in the proton can also be obtained,
free of nuclear effects, through the measurement of $W^\pm$ charge
asymmetries in $pp$ collisions at very high rapidity \cite{Brady:2011hb},
although this would require reconstruction of the $W^\pm$ distributions
themselves from the decay lepton distributions and, unlike in $p\bar{p}$
collisions at the Tevatron, would require precise knowledge of the
$\bar u$ and $\bar d$ distributions \cite{Aaltonen:2009ta}.

Another important area of investigation is the flavor separation of
the PDFs.  For example, currently our knowledge of the $s$ and $\bar s$
PDFs comes primarily from one neutrino experiment which used an iron
target \cite{Mason:2007zz}, thus requiring the use of model-dependent
nuclear corrections.  Other sources, for example weak vector boson
production at the LHC \cite{Aad:2012sb}, can provide additional
constraints, especially as the statistics of the measurements improve.
In a complementary effort, the HERMES Collaboration has measured
\cite{Airapetian:2008qf} $K^{\pm}$ production in semi-inclusive DIS
and used these data to place constraints on the strange quark PDF.
While an interesting analysis, it does depend on knowledge of the
$K^{\pm}$ fragmentation functions, and a recent, more complete analysis
\cite{Airapetian:2013} of the HERMES data suggests a somewhat smaller
$s$ PDF than that extracted in Ref.~\cite{Airapetian:2008qf}.
Better knowledge of the fragmentation functions would clearly help
in the interpretation of the semi-inclusive DIS measurements.
Ideally, a high statistics neutrino experiment on {\it hydrogen}
could further provide knowledge of both the $s$ and $\bar s$ PDFs
through the measurement of charm production (using the
$\nu s \to \mu^- c$ and $\bar \nu \bar s \to \mu^+ \bar c$
subprocesses) without having to make model-dependent nuclear
corrections.  Note that such an experiment could also yield
information on the $d/u$ ratio at high values of $x$.

Information on the SU(2) flavor asymmetry of the proton sea will be
provided by Experiment E906/SeaQuest \cite{SeaQuest} at Fermilab,
which is currently taking data on muon pair production using a
120~GeV proton beam on a variety of targets.  It is designed to
improve our knowledge of $\bar d - \bar u$ over the $x$ range out
to $x \sim 0.45$, whereas the currently available data run out of
statistics at $x \approx 0.3$.  The experiment will further enable
a study of the effects of nuclear binding on the sea quark PDFs.
There is also the possibility of follow-on measurements of lepton
pair production at the Nuclear and Particle Physics facility at
J-PARC \cite{J-PARC-P04, Kumano:2010qm} in Japan.

The question of charge symmetry violation in PDFs can be addressed
in future through study of Drell-Yan lepton pair production in
$\pi^\pm$ scattering from the deuteron, which could be measured
at the Fermilab Main Injector or with the COMPASS experiment at
CERN \cite{Londergan:2009kj}, or via measurement of $\pi^+/\pi^-$
electroproduction ratios in semi-inclusive DIS from isoscalar targets
\cite{E12-09-002}.  Another avenue to explore will be parity-violating
DIS on a deuterium target \cite{Prescott:1978tm, Wang:2013kkc},
for which the parity-violating asymmetry arising from $\gamma-Z$
interference is expected to be independent of hadronic structure
at large $x$ in the absence of CSV effects \cite{Brady:2011uy,
Bjorken:1978ry, Hobbs:2008mm, Mantry:2010ki}.
Beyond searches for CSV effects, a program of parity-violating DIS
studies at the 12~GeV Jefferson Lab \cite{SOLID} could open up a new
window on the vector $F_{1,2}^{\gamma Z}$ interference structure
functions, as well as provide glimpses into the parity-odd
$F_3^{\gamma Z}$ structure function.

For heavier quarks, the issue of intrinsic charm will require
measuring the charm structure function at large values of $x$,
which may be possible at a future Electron-Ion Collider facility
\cite{Accardi:2012hwp}.  The semi-inclusive production of prompt
photons and charmed jets in $pp$ collisions at the LHC,
$pp \to \gamma\, c\, X$, may produce an enhancement at large
transverse momenta from intrinsic charm \cite{Bednyakov:2013zta}.

Measurements of jet, dijet, isolated $\gamma$, and $\gamma$ + jet
production at the LHC will all play a role in helping to further
constrain the gluon PDF.  Of particular interest would be
$\gamma$ + jet production where both the $\gamma$ and the jet have
equal and opposite large rapidities.  The dominant $qg \to \gamma q$
subprocess would thus require both a high-$x$ gluon and a high-$x$
quark.  Furthermore, continued progress on resummation, especially
for the fragmentation contribution at fixed target energies may
provide additional information on the gluon at high values of $x$.

In the polarized sector, several experiments planned at Jefferson Lab
will measure the longitudinal polarization asymmetries for hydrogen,
deuterium and $^3$He to unprecedented large values of $x$
\cite{E12-06-109, E12-06-110, E12-06-122}.  These will constrain
the spin-dependent valence quark PDFs, especially the $\Delta d$
distribution, to values of $x \sim 0.8$.  This should help answer
long-standing questions about the behavior of the $\Delta q/q$
ratios as $x \to 1$, and reveal insights into the role of quark
orbital angular momentum in the nucleon.

The flavor asymmetry of the polarized light quark sea,
$\Delta \bar u - \Delta \bar d$, will be further probed via $W$ boson
production in polarized $pp$ scattering at RHIC at $\sqrt{s} = 510$~GeV.
Both the STAR and PHENIX experiments anticipate significantly improved
statistics and kinematic coverage in their 2012 and 2013 runs
\cite{Aschenauer:2013woa}.
The strange quark polarization, despite appearing to be smaller than
initially surmised from early polarized DIS experiments, is neverthess
important to pin down, as it allows a portal to nonperturbative QCD
effects in the nucleon.  Improved constraints on kaon fragmentation
functions should lead to more reliable determinations of $\Delta s$
in semi-inclusive DIS.

Finally, the polarized gluon distribution is being actively
investigated in semi-inclusive deep-inelastic hadron production
at COMPASS, and through double-spin asymmetries for pion and jet
production in polarized $pp$ collisions at RHIC.  Preliminary results
from STAR on jet and dijet cross sections hint at nonzero values for
$\Delta g$, which will need to be confirmed by further measurements
at small $x$.  Access to $\Delta g$ below $x \sim 0.05$ will be
possible with future running at $\sqrt{s} = 500$~GeV and at large
forward rapidity \cite{Aschenauer:2013woa}, although reaching values
down to $x \sim 10^{-4}$ will only be possible with an Electron-Ion
Collider \cite{Accardi:2012hwp}.

On the theory front, continued progress on NNLO calculations will
eventually result in global fits where all the processes are treated
consistently at NNLO, for both unpolarized and polarized observables.
The increased precision and decreased scale dependence will help to
reduce the uncertainty bands on the PDFs.
The parametrization dependence of global fits will continue to be
explored, allowing for greater flexibility and reducing bias
introduced with the use of specific forms, such as for the $d/u$
ratio at large $x$.
Improvements in electroweak radiative corrections are being sought,
incorporating, for example, PDFs of a photon with ${\cal O}(\alpha)$
$Q^2$ evolution.
Progress on computing moments of PDFs in lattice QCD is also
anticipated, with more reliable simulations performed on larger
lattice volumes and at quark masses near the physical limit.
A new approach \cite{Ji:2013dva} to calculating the $x$ dependence
of the PDFs directly in the infinite momentum frame may provide
complementary constraints on PDFs that are difficult to access
experimentally.

\ack

We thank A.~Accardi and K.~Griffioen for helpful comments
and a careful reading of the manuscript.
The work of P.J.-D. and W.M. was supported by DOE contract
No.~DE-AC05-06OR23177, under which Jefferson Science
Associates, LLC operates Jefferson Lab.  The work of J.F.O.
was supported by DOE contract No.~DE-FG02-13ER41942.



\section*{References}

\begin{thebibliography}{10}


\bibitem{Forte:2013wc} 
S.~Forte and G.~Watt,
Ann. Rev. of Nucl. Part. Sci. {\bf 63}, (2013),
arXiv:1301.6754 [hep-ph].

\bibitem{Blumlein:2012bf} 
J.~Bl\"umlein,
Prog. Part. Nucl. Phys. {\bf 69}, 28 (2013).

\bibitem{Accardi:2013}
A.~Accardi,
{\it Large-x connections of nuclear and high-energy physics},
to appear in Mod. Phys. Lett. A;
PoS (Confinement X) 227,	\\
{\tt http://pos.sissa.it/archive/conferences/171/227/Confinement\%20X\_227.pdf}.

\bibitem{Buras:1979yt} 
A.~J.~Buras,
Rev. Mod. Phys. {\bf 52}, 199 (1980).

\bibitem{Owens:1992hd} 
J.~F.~Owens and W.-K.~Tung,
Ann. Rev. Nucl. Part. Sci. {\bf 42}, 291 (1992).

\bibitem{Lampe:1998eu} 
B.~Lampe and E.~Reya,
Phys. Rep. {\bf 332}, 1 (2000).

\bibitem{deFlorian:2011ia} 
D.~de Florian, R.~Sassot, M.~Stratmann and W.~Vogelsang,
Prog. Part. Nucl. Phys. {\bf 67}, 251 (2012).

\bibitem{Aidala:2012mv} 
C.~A.~Aidala, S.~D.~Bass, D.~Hasch and G.~K.~Mallot,
Rev. Mod. Phys. {\bf 85}, 655 (2013).


\bibitem{PDG12}
J.~Beringer {\it et al.} [Particle Data Group],
Phys. Rev. D {\bf 86}, 010001 (2012).

\bibitem{Yndurain}
F.~J.~Yndurain,
{\it The Theory of Quark and Gluon Interactions},
Springer-Verlag (1983).

\bibitem{BargerPhillips}
V.~D.~Barger and R.~J.~N.~Phillips,
{\it Collider Physics}, Westview Press (1991).

\bibitem{Pink}
R.~K.~Ellis, W.~J.~Stirling and B.~R.~Webber,
{\it QCD and Collider Physics}, Cambridge University Press (1996).

\bibitem{Drell:1970yt} 
S.~D.~Drell and T.~M.~Yan,
Ann. Phys. {\bf 66}, 578 (1971).
  
\bibitem{Reya:1979zk}
E.~Reya,
Phys. Rep. {\bf 69}, 195 (1981).

\bibitem{Abramowicz:1900rp}
H.~Abramowicz {\it et al.}  [H1 and ZEUS Collaborations],  
Eur. Phys. J. C {\bf 73}, 2311 (2013).

\bibitem{Gluck:2006ju}
M.~Gl\"uck and E.~Reya,
Mod. Phys. Lett.  A {\bf 22}, 351 (2007).

\bibitem{Witten:1975bh}
E.~Witten,
Nucl. Phys. {\bf B104}, 445 (1976).

\bibitem{Babcock:1977fi}
J.~Babcock, D.~W.~Sivers and S.~Wolfram,
Phys. Rev. D {\bf 18}, 162 (1978).

\bibitem{Leveille:1978px}
J.~P.~Leveille and T.~J.~Weiler,
Nucl. Phys. {\bf B147}, 147 (1979).

\bibitem{Gluck:1980cp}
M.~Gl\"uck, E.~Hoffmann and E.~Reya,
Z. Phys. C {\bf 13}, 119 (1982).

\bibitem{Laenen:1992zk}
E.~Laenen, S.~Riemersma, J.~Smith and W.~L.~van Neerven,
Nucl. Phys. {\bf B392}, 162 (1993).

\bibitem{Riemersma:1994hv}
S.~Riemersma, J.~Smith and W.~L.~van Neerven,
Phys. Lett. B {\bf 347}, 143 (1995).

\bibitem{Laenen:1998kp}
E.~Laenen and S.-O.~Moch,
Phys. Rev. D {\bf 59}, 034027 (1999).

\bibitem{Kawamura:2012cr}
H.~Kawamura, N.~A.~Lo Presti, S.-O.~Moch and A.~Vogt,
Nucl. Phys. {\bf B864}, 399 (2012).

\bibitem{Bierenbaum:2009mv}
I.~Bierenbaum, J.~Bl\"umlein and S.~Klein,
Nucl. Phys. {\bf B820}, 417 (2009).

\bibitem{Bierenbaum:2009zt}
I.~Bierenbaum, J.~Bl\"umlein and S.~Klein,
Phys. Lett. B {\bf 672}, 401 (2009).

\bibitem{Ablinger:2012sm}
J.~Ablinger, J.~Bl\"umlein, A.~De Freitas, A.~Hasselhuhn, S.~Klein, 
C.~Schneider and F.~Wi\ss brock,
arXiv:1212.5950 [hep-ph].

\bibitem{Alekhin:2010sv} 
S.~Alekhin and S.~Moch,
Phys. Lett. B {\bf 699}, 345 (2011).

\bibitem{Buza:1996wv}
M.~Buza, Y.~Matiounine, J.~Smith and W.~L.~van Neerven,
Eur. Phys. J. C {\bf 1}, 301 (1998).

\bibitem{Gluck:2008gs}
M.~Gl\"uck, P.~Jimenez-Delgado, E.~Reya and C.~Schuck,
Phys. Lett. B {\bf 664}, 133 (2008).

\bibitem{Aivazis:1993pi}
M.~A.~G.~Aivazis, J.~C.~Collins, F.~I.~Olness and W.-K.~Tung,
Phys. Rev. D {\bf 50}, 3102 (1994).

\bibitem{Thorne:1997ga}
R.~S.~Thorne and R.~G.~Roberts,
Phys. Rev. D {\bf 57}, 6871 (1998).
  
\bibitem{Steffens:1999hx} 
F.~M.~Steffens, W.~Melnitchouk and A.~W.~Thomas,
Eur. Phys. J. C {\bf 11}, 673 (1999).

\bibitem{Forte:2010ta}
S.~Forte, E.~Laenen, P.~Nason and J.~Rojo,
Nucl. Phys. {\bf B834}, 116 (2010).

\bibitem{Collins:1998rz} 
J.~C.~Collins,
Phys. Rev. D {\bf 58}, 094002 (1998).

\bibitem{Gluck:1993dpa}
M.~Gl\"uck, E.~Reya and M.~Stratmann,
Nucl. Phys. {\bf B422}, 37 (1994).

\bibitem{Gluck:1994uf}
M.~Gl\"uck, E.~Reya and A.~Vogt,
Z. Phys. C {\bf 67}, 433 (1995).

\bibitem{Gluck:1998xa}
M.~Gl\"uck, E.~Reya and A.~Vogt,
Eur. Phys. J. C {\bf 5}, 461 (1998).

\bibitem{Kretzer:2003it}
S.~Kretzer, H.~L.~Lai, F.~I.~Olness and W.-K.~Tung,
Phys. Rev. D {\bf 69}, 114005 (2004).

\bibitem{Alekhin:2012ig}
S.~Alekhin, J.~Bl\"umlein and S.-O.~Moch,
Phys. Rev. D {\bf 86}, 054009 (2012).

\bibitem{Gluck:2007ck}
M.~Gl\"uck, P.~Jimenez-Delgado and E.~Reya,   
Eur. Phys. J. C {\bf 53}, 355 (2008).

\bibitem{JimenezDelgado:2009tv}
P.~Jimenez-Delgado and E.~Reya,
Phys. Rev. D {\bf 80}, 114011 (2009).

\bibitem{Martin:2009iq}
A.~D.~Martin, W.~J.~Stirling, R.~S.~Thorne and G.~Watt,  
Eur. Phys. J. C {\bf 63}, 189 (2009).

\bibitem{Lai:2010vv}
H.-L.~Lai, M.~Guzzi, J.~Huston, Z.~Li, P.~M.~Nadolsky, J.~Pumplin
and C.-P.~Yuan,
Phys. Rev. D {\bf 82}, 074024 (2010).

\bibitem{Gao:2013xoa} 
J.~Gao, M.~Guzzi, J.~Huston, H.~-L.~Lai, Z.~Li, P.~Nadolsky, J.~Pumplin,
D.~Stump and C.-P.~Yuan,
arXiv:1302.6246 [hep-ph].

\bibitem{Aaron:2009aa}
F.~D.~Aaron {\it et al.} [H1 and ZEUS Collaborations],	
JHEP {\bf 1001}, 109 (2010).

\bibitem{Ball:2010de}
R.~D.~Ball, L.~Del Debbio, S.~Forte, A.~Guffanti, J.~I.~Latorre,
J.~Rojo and M.~Ubiali,
Nucl. Phys. {\bf B838}, 136 (2010).

\bibitem{Melnitchouk:2005zr} 
W.~Melnitchouk, R.~Ent and C.~Keppel,
Phys. Rep. {\bf 406}, 127 (2005).

\bibitem{Ball:2012cx}
R.~D.~Ball {\it et al.},
Nucl. Phys. {\bf B867}, 244 (2013).

\bibitem{Alekhin:2009ni}
S.~Alekhin, J.~Bl\"umlein, S.~Klein and S.-O.~Moch,
Phys. Rev. D {\bf 81}, 014032 (2010).

\bibitem{CJ10}
A.~Accardi, M.~E.~Christy, C.~E.~Keppel, P.~Monaghan, W.~Melnitchouk,
J.~G.~Morfin and J.~F.~Owens,
Phys. Rev. D {\bf 81}, 034016 (2010).
        
\bibitem{CJ11}
A.~Accardi, W.~Melnitchouk, J.~F.~Owens, M.~E.~Christy, C.~E.~Keppel,
L.~Zhu and J.~G.~Morfin,   
Phys. Rev. D {\bf 84}, 014008 (2011).

\bibitem{CJ12}
J.~F.~Owens, A.~Accardi and W.~Melnitchouk,
Phys. Rev. D {\bf 87}, 094012 (2013).

\bibitem{Georgi:1976ve}
H.~Georgi and H.~D.~Politzer,
Phys. Rev. D {\bf 14}, 1829 (1976).

\bibitem{Matsuda:1979ad}
S.~Matsuda and T.~Uematsu,  
Nucl. Phys. {\bf B168}, 181 (1980).      
  
\bibitem{Piccione:1997zh}
A.~Piccione and G.~Ridolfi,  
Nucl. Phys. {\bf B513}, 301 (1998).

\bibitem{Blumlein:1998nv}
J.~Bl\"umlein and A.~Tkabladze,
Nucl. Phys. {\bf B553}, 427 (1999).  

\bibitem{Kretzer:2003iu}
S.~Kretzer and M.~H.~Reno,
Phys. Rev. D {\bf 69}, 034002 (2004).

\bibitem{Steffens:2012jx} 
F.~M.~Steffens, M.~D.~Brown, W.~Melnitchouk and S.~Sanches,
Phys. Rev. C {\bf 86}, 065208 (2012).

\bibitem{Ellis:1982cd} 
R.~K.~Ellis, W.~Furmanski and R.~Petronzio,
Nucl. Phys. {\bf B212}, 29 (1983).

\bibitem{Accardi:2008ne}
A.~Accardi and J.-W.~Qiu,
JHEP {\bf 0807}, 090 (2008).

\bibitem{Accardi:2009md}
A.~Accardi, T.~J.~Hobbs and W.~Melnitchouk,
JHEP {\bf 0911}, 084 (2009).

\bibitem{Accardi:2009qv} 
A.~Accardi, F.~Arleo, W.~K.~Brooks, D.~D'Enterria and V.~Muccifora,
Riv. Nuovo Cim. {\bf 32}, 439 (2010).

\bibitem{Schienbein:2007gr} 
I.~Schienbein {\it et al.},
J. Phys. G {\bf 35}, 053101 (2008).

\bibitem{Brady:2011uy}
L.~T.~Brady, A.~Accardi, T.~J.~Hobbs and W.~Melnitchouk,
Phys. Rev. D {\bf 84}, 074008 (2011)
[Erratum-ibid. D {\bf 85}, 039902 (2012)].

\bibitem{Bitar:1978cj}
K.~Bitar, P.~W.~Johnson and W.-K.~Tung,
Phys. Lett. B {\bf 83}, 114 (1979);
%
P.~W.~Johnson and \mbox{W.-K.~Tung}, 
Print-79-1018 (Illinois Tech),
{\it Contribution to Neutrino '79, Bergen, Norway} (1979).

\bibitem{Steffens:2006ds}
F.~M.~Steffens and W.~Melnitchouk,
Phys. Rev. C {\bf 73}, 055202 (2006).

\bibitem{DeRujula:1976ih}
A.~De R\'ujula, H.~Georgi and H.~D.~Politzer,
Phys. Rev. D {\bf 15}, 2495 (1977).

\bibitem{DeRujula:1976tz}
A.~De R\'ujula, H.~Georgi and H.~D.~Politzer,
Ann. Phys. {\bf 103}, 315 (1977). 

\bibitem{Virchaux:1991jc}
M.~Virchaux and A.~Milsztajn,
Phys. Lett. B {\bf 274}, 221 (1992). 

\bibitem{Alekhin:2003qq}
S.~I.~Alekhin, S.~A.~Kulagin and S.~Liuti,
Phys. Rev. D {\bf 69}, 114009 (2004).

\bibitem{Blumlein:2008kz} 
J.~Bl\"umlein and H. B\"ottcher,
Phys. Lett. B {\bf 662}, 336 (2008).

\bibitem{JR13} 
P.~Jimenez--Delgado and E.~Reya, 
in preparation.

\bibitem{AKP} 
S.~Alekhin, S.~A.~Kulagin and R.~Petti, 
Proceedings of the 15th International Workshop on Deep-Inelastic
Scattering and Related Subjects (DIS2007), 16-20 April, 2007, 
Munich, Germany.

\bibitem{Leader:2010rb}
E.~Leader, A.~V.~Sidorov and D.~B.~Stamenov,
Phys. Rev. D {\bf 82}, 114018 (2010).

\bibitem{Blumlein:2010rn}
J.~Bl\"umlein and H.~B\"ottcher,
Nucl. Phys. {\bf B841}, 205 (2010).

\bibitem{Blumlein:2012se} 
J.~Bl\"umlein and H.~B\"ottcher,
arXiv:1207.3170 [hep-ph].

\bibitem{Accardi:2009au} 
A.~Accardi, A.~Bacchetta, W.~Melnitchouk and M.~Schlegel,
JHEP {\bf 0911}, 093 (2009).

\bibitem{JAM13}
P.~Jimenez-Delgado {\it et al.} [JAM Collaboration],
in preparation (2013).

\bibitem{Ji:1993sv} 
X.~-D.~Ji and P.~Unrau,
Phys. Lett. B {\bf 333}, 228 (1994).

\bibitem{Stein:1995si} 
E.~Stein, P.~Gornicki, L.~Mankiewicz and A.~Schafer,
Phys. Lett. B {\bf 353}, 107 (1995).

\bibitem{Meziani:2004ne} 
Z.~E.~Meziani {\it et al.},
Phys. Lett. B {\bf 613}, 148 (2005).

\bibitem{Deur:2004ti} 
A.~Deur {\it et al.},
Phys. Rev. Lett. {\bf 93}, 212001 (2004).

\bibitem{Osipenko:2004xg} 
M.~Osipenko {\it et al.},
Phys. Lett. B {\bf 609}, 259 (2005).

\bibitem{Aubert:1983rq} 
J.~J.~Aubert {\it et al.} [European Muon Collaboration],
Phys. Lett. B {\bf 123}, 123 (1983).

\bibitem{Geesaman:1995yd} 
D.~F.~Geesaman, K.~Saito and A.~W.~Thomas,
Ann. Rev. Nucl. Part. Sci. {\bf 45}, 337 (1995).

\bibitem{Norton:2003cb} 
P.~R.~Norton,
Rept. Prog. Phys. {\bf 66}, 1253 (2003).

\bibitem{Melnitchouk:1993nk}
W.~Melnitchouk, A.~W.~Schreiber and A.~W.~Thomas,
Phys. Rev. D {\bf 49}, 1183 (1994). 

\bibitem{Kulagin:1994fz}
S.~A.~Kulagin, G.~Piller and W.~Weise,
Phys. Rev. C {\bf 50}, 1154 (1994).

\bibitem{Kulagin:2004ie}
S.~A.~Kulagin and R.~Petti,   
Nucl. Phys. {\bf A765}, 126 (2006).

\bibitem{Kahn:2008nq}
Y.~Kahn, W.~Melnitchouk and S.~A.~Kulagin,
Phys. Rev. C {\bf 79}, 035205 (2009).

\bibitem{Martin:2012da} 
A.~D.~Martin, A.~J.~T.~M.~Mathijssen, W.~J.~Stirling, R.~S.~Thorne, 
B.~J.~A.~Watt and G.~Watt,
Eur. Phys. J. C {\bf 73}, 2318 (2013).

\bibitem{AV18}
R.~B.~Wiringa, V.~G.~J.~Stoks and R.~Schiavilla,
Phys. Rev. C {\bf 51}, 38 (1995).

\bibitem{CDBonn}
R.~Machleidt,  
Phys. Rev. C {\bf 63}, 024001 (2001).

\bibitem{WJC}
F.~Gross and A.~Stadler,
Phys. Rev. C {\bf 78}, 014005 (2008);
{\it ibid.} C {\bf 82}, 034004 (2010).

\bibitem{Ball:2013gsa} 
R.~D.~Ball {\it et al.}, 	
Phys. Lett. B {\bf 723}, 330 (2013).

\bibitem{Arrington:2008zh}
J.~Arrington, F.~Coester, R.~J.~Holt and T.-S.~H.~Lee,
J. Phys. G {\bf 36}, 025005 (2009).

\bibitem{Arrington:2011qt}
J.~Arrington, J.~G.~Rubin and W.~Melnitchouk,
Phys. Rev. Lett. {\bf 108}, 252001 (2012).
        
\bibitem{Kuhlmann:1999sf}
S.~Kuhlmann {\it et al.},
Phys. Lett. B {\bf 476}, 291 (2000).

\bibitem{Brady:2011hb}
L.~T.~Brady, A.~Accardi, W.~Melnitchouk and J.~F.~Owens,
JHEP {\bf 1206}, 019 (2012).

\bibitem{Bazarko:1994tt}
A.~O.~Bazarko {\it et al.} [CCFR Collaboration],
Z. Phys. C {\bf 65}, 189 (1995).

\bibitem{Mason:2007zz}
D.~Mason {\it et al.} [NuTeV Collaboration],
Phys. Rev. Lett. {\bf 99}, 192001 (2007).

\bibitem{Kulagin:2007ju}
S.~A.~Kulagin and R.~Petti,
Phys. Rev. D {\bf 76}, 094023 (2007).

\bibitem{Schienbein:2009kk}
I.~Schienbein, J.~Y.~Yu, K.~Kovarik, C.~Keppel, J.~G.~Morfin, F.~Olness
and J.~F.~Owens,
Phys. Rev. D {\bf 80}, 094004 (2009).

\bibitem{Hirai:2007sx} 
M.~Hirai, S.~Kumano and T.~-H.~Nagai,
Phys. Rev. C {\bf 76}, 065207 (2007).

\bibitem{Eskola:2009uj} 
K.~J.~Eskola, H.~Paukkunen and C.~A.~Salgado,
JHEP {\bf 0904}, 065 (2009).

\bibitem{Kovarik:2010uv} 
K.~Kovarik, I.~Schienbein, F.~I.~Olness, J.~Y.~Yu, C.~Keppel,
J.~G.~Morfin, J.~F.~Owens and T.~Stavreva,
Phys. Rev. Lett. {\bf 106}, 122301 (2011).

\bibitem{Paukkunen:2013grz}
H.~Paukkunen and C.~A.~Salgado,
Phys. Rev. Lett. {\bf 110}, 212301 (2013).

\bibitem{deFlorian:2011fp} 
D.~de Florian, R.~Sassot, P.~Zurita and M.~Stratmann,
Phys. Rev. D {\bf 85}, 074028 (2012).

\bibitem{E12-06-109}
Jefferson Lab Experiment E12-06-109,	
{\it The Longitudinal Spin Structure of the Nucleon},
D.~Crabb {\it et al.}, spokespersons.

\bibitem{E12-06-110}
Jefferson Lab Experiment E12-06-110,	
{\it Measurement of the Neutron Spin Asymmetry $A_1^n$ in the Valence
Quark Region Using an 11~GeV Beam in Hall~C},
J.-P.~Chen {\it et al.}, spokespersons.

\bibitem{E12-06-122}
Jefferson Lab Experiment E12-06-122,	
{\it Measurement of the Neutron Asymmetry $A_1^n$ in the Valence Quark
Region Using 8.8 and 6.6~GeV Beam Energies and the BigBite Spectrometer
in Hall~A},
B.~Wojtsekhowski {\it et al.}, spokespersons.

\bibitem{Accardi:2012hwp} 
A.~Accardi {\it et al.},
{\it Electron Ion Collider: The Next QCD Frontier - Understanding the
glue that binds us all},
BNL-98815-2012-JA, JLAB-PHY-12-1652,
arXiv:1212.1701 [nucl-ex].

\bibitem{Bourrely:2001du} 
C.~Bourrely, J.~Soffer and F.~Buccella,
Eur. Phys. J. C {\bf 23}, 487 (2002).

\bibitem{Forte:2002fg} 
S.~Forte, L.~Garrido, J.~I.~Latorre and A.~Piccione,
JHEP {\bf 0205}, 062 (2002).

\bibitem{Honkanen:2008mb} 
H.~Honkanen, S.~Liuti, J.~Carnahan, Y.~Loitiere and P.~R.~Reynolds,
Phys. Rev. D {\bf 79}, 034022 (2009).

\bibitem{Close:1993mv} 
F.~E.~Close and R.~G.~Roberts,
Phys. Lett. B {\bf 316}, 165 (1993).

\bibitem{Pumplin:2001ct}
J.~Pumplin {\it et al.},
Phys. Rev. D {\bf 65}, 014013 (2001).

\bibitem{Stump:2001gu}
D.~Stump {\it et al.},
Phys. Rev. D {\bf 65}, 014012 (2001).

\bibitem{DelDebbio:2007ee}
L.~D.~Debbio {\it et al.},
JHEP {\bf 0703}, 039 (2007).

\bibitem{Stump:2003yu}  
D.~Stump {\it et al.},
JHEP {\bf 0310}, 046 (2003).

\bibitem{JimenezDelgado:2012zx} 
P.~Jimenez-Delgado,
Phys. Lett. B {\bf 714}, 301 (2012).

\bibitem{Pumplin:2002vw}
J.~Pumplin {\it et al.},
JHEP {\bf 0207}, 012 (2003).

\bibitem{Aaltonen:2009ta}
T.~Aaltonen {\it et al.} [CDF Collaboration], 
Phys. Rev. Lett. {\bf 102}, 181801 (2009).

\bibitem{Aad:2012sb}
G.~Aad {\it et al.} [ATLAS Collaboration],
Phys. Rev. Lett. {\bf 109}, 012001 (2012).

\bibitem{deFlorian:2005wf}
D.~de~Florian and W.~Vogelsang,
Phys. Rev. D {\bf 72}, 014014 (2005).

\bibitem{Aurenche:1998gv}
P.~Aurenche {\it et al.},
Eur. Phys. J. {\bf C9}, 107 (1999).

\bibitem{Aurenche:2006vj}
P.~Aurenche {\it et al.},
Phys. Rev. D {\bf 73}, 094007 (2006).

\bibitem{d'Enterria:2012yj}
D.~d'Enterria and J.~Rojo,
Nucl. Phys. {\bf B860}, 311 (2012).

\bibitem{Berger:1988ke} 
E.~L.~Berger and J.~-w.~Qiu,
Phys. Rev. D {\bf 40}, 778 (1989).


\bibitem{CTEQweb}
CTEQ Collaboration website,
{\tt http://www.cteq.org}.

\bibitem{Feyn72}
R.~P.~Feynman, {\em Photon Hadron Interactions}
(Benjamin, Reading, Massachusetts, 1972).
        
\bibitem{Close:1973xw}
F.~E.~Close,
Phys. Lett. B {\bf 43}, 422 (1973).

\bibitem{Melnitchouk:1995fc}
W.~Melnitchouk and A.~W.~Thomas,
Phys. Lett. B {\bf 377}, 11 (1996).

\bibitem{Holt:2010vj}
R.~J.~Holt and C.~D.~Roberts,
Rev. Mod. Phys. {\bf 82}, 2991 (2010).

\bibitem{Farrar:1975yb}
G.~R.~Farrar and D.~R.~Jackson,
Phys. Rev. Lett. {\bf 35}, 1416 (1975).

\bibitem{Blankenbecler:1974tm}
R.~Blankenbecler and S.~J.~Brodsky,
Phys. Rev. D {\bf 10}, 2973 (1974).

\bibitem{Gunion:1973nm}
J.~F.~Gunion,
Phys. Rev. D {\bf 10}, 242 (1974).

\bibitem{BrodskyLepage79}
S.~J.~Brodsky and G.~P.~Lepage,
Proceedings of the 1979 Summer Institute on Particle Physics, SLAC (1979).

\bibitem{Itow:2001ee} 
Y.~Itow {\it et al.} [T2K Collaboration],
KEK-REPORT-2001-4,
arXiv:hep-ex/0106019.

\bibitem{Ayres:2004js}
D.~S.~Ayres {\it et al.} [NOvA Collaboration],
FERMILAB-PROPOSAL-0929,
arXiv:hep-ex/0503053.

\bibitem{Raby:2008pd}
S.~Raby {\it et al.},
SLAC-PUB-14734,
arXiv:0810.4551 [hep-ph].

\bibitem{BONUS12}
Jefferson Lab Experiment E12-10-102 [BONUS12],
{\it The Structure of the Free Neutron at Large $x$-Bjorken},
S.~B\"ultmann, M.~E.~Christy, H.~Fenker, K.~Griffioen, C.~E.~Keppel,
S.~Kuhn and W.~Melnitchouk, spokespersons.

\bibitem{MARATHON}
Jefferson Lab Experiment E12-10-103 [MARATHON],
{\it Measurement of the $F_2^n/F_2^p$, $d/u$ Ratios and $A=3$ EMC Effect
in DIS off the Tritium and Helium Mirror Nuclei},
G.~G.~Petratos, J.~Gomez, R.~J.~Holt and R.~D.~Ransome,
spokespersons.

\bibitem{SOLID}
Jefferson Lab Experiment E12-10-007 [SoLID],
P.~Souder, spokesperson.

\bibitem{AbelleiraFernandez:2012ty} 
J.~L.~Abelleira Fernandez {\it et al.} [LHeC Study Group],
{\it On the Relation of the LHeC and the LHC},
arXiv:1211.5102 [hep-ex].

\bibitem{CJweb}
CTEQ-Jefferson Lab (CJ) Collaboration website,
{\tt http://www.jlab.org/cj}.

\bibitem{Drell:1970wh} 
S.~D.~Drell and T.~-M.~Yan,
Phys. Rev. Lett. {\bf 25}, 316 (1970)
[Erratum-ibid. {\bf 25}, 902 (1970)].

\bibitem{NA51}
A.~Baldit {\it et al.} [NA51 Collaboration],
Phys. Lett. B {\bf 332}, 244 (1994).

\bibitem{Hawker:1998ty}
E.~A.~Hawker {\it et al.} [FNAL E866/NuSea Collaboration],
Phys. Rev. Lett. {\bf 80}, 3715 (1998).

\bibitem{Towell:2001nh}
R.~S.~Towell {\it et al.} [FNAL E866/NuSea Collaboration],
Phys. Rev. D {\bf 64}, 052002 (2001).

\bibitem{Ross:1978xk} 
D.~A.~Ross and C.~T.~Sachrajda,
Nucl. Phys. {\bf B149}, 497 (1979).

\bibitem{Amaudruz:1991at} 
P.~Amaudruz {\it et al.} [New Muon Collaboration],
Phys. Rev. Lett. {\bf 66}, 2712 (1991).

\bibitem{Arneodo:1994sh} 
M.~Arneodo {\it et al.} [New Muon Collaboration],
Phys. Rev. D {\bf 50}, 1 (1994).

\bibitem{Gottfried:1967kk} 
K.~Gottfried,
Phys. Rev. Lett. {\bf 18}, 1174 (1967).

\bibitem{Ellis:1990ti} 
S.~D.~Ellis and W.~J.~Stirling,
Phys. Lett. B {\bf 256}, 258 (1991).

\bibitem{Field:1976ve} 
R.~D.~Field and R.~P.~Feynman,
Phys. Rev. D {\bf 15}, 2590 (1977).

\bibitem{Schreiber:1991tc} 
A.~W.~Schreiber, A.~I.~Signal and A.~W.~Thomas,
Phys. Rev. D {\bf 44}, 2653 (1991).

\bibitem{Signal:1991ug} 
A.~I.~Signal, A.~W.~Schreiber and A.~W.~Thomas,
Mod. Phys. Lett. A {\bf 6}, 271 (1991).

\bibitem{Steffens:1996bc} 
F.~M.~Steffens and A.~W.~Thomas,
Phys. Rev. C {\bf 55}, 900 (1997).

\bibitem{Thomas:1983fh} 
A.~W.~Thomas,
Phys. Lett. B {\bf 126}, 97 (1983).

\bibitem{Speth:1996pz}  
J.~Speth and A.~W.~Thomas,
Adv. Nucl. Phys. {\bf 24}, 83 (1997).

\bibitem{Kumano:1997cy}
S.~Kumano,
Phys. Rep. {\bf 303}, 183 (1998).

\bibitem{Thomas:2000ny}
A.~W.~Thomas, W.~Melnitchouk and F.~M.~Steffens,
Phys. Rev. Lett. {\bf 85}, 2892 (2000).

\bibitem{Burkardt:2012hk}
M.~Burkardt, K.~S.~Hendricks, C.~-R.~Ji, W.~Melnitchouk and A.~W.~Thomas,
Phys. Rev. D {\bf 87}, 056009 (2013).

\bibitem{SeaQuest}
Fermilab E906 experiment,
{\it Drell-Yan Measurements of Nucleon and Nuclear Structure with the
Fermilab Main Injector},
D.~F.~Geesaman and P.~E.~Reimer, spokespersons;
{\tt http://www.phy.anl.gov/mep/SeaQuest/index.html}.

\bibitem{J-PARC-P04}
J-PARC proposal P04,
{\it Measurement of High-Mass Dimuon Production at the 50-GeV Proton
Synchrotron},
J.~C.~Peng and S.~Sawada spokespersons;
{\tt http://j-parc.jp/index-e.html}.

\bibitem{Kumano:2010qm} 
S.~Kumano,
J. Phys. Conf. Ser. {\bf 312}, 032005 (2011).

\bibitem{Sather:1991je} 
E.~Sather,
Phys. Lett. B {\bf 274}, 433 (1992).

\bibitem{Rodionov:1994cg} 
E.~N.~Rodionov, A.~W.~Thomas and J.~T.~Londergan,
Mod. Phys. Lett. A {\bf 9}, 1799 (1994).

\bibitem{Londergan:2003pq}
J.~T.~Londergan and A.~W.~Thomas,
Phys. Lett. B {\bf 558}, 132 (2003).

\bibitem{Londergan:2009kj} 
J.~T.~Londergan, J.~C.~Peng and A.~W.~Thomas,
Rev. Mod. Phys. {\bf 82}, 2009 (2010).

\bibitem{Gluck:2005xh} 
M.~Gluck, P.~Jimenez-Delgado and E.~Reya,
Phys. Rev. Lett. {\bf 95}, 022002 (2005).
 
\bibitem{Martin:2004dh}
A.~D.~Martin, R.~G.~Roberts, W.~J.~Stirling and R.~S.~Thorne,
Eur. Phys. J. C {\bf 39}, 155 (2005).

\bibitem{Zeller:2001hh} 
G.~P.~Zeller {\it et al.} [NuTeV Collaboration],
Phys. Rev. Lett. {\bf 88}, 091802 (2002)
[Erratum-ibid. {\bf 90}, 239902 (2003)].

\bibitem{Paschos:1972kj} 
E.~A.~Paschos and L.~Wolfenstein,
Phys. Rev. D {\bf 7}, 91 (1973).

\bibitem{JimenezDelgado:2010pc} 
P.~Jimenez-Delgado,
Phys. Lett. B {\bf 689}, 177 (2010).

\bibitem{Signal:1987gz}
A.~I.~Signal and A.~W.~Thomas,
Phys. Lett. B {\bf 191}, 205 (1987).

\bibitem{Burkardt:1991di} 
M.~Burkardt and B.~Warr,
Phys. Rev. D {\bf 45}, 958 (1992).

\bibitem{Melnitchouk:1996fj}
W.~Melnitchouk and M.~Malheiro,
Phys. Rev. C {\bf 55}, 431 (1997).

\bibitem{Alwall:2004rd}
J.~Alwall and G.~Ingelman,
Phys. Rev. D {\bf 70}, 111505 (2004).

\bibitem{Catani:2004nc} 
S.~Catani, D.~de Florian, G.~Rodrigo and W.~Vogelsang,
Phys. Rev. Lett. {\bf 93}, 152003 (2004).

\bibitem{Gottschalk:1980rv}
T.~Gottschalk,
Phys. Rev. D {\bf 23}, 56 (1981).

\bibitem{Gluck:1996ve}
M.~Gl\"uck, S.~Kretzer and E.~Reya,
Phys. Lett. B {\bf 380}, 171 (1996)
[Erratum-ibid. B {\bf 405}, 391 (1997)].

\bibitem{Kretzer:2001tc}
S.~Kretzer, D.~Mason and F.~Olness,
Phys. Rev. D {\bf 65}, 074010 (2002).

\bibitem{Olness:2003wz} 
F.~Olness {\it et al.},
Eur. Phys. J. C {\bf 40}, 145 (2005).

\bibitem{deFlorian:2003qf}
D.~de Florian and R.~Sassot,
Phys. Rev. D {\bf 69}, 074028 (2004).

\bibitem{Harris:1997zq}
B.~W.~Harris and J.~Smith,
Phys. Rev. D {\bf 57}, 2806 (1998).

\bibitem{Brodsky:1980pb}
S.~J.~Brodsky, P.~Hoyer, C.~Peterson and N.~Sakai,
Phys. Lett. B {\bf 93}, 451 (1980).

\bibitem{Vogt:1995dn} 
R.~Vogt,
Nucl. Phys. {\bf B446}, 159 (1995).

\bibitem{Navarra:1995rq}
F.~S.~Navarra, M.~Nielsen, C.~A.~A.~Nunes and M.~Teixeira,  
Phys. Rev. D {\bf 54}, 842 (1996).   

\bibitem{Vogt:2000sk}
R.~Vogt,  
Prog. Part. Nucl. Phys. {\bf 45}, S105 (2000).        
  
\bibitem{Pumplin:2005yf}
J.~Pumplin,
Phys. Rev. D {\bf 73}, 114015 (2006).

\bibitem{Aubert:1981ix}
J.~J.~Aubert {\it et al.} [European Muon Collaboration],  
Phys. Lett. B {\bf 110}, 73 (1982).    
  
\bibitem{Aubert:1982tt}
J.~J.~Aubert {\it et al.} [European Muon Collaboration],  
Nucl. Phys. {\bf B213}, 31 (1983).     
  
\bibitem{Harris:1995jx} 
B.~W.~Harris, J.~Smith and R.~Vogt,
Nucl. Phys. {\bf B461}, 181 (1996).

\bibitem{Hobbs:2013}
T.~J.~Hobbs, J.~T.~Londergan and W.~Melnitchouk,
in preparation.

\bibitem{J-PARC13}
Workshop on Future Prospects of Hadron Physics at J-PARC
and Large Scale Computational Physics in 2013,
{\tt http://j-parc-th.kek.jp/html/English/JPARCFeb2013.html}.

\bibitem{Riedl:2007sv}
J.~Riedl, A.~Sch\"afer and M.~Stratmann,
Eur. Phys. J. C {\bf 52}, 987 (2007).

\bibitem{PR12-07-106}
Jefferson Lab proposal PR12-07-106,
{\it The $A$-dependence of $J/\psi$ photoproduction near threshold},
E.~Chudakov {\it et al.}, spokespersons.

\bibitem{Brambilla:2010cs}
N.~Brambilla {\it et al.},
Eur. Phys. J. C {\bf 71}, 1534 (2011).

\bibitem{Brodsky:2012vg}
S.~J.~Brodsky, F.~Fleuret, C.~Hadjidakis and J.~P.~Lansberg,
Phys. Rep. {\bf 522}, 239 (2013).

\bibitem{Monaghan:2012et} 
P.~Monaghan, A.~Accardi, M.~E.~Christy, C.~E.~Keppel, W.~Melnitchouk and 
L.~Zhu,
Phys. Rev. Lett. {\bf 110}, 152002 (2013).

\bibitem{Berger98:lpp}
E.~L.~Berger, L.~E.~Gordon and M.~Klasen,
Phys. Rev. D {\bf 58}, 074012 (1998).

\bibitem{Carminati:2013}
L.~Carminati {\it et al.},
Europhys. Lett. {\bf 101}, 61002 (2013),

\bibitem{Czakon:2013}
M.~Czakon {\it et al.},
arXiv:1303.7215 [hep-ph].

\bibitem{Mangano:2013er}
M.~Mangano and J.~Rojo,
JHEP {\bf 1208}, 010 (2012).

\bibitem{Ridder:2013mf} 
A.~G.-D.~Ridder, T.~Gehrmann, E.~W.~N.~Glover and J.~Pires,
Phys. Rev. Lett. {\bf 110}, 162003 (2013).

\bibitem{JimenezDelgado:2008hf}		
P.~Jimenez-Delgado and E.~Reya,	
Phys. Rev. D {\bf 79}, 074023 (2009).

\bibitem{Detmold:2001dv} 
W.~Detmold, W.~Melnitchouk and A.~W.~Thomas,
Eur. Phys. J. direct C {\bf 3}, 1 (2001).

\bibitem{Bratt:2010jn}
J.~D.~Bratt {\it et al.} [LHP Collaboration],
Phys. Rev. D {\bf 82}, 094502 (2010).

\bibitem{Pleiter:2011gw} 
D.~Pleiter {\it et al.} [QCDSF/UKQCD Collaboration],
PoS LATTICE {\bf 2010}, 153 (2010).

\bibitem{Detmold:2001jb} 
W.~Detmold, W.~Melnitchouk, J.~W.~Negele, D.~B.~Renner and A.~W.~Thomas,
Phys. Rev. Lett. {\bf 87}, 172001 (2001).

\bibitem{Procura:2006gq} 
M.~Procura, B.~U.~Musch, T.~R.~Hemmert and W.~Weise,
Phys. Rev. D {\bf 75}, 014503 (2007).

\bibitem{Shanahan:2013xw} 
P.~E.~Shanahan, A.~W.~Thomas and R.~D.~Young,
arXiv:1301.6861 [nucl-th].

\bibitem{Detmold:2002nf}
W.~Detmold, W.~Melnitchouk and A.~W.~Thomas,
Phys. Rev. D {\bf 66}, 054501 (2002).


\bibitem{Ashman:1989ig} 
J.~Ashman {\it et al.} [European Muon Collaboration],
Phys. Lett. B {\bf 206}, 364 (1988);
Nucl. Phys. {\bf B328}, 1 (1989).

\bibitem{SMC}
B.~Adeva {\it et al.} [Spin Muon Collaboration],
Phys. Rev. D {\bf 58}, 112001 (1998);
Phys. Rev. D {\bf 60}, 072004 (1999)
[Erratum-ibid. D {\bf 62}, 079902 (2000)].

\bibitem{COMPASS07}
V.~Yu.~Alexakhin {\it et al.} [COMPASS Collaboration],
Phys. Lett. B {\bf 647}, 8 (2007).

\bibitem{COMPASS10}
M.~G.~Alekseev {\it et al.} [COMPASS Collaboration],
Phys. Lett. B {\bf 690}, 466 (2010).

\bibitem{Adeva:1997qz} 
B.~Adeva {\it et al.}  [Spin Muon Collaboration],
Phys. Lett. B {\bf 420}, 180 (1998).

\bibitem{Alekseev:2010ub} 
M.~G.~Alekseev {\it et al.} [COMPASS Collaboration],
Phys. Lett. B {\bf 693}, 227 (2010).

\bibitem{SLAC:E80/E130}
G.~Baum {\it et al.},
Phys. Rev. Lett. {\bf 51}, 1135 (1983).

\bibitem{SLAC:E142}
P.~L.~Anthony {\it et al.} [SLAC E142 Collaboration],
Phys. Rev. D {\bf 54}, 6620 (1996).

\bibitem{SLAC:E154}
K.~Abe {\it et al.} [SLAC E154 Collaboration],
Phys. Rev. Lett. {\bf 79}, 26 (1997).

\bibitem{SLAC:E143}
K.~Abe {\it et al.} [SLAC E143 Collaboration],
Phys. Rev. D {\bf 58}, 112003 (1998).

\bibitem{SLAC:E155}
P.~L.~Anthony {\it et al.} [SLAC E155 Collaboration],
Phys. Lett. B {\bf 458}, 529 (1999);
Phys. Lett. B {\bf 463}, 339 (1999);
Phys. Lett. B {\bf 493}, 19 (2000).

\bibitem{SLAC:E155x}
P.~L.~Anthony {\it et al.} [SLAC E155x Collaboration],
Phys. Lett. B {\bf 553}, 18 (2003).

\bibitem{HERMES97}
K.~Ackerstaff {\it et al.} [HERMES Collaboration],
Phys. Lett. B {\bf 404}, 383 (1997).

\bibitem{HERMES07}
A.~Airapetian {\it et al.} [HERMES Collaboration],
Phys. Rev. D {\bf 75}, 012007 (2007).

\bibitem{HERMES12}
A.~Airepetian {\it et al.} [HERMES Collaboration],
Eur. Phys. J. C {\bf 72}, 1921 (2012).

\bibitem{Airapetian:2004zf} 
A.~Airapetian {\it et al.}  [HERMES Collaboration],
Phys. Rev. D {\bf 71}, 012003 (2005).

\bibitem{Airapetian:2008qf} 
A.~Airapetian {\it et al.}  [HERMES Collaboration],  
Phys. Lett. B {\bf 666}, 446 (2008).

\bibitem{E99-117}
X.~Zhang {\it et al.} [Jefferson Lab E99-117 Collaboration],
Phys. Rev. Lett. {\bf 92}, 012004 (2004);
Phys. Rev. C {\bf 70}, 065207 (2004).

\bibitem{E97-103}
K.~Kramer {\it et al.} [Jefferson Lab E97-103 Collaboration],
Phys. Rev. Lett. {\bf 95}, 142002 (2005).

\bibitem{E01-012}
P.~Solvignon {\it et al.} [Jefferson Lab E01-012 Collaboration],
Phys. Rev. Lett. {\bf 101}, 182502 (2008);
arXiv:1304.4497 [nucl-ex].

\bibitem{EG1a}
K.~V.~Dharmwardane {\it et al.} [CLAS Collaboration],
Phys. Lett. B {\bf 641}, 11 (2006).

\bibitem{EG1b}
Y.~Prok {\it et al.} [CLAS Collaboration],
Phys. Lett. B {\bf 672}, 12 (2009).

\bibitem{Adams:1994bg} 
D.~L.~Adams {\it et al.} [FNAL E581/704 Collaboration],
Phys. Lett. B {\bf 336}, 269 (1994).

\bibitem{Adare:2008aa} 		
A.~Adare {\it et al.} [PHENIX Collaboration],
Phys. Rev. Lett. {\bf 103}, 012003 (2009).

\bibitem{Manion:2011zz} 
A.~Manion [PHENIX Collaboration],
J. Phys. Conf. Ser. {\bf 295}, 012070 (2011).

\bibitem{Aggarwal:2010vc}	
M.~M.~Aggarwal {\it et al.} [STAR Collaboration],
Phys. Rev. Lett. {\bf 106}, 062002 (2011).

\bibitem{Adamczyk:2012qj} 	
L.~Adamczyk {\it et al.} [STAR Collaboration],
Phys. Rev. D {\bf 86}, 032006 (2012).

\bibitem{collaboration:2011fga} 
P.~Djawotho [STAR Collaboration],
XIX International Workshop on Deep-Inelastic Scattering and Related 
Subjects (DIS 2011),
arXiv:1106.5769 [nucl-ex].

\bibitem{deFlorian:2009vb}
D.~de Florian, R.~Sassot, M.~Stratmann and W.~Vogelsang,
Phys. Rev. D {\bf 80}, 034030 (2009).

\bibitem{Hirai:2008aj}
M.~Hirai and S.~Kumano,
Nucl. Phys. {\bf B813}, 106 (2009).

\bibitem{Ball:2013lla}
R.~D.~Ball {\it et al.},
arXiv:1303.7236 [hep-ph].

\bibitem{Adolph:2012ca} 
C.~Adolph {\it et al.} [COMPASS Collaboration],
Phys. Rev. D {\bf 87}, 052018 (2013).

\bibitem{Arrington:2006zm} 
J.~Arrington, C.~D.~Roberts and J.~M.~Zanotti,
J. Phys. G {\bf 34}, S23 (2007).

\bibitem{Melnitchouk:2001eh} 
W.~Melnitchouk,
Phys. Rev. Lett. {\bf 86}, 35 (2001)
[Erratum-ibid. {\bf 93}, 199901 (2004)].

\bibitem{Bloom:1970xb} 
E.~D.~Bloom and F.~J.~Gilman,
Phys. Rev. Lett. {\bf 25}, 1140 (1970).

\bibitem{Avakian:2007xa} 
H.~Avakian, S.~J.~Brodsky, A.~Deur and F.~Yuan,
Phys. Rev. Lett. {\bf 99}, 082001 (2007)

\bibitem{E12-06-121}
Jefferson Lab Experiment E12-06-121,	
{\it A Path to Color Polarizabilities},
Z.-E.~Meziani {\it et al.}, spokespersons.

\bibitem{Wakamatsu:1991yu} 
M.~Wakamatsu,
Phys. Rev. D {\bf 44}, 2631 (1991).

\bibitem{Pobylitsa:1998tk} 
P.~V.~Pobylitsa, M.~V.~Polyakov, K.~Goeke, T.~Watabe and C.~Weiss,
Phys. Rev. D {\bf 59}, 034024 (1999).

\bibitem{Bourrely:1993dd} 
C.~Bourrely and J.~Soffer,
Phys. Lett. B {\bf 314}, 132 (1993).

\bibitem{deFlorian:2010aa} 
D.~de Florian and W.~Vogelsang,
Phys. Rev. D {\bf 81}, 094020 (2010).

\bibitem{Aschenauer:2013woa} 
E.~C.~Aschenauer {\it et al.},
arXiv:1304.0079 [nucl-ex].

\bibitem{Walker:2011vs} 
M.~Walker [STAR Collaboration],
XIX International Workshop on Deep-Inelastic Scattering and Related 
Subjects (DIS 2011),
arXiv:1107.0917 [hep-ex].


\bibitem{Airapetian:2013}
A.~Airapetian {\it et al.} [HERMES Collaboration],
in print;
H.~Jackson, talk given at the {\it XXI International Workshop on 
Deep--Inelastic Scattering and Related Subjects (DIS2013)},
Marseilles, France, April 22-26, 2013.

\bibitem{E12-09-002}
Jefferson Lab Experiment E12-09-002,	
{\it Precise Measurement of $\pi^+/\pi^-$ Ratios Semi-Inclusive DIS:
Charge Symmetry Violating Quark Distributions},
K.~Hafidi {\it et al.}, spokespersons.

\bibitem{Prescott:1978tm} 
C.~Y.~Prescott {\it et al.},
Phys. Lett. B {\bf 77}, 347 (1978).

\bibitem{Wang:2013kkc} 
D.~Wang {\it et al.} [Jefferson Lab Hall A Collaboration],
arXiv:1304.7741 [nucl-ex].

\bibitem{Bjorken:1978ry} 
J.~D.~Bjorken,
Phys. Rev. D {\bf 18}, 3239 (1978).

\bibitem{Hobbs:2008mm} 
T.~Hobbs and W.~Melnitchouk,
Phys. Rev. D {\bf 77}, 114023 (2008).

\bibitem{Mantry:2010ki} 
S.~Mantry, M.~J.~Ramsey-Musolf and G.~F.~Sacco,
Phys. Rev. C {\bf 82}, 065205 (2010).

\bibitem{Bednyakov:2013zta} 
V.~A.~Bednyakov, M.~A.~Demichev, G.~I.~Lykasov, T.~Stavreva and 
M.~Stockton,
arXiv:1305.3548 [hep-ph].

\bibitem{Ji:2013dva} 
X.~Ji,
Phys. Rev. Lett. {\bf 110}, 262002 (2013).

\end{thebibliography}
\end{document}